\documentclass[12pt]{article}
\pdfoutput=1
\usepackage{float}
\usepackage[final]{graphicx}
\usepackage{amssymb,amsmath}
\numberwithin{equation}{section}
\usepackage{commath}
\usepackage{epsfig}
\usepackage{epstopdf}
\usepackage{latexsym}
\usepackage{graphicx}
\usepackage{subfigure}
\usepackage{booktabs}
\usepackage{hyperref}
\usepackage{tkz-euclide}

\usepackage{makeidx}
\usepackage{cite}
\usepackage{bm}
\usepackage{geometry}
\usepackage{braket}
\usepackage[margin=20pt,small]{caption}

\usepackage[toc]{appendix}

\usepackage{tikz}
\usetikzlibrary{calc,decorations.markings,arrows.meta,shapes.misc,decorations.pathmorphing,calc,bending}

\usetikzlibrary{arrows.meta,shapes.misc,decorations.pathmorphing,calc,bending}

\tikzset{
  branch point/.style={cross out,draw=black,fill=none,minimum size=2*(#1-\pgflinewidth),inner sep=0pt,outer sep=0pt}, 
  branch point/.default=5
}
\tikzset{
  branch cut/.style={
    decorate,decoration=snake,
    to path={
      (\tikztostart) -- (\tikztotarget) \tikztonodes
    },
    }
  }

\usepackage{color}
\usepackage{datetime}
  
\ifpdf
\fi

\definecolor{cardinal}{rgb}{0.6,0,0}
\definecolor{darkgreen}{rgb}{0,0.5,0}
\definecolor{golden}{rgb}{0.92, 0.7, 0}
\definecolor{midnight}{rgb}{0, 0, 0.5}
\definecolor{darkblue}{rgb}{0.2, 0, 0.8}

\newcommand{\be}{\begin{equation}}
\newcommand{\ee}{\end{equation}}
\newcommand{\bea}{\begin{eqnarray}}
\newcommand{\eea}{\end{eqnarray}}

\begin{document}

\begin{titlepage}

\bigskip
\bigskip
\bigskip
\centerline{{\bf Modular quantization and black holes}}

\bigskip
\begin{center}
\ Suchetan Das\footnote{\texttt{suchetan1993@gmail.com }} 

\end{center}

\renewcommand{\thefootnote}{\arabic{footnote}}

\begin{center}
{School of Physical Sciences, Indian Association for the Cultivation of Science,\\
2A and 2B Raja S. C. Mullick Road, Jadavpur Kolkata-700032, India.}


\end{center}
\bigskip
\bigskip
\vspace{-0.20in}
\begin{abstract}

\noindent 
Witten recently proposed a background-independent algebraic framework for quantum gravity, wherein an observer endowed with a Hamiltonian defines a diffeomorphism invariant worldline algebra manifested by the modified Hamiltonian constraint. In the limit of decoupling gravity, this construction admits a lift to a von Neumann algebra acting on a Hilbert space defined by geodesic in a fixed background. Motivated by this, we revisit quantization of certain class of SL(2,$\mathbb{R})$ deformed CFT Hamiltonian on a cylinder to capture non-perturbative aspects of black holes. We construct a type-I Von-Neuman algebra by imposing conformal boundary conditions on cut-offs near \textit{fixed points} of Hamiltonian flow, acting on a GNS Hilbert space built from highest-weight representation of \textit{`emergent modular Virasoro algebra'}. Upon identifying the Hamiltonian with the modular Hamiltonian of a sharp subregion associated to a fixed reference KMS (conformal vacuum) state, the algebra changes to type-III$_{1}$ factor. We also discuss the structure of \textit{emergent} Hilbert spaces using `open-closed string' duality after incorporating an \textit{emergent non-trivial center} made out of scalars at fixed points. We further employ this \textit{modular quantization} of a \textit{single holographic CFT} to demonstrate how the boundary limit of \textit{exact}  Hartle-Hawking correlator of \textit{smooth} BTZ background emerge in the strict semiclassical limit in an alternative dual description, while at finite $G_{N}$, the corresponding description is intrinsically non-smooth, featuring both a stretched horizon and a boundary cutoff. The exact same boundary correlator has also been reproduced from the thermal correlators in modular quantization. We further discuss the effect of incorporating gravity by including the center via AdS/CFT on boundary correlators and the fate of the smooth horizon in the context of preserving unitarity. The description of a smooth horizon is replaced by a (stretched) horizon containing explicit microstructures embedded within it. In this description, unitarity is restored without any `wormhole' construction or ensemble averaging.
\end{abstract}

\newpage


\end{titlepage}
\tableofcontents


\rule{\textwidth}{.5pt}\\

\section{Introduction}\label{sec1}
It is usually believed in a theory of quantum gravity, the fundamental constants of nature is \textit{not} random. By definition, \textit{at least} such theory must possess a semiclassical limit in which low-energy effective field theory (EFT) on a curved background emerges, where the background is a solution of an appropriate low-energy effective theory of gravity. However, in the presence of black hole horizon, the low energy EFT is known to exhibit unusual paradoxes challenging Unitarity \cite{Hawking:1976ra}-\cite{Harlow:2014yka}. While the black hole information paradox was originally discussed in the context of evaporating black holes, Maldacena recast and sharpened the problem for eternal AdS black holes within the framework of the AdS/CFT correspondence \cite{Maldacena:2001kr}, where the dual unitary CFT provides a UV-complete description of quantum gravity in AdS spacetime. Furthermore, Mathur's `small correction theorem' \cite{Mathur:2009hf} indicated that the failure of the unitary EFT description near the black hole horizon cannot be resolved through small perturbative corrections in $G_{N}$. This line of reasoning subsequently led to the formulation of the `firewall paradox' \footnote{The concerns regarding Hilbert space factorization to use entanglement inequality, can be avoided while working purely on EFT framework \cite{Witten:2018zxz}, \cite{Witten:2021jzq}.} \cite{Almheiri:2012rt}, \cite{Almheiri:2013hfa}. In this fuzzball/firewall paradigm, the central theme is to replace the `smooth' nature of semiclassical horizon to restore unitarity.

Over the past decade, remarkable progress has been achieved through the study of Euclidean gravitational path integrals \cite{Gibbons:1976ue}- \cite{Geng:2024xpj}. In particular, certain Euclidean wormhole saddles have played a central role in reproducing the Page curve via the quantum extremal surface (QES) prescription \cite{Engelhardt:2014gca}. Despite their striking success, the precise physical origin and interpretation of these wormholes remain poorly understood within the Euclidean gravitational path integral framework itself\footnote{See also the related challenges in the set-up while attaching a non-gravitational bath \cite{Geng:2020qvw}-\cite{Geng:2023zhq}, \cite{Das:2025cuq}.}. Furthermore, the appearance of such wormholes seems to be in tension with expectations from fundamental theories of quantum gravity, such as string theory and holographic AdS/CFT, where the microscopic couplings are fixed rather than drawn from an ensemble of random theories \footnote{See some explicit construction of replica wormholes in the ensemble of CFTs \cite{Geng:2025efs},\cite{Geng:2025efs}.}. However, the underlying theme of such proposals may also be understood through the lens of ER=EPR \cite{Maldacena:2013xja}.

On the other hand, within the recent progress of algebraic formulation of AdS/CFT, the issue has been addressed by the emergence of type-III algebra by Leutheusser and Liu \cite{Leutheusser:2021qhd}, \cite{Leutheusser:2021frk}. Subsequently, Witten recast how the algebra can be changed from type-III to type-II by incorporating \textit{gravity} in the algebra\footnote{See also related discussion on gravitational dressing and holography of information \cite{Laddha:2020kvp}, \cite{Raju:2024gvc}}. One of the major success of such algebraic formulations is the derivation of generalized entropy \footnote{Even though, it is not the exact QES formula involving extremization of entropy.} purely from type-II algebra even in the absence of pure states \cite{Witten:2021unn}- \cite{Chandrasekaran:2022eqq}. From this line of thoughts, a proposal of `background independent algebra of quantum gravity' has been made by Witten \cite{Witten:2023xze}, where an `observer' plays a central role to define the algebra. The central proposal proceeds in two stages. The first stage is to associate a set of degrees of freedom with an `observer', described by a Hamiltonian $H_{obs}$ in the full theory, such that the gravitational constraints are modified by considering the total Hamiltonian
$H_{tot} = H_{obs} + H_{bulk}$.
As argued in \cite{Witten:2023xze}, this construction allows one to effectively describe gravitational dynamics by dressing operators to the observer's worldline determined by $H_{obs}$. At this stage, the resulting operator algebra is defined without reference to any particular state or background and may therefore be regarded as \textit{background independent}. This is exactly related to the study of `quantum reference frame' in the context of quantum gravity\cite{DeVuyst:2024khu}-\cite{Wei:2025guh}\footnote{See also discussion on observer from the point of view of spontaneous symmetry breaking \cite{Geng:2024dbl}-\cite{Geng:2025gqu}.}. The main outline is:

\textit{Background independence $\implies$ `observer' dependence}.

The second stage involves selecting a specific background geometry together with a corresponding timelike geodesic, which may be viewed as a decoupling limit of gravity or the strict semiclassical limit $G_{N} \rightarrow 0$. In this limit, the algebra acts on a Hilbert space associated with the chosen background and geodesic. One may think of this procedure as lifting the original background-independent algebra to a von Neumann algebra acting on a reference state. Although the motivation for introducing such observer-dependent algebras originally arose from the problem of defining operator algebras in closed universes such as de Sitter space \cite{Chandrasekaran:2022cip}, the same framework can also be applied to asymptotically flat and asymptotically AdS spacetimes including subregions \cite{Leutheusser:2022bgi}-\cite{Kudler-Flam:2023qfl}.

An interesting feature of asymptotically AdS(AAdS) spacetimes with boundaries is that the observer can naturally be identified with an external observer located at the asymptotic boundary. Within the AdS/CFT correspondence, this perspective is particularly transparent because $H_{bulk}$ acquires a boundary contribution, $H_{ADM}$, which is dual to the CFT Hamiltonian $H_{CFT}$. Consequently, the entire observer-based construction can be formulated directly in terms of the Hamiltonian structure of the dual CFT. In particular, the boundary $H_{CFT}$ determines the notion of bulk time evolution. Consequently, different asymptotically AdS (AAdS) geometries are naturally associated with different notions of time set by ADM clock $H_{ADM}$. Hence in this case, we have $H_{tot}=H_{bulk}+H_{ADM}=H_{bulk}+H_{CFT}$.

A possible way to reinterpret this observer-dependent, or equivalently background-independent framework within AdS/CFT, is through the language of Floquet CFTs \cite{Wen:2018agb}-\cite{Gill:2026wcj}. The original Floquet construction concerns a periodically driven CFT examined in stroboscopic time. In this setting, a time-dependent out-of-equilibrium system can be reformulated in terms of a Floquet Hamiltonian $H_{F}$, which is effectively time independent when viewed stroboscopically. A notable feature of Floquet Hamiltonians is their rich landscape of dynamical phases, which can be accessed by varying the parameters of the driving protocol.

In \cite{Han:2020kwp}, it was shown that a class of SL(2,$\mathbb{R}$) deformed Floquet Hamiltonians exhibits three distinct dynamical phases. Upon choosing the conformal vacuum as the reference state, these phases were found to correspond to an analogue of the Hawking–Page transition, relating to three different bulk AdS${_3}$ geometries \cite{Das:2022pez}, \cite{Das:2024vqe}. This observation suggests a broader possibility: a sufficiently general Floquet Hamiltonian $H_{F}$ may provide a framework for generating the notion of time associated with arbitrary large diffeomorphisms in asymptotically AdS$_3$ spacetimes \cite{Das:2024mlx},\cite{Das:2018ojl} which is consistent with the observer proposal. In this way, we may identify $H_{ADM}=H_{CFT}=H_{F}$.

Following this line of reasoning, the natural CFT Hamiltonian associated with the BTZ black hole should be identified with the modular Hamiltonian $H_{CFT}=H_{mod}=H_{R}-H_{L}$, as argued in \cite{Das:2024mlx}. In the Floquet-CFT framework, this Hamiltonian lies within the so-called \textit{heating phase}. Motivated by this connection, and by the desire to better understand certain aspects of black-hole unitarity paradoxes that can already be explored within the BTZ setting, we revisit the quantization of the modular Hamiltonian in this note. However, as discussed above, selecting a particular background amounts to restricting $H_{F}$ to a specific dynamical phase by tuning external parameters such as the driving frequency and amplitude. This is the essence of observer dependent mechanism, which of course provides the background independence in the sense it is argued.

The main point we wish to emphasize, however, is that in the context of quantization\footnote{See also \cite{Chikazawa:2026iro}, \cite{Okunishi:2019dmv}for different way of Rindler quantization with cut-off.}, the transition to a fixed background exhibits \textit{additional structure}. In \cite{Das:2024mlx}, following the approach of Tada \cite{Tada:2019rls}, we argued that the quantization of the modular Hamiltonian requires the introduction of conformal cutoffs around the fixed points of the modular flow. Once these cutoffs are implemented, one obtains a single copy of an \textit{emergent Virasoro algebra}, which in turn defines a highest-weight representation. We shall revisit this construction in the present note and show that this highest-weight representation is isomorphic to a GNS representation generated by the action of local operators on its ground state representation.

From this perspective, one may view a BCFT observer equipped with a one-sided cutoff Hamiltonian $H^{R,L}_{\epsilon}$ as defining a type-I von Neumann algebra acting on the corresponding highest-weight representation. In the limit where the cutoff is removed, the algebra becomes a type-III$_1$ factor \cite{Das:2024vqe} acting on the conformal vacuum $|0\rangle$ and the total Hamiltonian is $H_{mod}=\lim_{\epsilon \rightarrow 0}(H^{R}_{\epsilon}-H^{L}_{\epsilon})$. This limiting procedure is precisely analogous to selecting a fixed background for the modular Hamiltonian relative to a chosen reference state. In this way, the construction parallels Witten's notion of a background-independent algebra within CFT.

As we shall see, the novel feature of this framework is the emergence of a \textit{center} in the same limit. Enlarging the algebra to include this center leads to a new Hilbert-space structure. In fact, the resulting construction admits two isomorphic Hilbert-space realizations corresponding to two observers via an `open-closed string' duality, whose relation we will describe in detail. Correspondingly, the Lorentzian description is precisely a two-sided `Rindler-like' geometry, while adding the center amounts to detect a non-smooth (stretched) horizon capped-off just outside the horizon.

To understand the corresponding bulk interpretation, we employ the AdS/CFT correspondence at the level of the identification
$H_{ADM}=H_{mod}=H_{obs}$.
In parallel with boundary modular quantization, we adopt a \textit{non-standard} bottom-up description of the BTZ black hole background following \cite{Burman:2023kko}, \cite{Burman:2024egy}, \cite{Das:2024mlx} incorporating both a stretched horizon and a boundary cutoff. The unfixed parameters of the construction, including the location of the stretched horizon, are chosen such that the resulting framework provides a \textit{non-perturbative} description, guided by Witten's proposal of a background-independent algebra.

Following the modular quantization program, we consider a bulk massless scalar field in the Dirichlet stretched-horizon background, precisely the same setup analyzed in detail in \cite{Burman:2023kko}. After constructing the BTZ microstates from the scalar degrees of freedom in the Dirichlet background, we demonstrate that the corresponding microstate correlators, in the boundary limit, reproduce exactly the boundary Hartle-Hawking correlator, following the same line of reasoning advocated in \cite{Burman:2023kko}. This result emerges in a bulk \textit{semiclassical limit}, defined by taking the stretched horizon arbitrarily close to the event horizon. In this limit, the resulting correlator can be identified with the boundary correlator evaluated in the conformal vacuum state, whose contribution dominates the partition function, as we shall argue. Consequently, boundary limit of Hartle-Hawking correlators of \textit{smooth} BTZ background at the exact semiclassical limit arises naturally from this non-standard bulk construction, in close analogy with the boundary modular quantization framework.

We also comment on the next step of incorporating gravity into this picture, following \cite{Witten:2021unn}\footnote{See also \cite{Sorce:2023fdx}, \cite{Liu:2025krl} for a review.}. Through the AdS/CFT correspondence \cite{Maldacena:1997re}, this amounts to including the eigenmodes of the modular Hamiltonian, or equivalently the center of the algebra in the context of modular quantization. Using the enlarged Hilbert spaces obtained after the inclusion of the center, we show that certain heavy states exhibit \textit{quasi-periodic behavior} in time in the strict large-$c$ limit using a semiclassical heavy-heavy-light-light(HHLL) Virasoro identity block dominance \cite{Fitzpatrick:2014vua}. From the Lorentzian boundary perspective, the inclusion of the center in the algebra can be interpreted as introducing microstructure at the (stretched) `Rindler' horizon. In the bulk description, it corresponds to taking all possible admissible boundary conditions in the stretched horizon framework, to find all these emergent structures at the (stretched) horizon in the semiclassical limit\footnote{From holographic perspective, we must expect that smooth limits corresponding to all such states do exist. See, however \cite{Krishnan:2026mpa}, \cite{Gayari:2026jwn} for related comments.}. Equivalently, in the Dirichlet stretched horizon picture, this \textit{may} correspond to a regime in which the effects of the stretched horizon do not vanish exactly, or equivalently, a limit in which $G_N$ is very small but remains finite. From this viewpoint, the preservation of unitarity becomes manifest in the bulk after incorporating first order non-perturbative corrections of
$e^{- \frac{l_{\mathrm{AdS}}}{G_N}}$ \cite{Burman:2023kko}. In a related context, in \cite{Das:2025cuq} with Dinda, we have constructed a two-dimensional CFT coupled to JT gravity set-up, inspired from island paradigm, where a unitary `Page curve' has been obtained in which the `Page time' is determined by conformal stretched horizon cut-off location for one-sided AdS$_{2}$ black hole. This also makes a parallel to the firewall description as proposed in \cite{Almheiri:2012rt}.  

Nevertheless, we would like to emphasize that our results, which are motivated by Witten's proposal of a background-independent algebra in AdS/CFT, bear a close conceptual as well as technical connection to the fuzzball or firewall paradigms discussed in the literature \cite{Lunin:2001jy}, \cite{Mathur:2005zp}, \cite{Almheiri:2012rt}. In our framework, the emergence of smooth Hartle-Hawking correlator in the BTZ background is a feature of the exact semiclassical limit. At the same time, the preservation of unitarity reflects the breakdown of the EFT description in the near-horizon region once (non-perturbative) gravitational effects are taken into account. Hence, restoration of unitarity directly suggests to withdraw the notion of `smooth horizon' or an independent interior in our framework. To realize this unitary scenario, from our analysis, one must consider the full set of admissible boundary conditions within the stretched-horizon framework in order to identify possible semiclassical deviations from bulk EFT with smooth horizon.

\textit{\textbf{Plan of the paper:}} 
In Section~(\ref{sec2}), we revisit modular quantization in detail, emphasizing the underlying algebraic structure and its modification upon the inclusion of the center. In Section~(\ref{sec6}), we present the bulk interpretation and demonstrate the exact matching between the Hartle--Hawking correlator and the boundary vacuum correlator in the semiclassical limit. We further discuss, through the AdS/CFT correspondence, how the correlators are modified when the center is incorporated into the algebra. In Section~(\ref{sec7}), we explore several implications of our results and outline a number of promising directions for future investigation. We also include two appendices. In Appendix~(\ref{app1}), we discuss a closed-contour prescription without a cutoff, where the vanishing of the central extension corresponds to a vanishing Cardy entropy contribution. In Appendix~(\ref{B}), we provide details of the boundary correlator exhibiting quasi-periodic decay, obtained from a particular semiclassical HHLL Virasoro block.

\textit{\textbf{Terminology:}} Here, we have used non-standard terms like:

`Modular quantization':= Quantizing one-sided modular Hamiltonian with the contour given by constant modular time trajectory with conformal cut-off.

`Modular Virasoro algebra':= The similar structure to usual Virasoro algebra, obtained explicitly from the `modular Virasoro generators' or the stress tensor eigenmodes of the modular Hamiltonian with finite conformal cut-off.

\section{Revisiting modular quantization on cylinder}\label{sec2}

In this section, we review the relevant aspects of \textit{modular quantization} as developed in \cite{Tada:2019rls}, \cite{Das:2024mlx}, with an improved discussion of the associated operator algebra. To begin, we consider the following class of SL(2,$\mathbb{R}$)-deformed Hamiltonians in the Floquet heating phase:
\begin{align}\label{Deformed H}
H=\alpha (L_{0}+\bar{L}_{0})+\beta(L_{1}+L_{-1}+\bar{L}_{1}+\bar{L}_{-1}), ; (\alpha,\beta\in \mathbb{R}, ; d\equiv \alpha^{2}-4\beta^{2}<0).
\end{align}

Here, the definition of $L_n$ depends on whether the conformal field theory is formulated on the cylinder or on the strip. In both cases, the Hamiltonian is Hermitian as $L_{n}^{\dagger}=L_{-n}$ in radial quantization. A key nontrivial feature of the flow generated by this class of Hamiltonians is the emergence of spatial fixed point(s), accompanied by contractible Euclidean time circles which shrink toward them. Quantization along such spatial contours is subtle and, in some cases, divergent, as noted in \cite{Tada:2019rls} \cite{Agia:2022srj}. 

To construct a well-defined Hilbert space, one must either cap off the contour near the fixed points using appropriate cutoff boundary conditions or deform the contour to include these points within the Hilbert space. Following \cite{Tada:2019rls}, we refer to the first kind of procedure by using a cut-off as \textit{modular quantization}. As observed in \cite{Das:2025cuq}, the stress tensor receives nonzero contributions at the fixed points arising purely from the vacuum sector. Therefore, the cutoff boundary conditions must be chosen so as to reproduce these contributions when the cutoff is removed.

It is observed that conformal boundary conditions provide a consistent prescription, recovering the correct divergent behavior of the entanglement entropy in this limit \cite{Ohmori:2014eia}, \cite{Cardy:2016fqc}, \cite{Agia:2022srj}. However, both the structure of the Hilbert space and the operator algebra differ significantly before and after taking the cutoff to zero. We will explore these issues by first analyzing modular quantization on the cylinder.

On a cylinder($x,\sigma$), the deformed Hamiltonian (\ref{Deformed H}) is mapped to the following:
\begin{align}
    H= \int^{2\pi}_{0}dx (\alpha + 2\beta\cos x)T_{00}(x) + \frac{c\alpha}{12}
\end{align}
Here,  $T_{00}(x)=\frac{1}{2\pi}(T(x)+\bar{T}(x))$. We can now use the usual Fourier mode expansion of stress tensors in cylinder which can be transformed back to complex plane as:
\begin{align}
    L_{n} = \frac{c}{24}\delta_{n,0} \;+ \frac{1}{2\pi}\int^{2\pi}_{0}dx\; e^{inx}T(x)=\frac{1}{2\pi i}\oint dz\; z^{n+1}T(z)
\end{align}
A similar expression also holds for $\bar{L}_{n}$ with a opposite directed closed contour in complex plane. Following the notion of `dipolar quantization' \footnote{I should acknowledge Bartek Czech for bringing to our attention the papers on “dipolar quantization” many years ago.} introduced in \cite{Ishibashi:2015jba}, \cite{Ishibashi:2016bey}, we have constructed the contours$(s,\theta)$ generated by the flow of $H=(\mathcal{L}_{0}+\bar{\mathcal{L}}_{0})$ and it's momentum conjugate $P=i(\mathcal{L}_{0}-\bar{\mathcal{L}}_{0})$ respectively in \cite{Das:2024mlx}, where
\begin{align}
    \mathcal{L}_{0} = \alpha L_{0}+\beta(L_{1}+L_{-1}), \; \bar{\mathcal{L}}_{0}=\alpha \bar{L}_{0}+\beta(\bar{L}_{1}+\bar{L}_{-1}).
\end{align}
From now on, we will only present the relevant formula which was already derived in \cite{Tada:2019rls},\cite{Das:2024mlx}. In terms of $(\omega=s+i\theta,\bar{\omega}=s-i\theta)$, we can find the following relations by solving the curve equations generated by $H,P$:
\begin{align}\label{no cutoff curve}
    \frac{z-z_{+}}{z-z_{-}}=e^{i\omega\sqrt{d}}, \; \frac{\bar{z}-z_{+}}{\bar{z}-z_{-}}=e^{i\bar{\omega}\sqrt{d}}
\end{align}
Here, $z_{\pm}$ are two \textit{fixed points} of the flow:
\begin{align}
    z_{\pm} = \frac{-\alpha\pm i\sqrt{d}}{2\beta}, \; \sqrt{d}\equiv \sqrt{|d|} = \sqrt{|\alpha^{2}-4\beta^{2}|}
\end{align}
Also from now on, we will denote $|d|$ simply as $d$. The presence of fixed points $z_{\pm}$ restricts the contour as in Figure (3) of \cite{Das:2024vqe}. In particular, in stead of getting closed, the $\theta$ curves are restricted to  arcs with the range $\theta:(-\infty,\infty)$, which translates to $z:(z_{-},z_{+})$ in the complex plane. As in Figure (3) of \cite{Das:2024vqe}, we can choose either one of the sides(left or right) for this $\theta$ or constant time ($s$) contour which formalizes one-sided quantization as we will do subsequently. On the other hand, $s$ contours form contractible Euclidean time circles around both fixed points. They shrink to fixed points: $(\theta=\infty,-\infty)$. Notably, at $\theta =0$ or at $z=\bar{z}$, they become circles of infinite radius. Hence the effective geometry in $(\omega,\bar{\omega})$ plane can be viewed as joining two semi-infinite cigar, where the tips correspond to fixed points and the joining part has infinite circumference. 

Now, we can write $H$ as:
\begin{align}
    \mathcal{L}_{0} = \frac{\beta}{2\pi i}\int_{C}dz (z-z_{+})(z-z_{-})T(z), \; \bar{\mathcal{L}}_{0} = \frac{\beta}{2\pi i}\int_{\bar{C}}d\bar{z} (\bar{z}-z_{+})(\bar{z}-z_{-})\bar{T}(\bar{z})
\end{align}
Here $C$ and $\bar{C}$ correspond to constant time($s$) contours $\theta:(-\infty,\infty)$ and $\theta:(\infty,-\infty)$ respectively. Quantizing the Hamiltonian $H$ with this contour is what we dubbed as \textit{modular quantization}. 

\subsection{Modular Virasoro algebra and eigenmodes}
The spacetime representation $(l_{0}+\bar{l}_{0})$\footnote{Here $l_{0}=(\alpha z+\beta(z^{2}+1))\partial_{z}\equiv g(z)\partial_{z}$.} of the Hamiltonian $H$ has eigenmodes $f_{k}(z),\bar{f}_{k}(\bar{z})$ with eigenvalues $k$. Here 
\begin{align}
    f_{k}(z) = \left(\frac{z-z_{+}}{z-z_{-}}\right)^{-\frac{ik}{\sqrt{d}}}.
\end{align}
At this point, $k$ would be any real number and could be continuous too. One now construct $l_{k}\equiv g(z)f_{k}(z)\partial_{z}$ (and similarly $\bar{l}_{k}(\bar{z})$ from anti-holomorphic part) which can be shown to satisfy Witt algebra \cite{Tada:2019rls} $[l_{k},l_{k'}]=(k-k')l_{k+k'}$. Hence it is natural to find it's Hilbert space representation $\mathcal{L}_{k}$ constructed out of the conformal Noether charges generated by $l_{k}$. Following \cite{Tada:2019rls}, we have
\begin{align}\label{Vir modes}
  \mathcal{L}_{k} = \frac{\beta}{2\pi i} \int_{C} dz (z-z_{+})^{-\frac{ik}{\sqrt{d}}+1}(z-z_{-})^{\frac{ik}{\sqrt{d}}+1} T(z), \; \bar{\mathcal{L}}_{k} = \frac{\beta}{2\pi i} \int_{\bar{C}} d\bar{z} (\bar{z}-z_{+})^{-\frac{ik}{\sqrt{d}}+1}(\bar{z}-z_{-})^{\frac{ik}{\sqrt{d}}+1} \bar{T}(\bar{z})
\end{align}
As in \cite{Tada:2019rls}, we can use $TT$ ope for CFT with central charge $c$ by deforming the contour along with a defined time ordering \cite{Das:2025cuq} to find the following:
\begin{align}\label{mva1}
    [\mathcal{L}_{k},\mathcal{L}_{k'}]=(k-k')\mathcal{L}_{k+k'}+ \frac{c}{12\beta}\frac{k(k^{2}+d)}{2\pi i} \int_{C} dz (z-z_{+})^{-\frac{i(k+k')}{\sqrt{d}}-1}(z-z_{-})^{\frac{i(k+k')}{\sqrt{d}}-1}
\end{align}
Here $\int_{C} = \int^{z_{+}}_{z_{-}}$ along a constant $s$ contour. For the above integration, we can shift the branch cut along a straight line connecting $z_{+}$ to $z_{-}$, while $z_{+}$ and $z_{-}$ are the branch points. Due to these branch points, we can see the integral to be divergent. To see it more directly, we can choose the following identity:
\begin{align}\label{total der identity}
    \partial_{z}\left[(z-z_{+})^{-\frac{i(k+k')}{\sqrt{d}}}(z-z_{-})^{\frac{i(k+k')}{\sqrt{d}}}\right] = \frac{(k+k')}{\beta}(z-z_{+})^{-\frac{i(k+k')}{\sqrt{d}}-1}(z-z_{-})^{\frac{i(k+k')}{\sqrt{d}}-1}
\end{align}
By using the above identity, we obtain:
\begin{align}\label{mva2}
  [\mathcal{L}_{k},\mathcal{L}_{k'}]=(k-k')\mathcal{L}_{k+k'}+ \frac{c}{12}\frac{k(k^{2}+d)}{2\pi i(k+k')} \left[ \left(\frac{z-z_{-}}{z-z_{+}}\right)^{\frac{i(k+k')}{\sqrt{d}}}\right]^{z=z_{+}}_{z=z_{-}}
\end{align}
We can easily see the central extension term to be divergent. This divergence clearly implies one can not construct a Hilbert space using the eigenmodes of the Hamiltonian. The remedy of this problem as given by Tada is to introduce cut-offs around the fixed points. Also there is another possibility to deform the contour around fixed points and make it close by avoiding the branch cuts. We will discuss this possibility in the appendix (\ref{app1})\footnote{I thank Koushik Ray for emphasizing this possibility.}. For now, we will follow \cite{Tada:2019rls} to find a Virasoro algebra with finite central extension term. We put $\epsilon$ radius cut-offs  at 
\begin{align}\label{cut-off}
    z:[z_{+}+\epsilon e^{i\phi_{\epsilon}},z_{-}+\epsilon e^{-i\phi_{\epsilon}}], \; \theta\in [\Lambda,-\Lambda], \; \text{where} \; \Lambda\equiv\frac{1}{\sqrt{d}}\log\left(\frac{\sqrt{d}}{\beta\epsilon}\right)
\end{align}
We note that, in principle, one may use two different cut-off lengths $\epsilon,\epsilon'$ at two different fixed points. However, as we will see, the same length of the cut-offs is crucial to get a real central term satisfying Jacobi identity\footnote{Otherwise, the central term would be complex and not proportional to a Kronecker delta which is needed to close the algebra consistently. Hence, we can not find a Hilbert space representation of them.}. After using the equal length cut-offs, (\ref{mva2}) yields:
\begin{align}\label{mva3}
    [\mathcal{L}_{k}^{\epsilon},\mathcal{L}_{k'}^{\epsilon}]=(k-k')\mathcal{L}_{k+k'}^{\epsilon}+ \frac{c}{12\pi}\frac{k(k^{2}+d)}{(k+k')}e^{\frac{(k+k')}{\sqrt{d}}\left(\phi_{\epsilon}-\frac{\pi}{2\sqrt{d}}\right)} \sin\left[ \frac{(k+k')}{\sqrt{d}}\log\left(\frac{\sqrt{d}}{\beta\epsilon}\right)\right]
\end{align}
Here $\mathcal{L}_{k}^{\epsilon},\bar{\mathcal{L}}_{k}^{\epsilon}$ are defined by (\ref{Vir modes}) with a slight modification of contours $C^{\epsilon},\bar{C}^{\epsilon}$, bounded by cut-offs as in (\ref{cut-off}). As the central term is symmetric in $k,k'$, the above form of the commutator is not well-defined. To make it well-defined or to close it under Jacobi identity, we need $\delta_{k+k',0}$. This is guaranteed once we make the following choice:
\begin{align}\label{k quant}
    k(n)=\frac{n\pi\sqrt{d}}{\log\left(\frac{\sqrt{d}}{\beta\epsilon}\right)}, \; n\in \frac{\mathbb{Z}}{2}
\end{align}
Hence we will end up with:
\begin{align}\label{mva4}
   [\mathcal{L}_{k(n)}^{\epsilon},\mathcal{L}_{k(n')}^{\epsilon}]=(k(n)-k(n'))\mathcal{L}_{k(n)+k(n')}^{\epsilon}+ \frac{c}{12\pi\sqrt{d}}\log\left(\frac{\sqrt{d}}{\beta\epsilon}\right)k(n)(k(n)^{2}+d)\delta_{(k(n),-k(n'))} 
\end{align}
We note that, we obtain (\ref{mva4}) from (\ref{mva3}) at a fixed $\epsilon$. If we want to understand the nature of the algebra in $\epsilon \rightarrow 0$ limit, we first take the limit in (\ref{mva3}) before taking $(k+k')\rightarrow 0$. This yields a Dirac delta function of $\delta(k+k')$. To find it, one can use Dirichlet Kernel as an approximation of Dirac delta function: $\lim_{\Lambda\rightarrow \infty} \frac{\sin(\Lambda(k+k')))}{\pi(k+k')} = \delta(k+k')$, where $\Lambda\equiv\frac{1}{\sqrt{d}}\log\left(\frac{\sqrt{d}}{\beta\epsilon}\right)$. The continuous label of $k$ is ensured from (\ref{k quant}) in $\epsilon\rightarrow 0$ limit. Hence the emergent \textit{continuous modular Virasoro algebra} in $\epsilon \rightarrow 0$ limit has the following form:
\begin{align}\label{continuous mva}
 [\mathcal{L}_{k},\mathcal{L}_{k'}]=(k-k')\mathcal{L}_{k+k'}+\frac{c}{12}k(k^{2}+d)\delta(k+k')   
\end{align}
Here we have used $\lim_{\epsilon\rightarrow 0}\mathcal{L}^{\epsilon}_{k}=\mathcal{L}_{k}$. We will come back to it later. For now, we will stick to finite(yet small) $\epsilon$ algebra. Using (\ref{k quant}), we can rewrite (\ref{mva4}) in terms of integer-valued $\mathcal{L}_{n}^{\epsilon}$ with $n\in \mathbb{Z}$:
\begin{align}
    [\mathcal{L}^{\epsilon}_{n},\mathcal{L}^{\epsilon}_{n'}]=(n-n')\frac{\pi\sqrt{d}}{\log\left(\frac{\sqrt{d}}{\beta\epsilon}\right)}\mathcal{L}^{\epsilon}_{n+n'}+\frac{c}{12}n\left(n^{2}+\frac{\left(\log\left(\frac{\sqrt{d}}{\beta\epsilon}\right)\right)^{2}}{\pi^{2}}\right)\frac{\pi^{2}d}{\left(\log\left(\frac{\sqrt{d}}{\beta\epsilon}\right)\right)^{2}}\delta_{n+n',0}
\end{align}
We now use the following redefinition of modular Virasoro modes:
\begin{align}
    \tilde{\mathcal{L}}_{n}^{\epsilon}\equiv\frac{\log\left(\frac{\sqrt{d}}{\beta\epsilon}\right)}{\pi\sqrt{d}}\mathcal{L}^{\epsilon}_{n}=\frac{\Lambda}{\pi}\mathcal{L}^{\epsilon}_{n}, \; (\alpha,\beta) \rightarrow\frac{\Lambda}{\pi}(\alpha,\beta), \; H^{\epsilon}\rightarrow \tilde{H}^{\epsilon}=\frac{\Lambda}{\pi}H^{\epsilon}
\end{align}
Using this modified modes, we obtain the following:
\begin{align}\label{mva fin}
    [\tilde{\mathcal{L}}^{\epsilon}_{n},\tilde{\mathcal{L}}^{\epsilon}_{n'}]=(n-n')\tilde{\mathcal{L}}^{\epsilon}_{n+n'}+\frac{c}{12}n\left(n^{2}+\frac{\left(\log\left(\frac{\sqrt{d}}{\beta\epsilon}\right)\right)^{2}}{\pi^{2}}\right)\delta_{n+n',0}
\end{align}
This algebra is defined as the \textit{modular Virasoro algebra} with integer-valued modes. This is not strictly a Virasoro algebra due to lack of SL(2,$\mathbb{R}$) sub-algebra. To obtain a meaningful SL(2,$\mathbb{R}$) sub-algebra, we need to absorb the central term into the zero mode term $\tilde{\mathcal{L}}^{\epsilon}_{0}$ as
\begin{align}\label{shift}
 \tilde{\mathcal{L}}^{\epsilon}_{0} \rightarrow \tilde{\mathcal{L}}^{\epsilon}_{0} + \frac{c}{24}\left(1+\frac{\left(\log\left(\frac{\sqrt{d}}{\beta\epsilon}\right)\right)^{2}}{\pi^{2}}\right)
\end{align}
Using this shift in $\tilde{\mathcal{L}}^{\epsilon}_{0}$, the algebra of $\tilde{\mathcal{L}}^{\epsilon}_n$s will follow the usual Virasoro algebra with exactly similar central term. We note that, the redefinition of modes as well as the Hamiltonian have changed the curve equation and thus the functional form of modes in the following way:
\begin{align}\label{cutoff curve}
  \tilde{\mathcal{L}}^{\epsilon}_{n} = \frac{\beta\Lambda}{2\pi^{2}i}\int_{C^{\epsilon}} dz' (z'-z_{+})^{-\frac{in\pi}{\sqrt{d}\Lambda}+1}(z'-z_{-})^{\frac{in\pi}{\sqrt{d}\Lambda}+1}\;T(z'), \;\text{where} \;  \frac{z'-z_{+}}{z'-z_{-}}=e^{i\frac{\sqrt{d}\Lambda}{\pi}\omega}  
\end{align}
 Similarly, one can follow the same construction for antiholomorphic modes and obtain the same form of the algebra in terms of $\bar{\mathcal{L}}^{\epsilon}_{n}$. For that, $\bar{C}^{\epsilon}:(z_{+}+\epsilon e^{i\phi_{\epsilon}},z_{-}+\epsilon e^{-i\phi_{\epsilon}})$, since $\bar{z}_{\pm}=z_{\mp}$. However, the boundary conditions at the cut-offs may significantly affect the holomorphic factorization as we will discuss now.

\textbf{\textit{Boundary condition at cut-offs:}} The curve equations (\ref{no cutoff curve}) and (\ref{cutoff curve}) provides a map from $(z,\bar{z})\;\text{or}\;(z',\bar{z}') \rightarrow (\omega,\bar{\omega})$. More explicitly:
\begin{align}\label{map}
    z = -\frac{\alpha}{2\beta}-\frac{\sqrt{d}}{2\beta}\cot\left(\frac{\sqrt{d}}{2}\omega\right); \;  \; z' =-\frac{\alpha}{2\beta}-\frac{\sqrt{d}}{2\beta}\cot\left(\frac{\Lambda\sqrt{d}}{2\pi}\omega\right)
\end{align}
In \cite{Das:2025cuq}, it was observed that, the appearance of fixed points affects the value of stress tensors. In particular, using (\ref{map}) and the following
\begin{align}
    T(\omega) = \left(\frac{\partial z}{\partial \omega}\right)^{2}T(z)+\frac{c}{12}Sch\{z,\omega\},
\end{align}
one can see that
\begin{align}\label{fixed point bc}
    T(\omega)|_{\theta=\pm\infty} =\bar{T}(\bar{\omega})|_{\theta=\pm \infty} =\frac{c}{24}d .
\end{align}
This clearly shows the stress tensor at fixed point is \textit{non-vanishing}! Here we have assumed finite and non-vanishing value of $T(z_{\pm})$. Hence in the presence of the cut-off at $\theta=\pm \Lambda$, the boundary conditions should be given in a way such that they would reproduce (\ref{fixed point bc}) in $\Lambda \rightarrow \infty$ limit. Since we are only interested in local B.C \footnote{Mathematically, one may still construct a non-local boundary condition that relates $T$ and $\bar{T}$ of two different fixed points which solves the fixed point criterion. However in a local QFT set-up it is not well-understood on how to deal with such non-locality.}, there will be two independent possible choices:
\begin{align}
    &\text{B.C-I:} \; \left(T(\omega)-\frac{c\Lambda^{2}d}{24\pi^{2}}\right)|_{\theta=\pm \Lambda} =  \left(\bar{T}(\bar{\omega})-\frac{c\Lambda^{2}d}{24\pi^{2}}\right)|_{\theta=\pm \Lambda} = 0; \\
  &  \text{B.C-II:} \; \left(T(\omega)-\bar{T}(\bar{\omega})\right)|_{\theta=\pm \Lambda} = 0
\end{align}
The solutions of (B.C-I) should respect holomorphic factorization. However, this translates to the following boundary condition in complex $z$ plane:
\begin{align}
    T(z)|_{z=z_{\pm}+\epsilon e^{\pm i\phi_{\epsilon}}} = 0 = \bar{T}(\bar{z})|_{\bar{z}=z_{\pm}+\epsilon e^{\pm i\phi_{\epsilon}}}
\end{align}
Since $T(z),\bar{T}(\bar{z})$ are analytic homolorphic functions in the complex plane, the above condition implies $T,\bar{T}$ and all of it's derivatives are vanishing at $z,\bar{z}=z_{\pm}$. Using Taylor expansion, one can immediately see that this implies $T(z)=\bar{T}(\bar{z})=0$ identically for all $z,\bar{z}$. Thus (B.C-I) would not give any physically meaningful solution. The only option we are left with the conformal boundary condition (B.C-II), for which the mode expansions are not holomorphically factorized as in BCFT. Thus we expect a single copy of Virasoro modular algebra consistent with the conformal boundary condition as noticed in \cite{Das:2024vqe}. To construct the single copy, it will be useful to write them in $(\omega,\bar{\omega})$ coordinate. Using (\ref{Vir modes}) and (\ref{no cutoff curve}), we can write
\begin{align}\label{modes in omega}
    \mathcal{L}_{n}^{\epsilon} =\frac{e^{\frac{n\pi s}{\Lambda}}}{2\pi}\int^{\Lambda}_{-\Lambda}d\theta e^{i\frac{n\pi}{\Lambda}\theta}T(\omega)-\frac{c\Lambda d}{24\pi}\delta_{n,0}; \; \bar{\mathcal{L}}_{n}^{\epsilon} =-\frac{e^{\frac{n\pi s}{\Lambda}}}{2\pi}\int^{-\Lambda}_{\Lambda}d\theta e^{i\frac{n\pi}{\Lambda}\theta}\bar{T}(\bar{\omega})-\frac{c\Lambda d}{24\pi}\delta_{n,0}
\end{align}
In the above expression, we have taken a constant $s$ path. Similarly we can construct $\tilde{\mathcal{L}}_{n}^{\epsilon}$ and $\tilde{\bar{\mathcal{L}}}_{n}^{\epsilon}$ as the previous method by multiplying $\frac{\Lambda}{\pi}$. We now define a generator by combining them as
\begin{align}\label{redefined Vir}
    \hat{\mathcal{L}}_{n}^{\epsilon} =\frac{\Lambda}{2\pi^{2}} \oint_{R}d\hat{\omega} \;e^{i\frac{n\pi}{\Lambda}\hat{\omega}} \; \hat{T}(\hat{\omega})
\end{align}
Here $R$ is a closed path, consisting of four cycles: $R=C^{\epsilon}_{\phi_{\epsilon}^{1}}\cup C^{1,2}_{\{\phi^{1}_{\epsilon},\phi^{2}_{\epsilon}\}}\cup \bar{C}^{\epsilon}_{\phi^{2}_{\epsilon}}\cup C^{1,2}_{\{\phi^{2}_{\epsilon},\phi^{1}_{\epsilon}\}}$. We choose the path such that,
\begin{align}
  \int_{C^{\epsilon}_{\phi^{1}_{\epsilon}}}+  \int_{C^{\epsilon}_{\phi^{2}_{\epsilon}}} =\tilde{\mathcal{L}}^{\epsilon}_{n}+\tilde{\bar{\mathcal{L}}}^{\epsilon}_{n}
\end{align}
The integral contributions from other two paths are canceled due to conformal boundary condition at the cut-off. Since $C^{1,2}_{\{\phi^{1}_{\epsilon},\phi^{2}_{\epsilon}\}}$ is a constant $\theta=\Lambda$ path, the contribution from (\ref{redefined Vir}) has the following terms:
\begin{align}
 \int_{C^{1,2}_{\{\phi^{1}_{\epsilon},\phi^{2}_{\epsilon}\}}}   d\hat{\omega} \;e^{i\frac{n\pi}{\Lambda}\hat{\omega}} \; \hat{T}(\hat{\omega}) = e^{in\pi}\int^{\phi^{1}_{\epsilon}}_{\phi^{2}_{\epsilon}}ds \; e^{\frac{n\pi s}{\Lambda}}T(s+i\Lambda) + e^{-in\pi}\int^{\phi^{2}_{\epsilon}}_{\phi^{1}_{\epsilon}} ds \; e^{\frac{n\pi s}{\Lambda}}\bar{T}(s-i\Lambda) = 0
\end{align}
Here we have used $T(s+i\Lambda)=\bar{T}(s-i\Lambda)$. Similarly $\int_{C^{1,2}_{\{\phi^{2}_{\epsilon},\phi^{1}_{\epsilon}\}}}d\hat{\omega} \;e^{i\frac{n\pi}{\Lambda}\hat{\omega}} \; \hat{T}(\hat{\omega}) =0$ due to the boundary condition $T(s-i\Lambda)=\bar{T}(s+i\Lambda)$ at $\theta=-\Lambda$. Thus the only non-vanishing contribution of this closed path comes from $C^{\epsilon}_{\phi^{1}_{\epsilon}}\cup C^{\epsilon}_{\phi^{2}_{\epsilon}}$. Thus along this path we have,
\begin{align}
   \hat{\mathcal{L}}_{n}^{\epsilon} =  \tilde{\mathcal{L}}^{\epsilon}_{n}+\tilde{\bar{\mathcal{L}}}^{\epsilon}_{n}
\end{align}
Hence, the one copy of Virasoro algebra has the following form\footnote{Since the algebra does not depend on the $\phi^{\epsilon}$ time at the cut-off, they will simply add.}:
\begin{align}\label{Final form of mod Vir algebra}
  [\hat{\mathcal{L}}_{n}^{\epsilon},\hat{\mathcal{L}}_{n'}^{\epsilon}]=(n-n')\hat{\mathcal{L}}_{n+n'}^{\epsilon}  + \frac{c}{6}n(n^{2}-1)\delta_{n+n',o}
\end{align}
Here we have absorbed the central term into $\hat{\mathcal{L}}_{0}^{\epsilon}$. From the expression (\ref{modes in omega}), this is naturally happened once we write the modes in $\omega$ plane.

\textbf{\textit{Non-vacuum eigenmodes:}} From the modular Virasoro algebra, we have constructed stress tensor eigenmodes $\hat{\mathcal{L}}^{\epsilon}_{n}$. A natural extension is to construct a possible generalization for primary operators. Before doing that, let us compute the action of Hamiltonian on local primary fields. Here we work with $\mathcal{L}^{\epsilon}_{n}$, since the translation to $\hat{\mathcal{L}}^{\epsilon}_{n}$ is straightforward from previous discussion. To compute $[\mathcal{L}^{\epsilon}_{0},\mathcal{O}(z,\bar{z})]$, we use $T(z')\mathcal{O}(z,\bar{z})$ OPE and get:
\begin{align}
    [\mathcal{L}^{\epsilon}_{0},\mathcal{O}(z,\bar{z})] &= \frac{\beta}{2\pi i}\oint_{z}dz' \; (z'-z_{+})(z'-z_{-})\left[\frac{h\mathcal{O}(z,\bar{z})}{(z'-z)^{2}}+\frac{\partial_{z}\mathcal{O}(z,\bar{z})}{(z'-z)}\right] \nonumber \\
    & = (2\beta z+\alpha)h\mathcal{O}(z,\bar{z})+\beta(z-z_{+})(z-z_{-})\partial_{z}\mathcal{O}(z,\bar{z})
\end{align}
This implies 
\begin{align}\label{action on local op}
   & [\hat{\mathcal{L}}_{0}^{\epsilon},\mathcal{O}(z,\bar{z})] \nonumber \\
 &   = \frac{\Lambda}{\pi}[(2\beta z+\alpha)h+(2\beta \bar{z}+\alpha)\bar{h}]\mathcal{O}(z,\bar{z})+\frac{\beta\Lambda}{\pi}[(z-z_{+})(z-z_{-})\partial_{z}+(\bar{z}-z_{+})(\bar{z}-z_{-})\partial_{\bar{z}}]\mathcal{O}(z,\bar{z})
\end{align}
If we want to construct local field as eigenmode, the coefficients of $\partial_{z}\mathcal{O}$ and $\partial_{\bar{z}}\mathcal{O}$ must vanish. However, this is only possible when the primaries are located at $z,\bar{z}=z_{\pm}$. Since at finite $\epsilon$, we cut-off the fixed points, those operator can not live in the algebra. Naively, this may suggest a state-operator correspondence \textit{does not} exist in modular quantization in the presence of a cut-off. We will see in later section that, those fixed point primaries would play a vital role in the operator algebras after taking $\epsilon \rightarrow 0$ limit. 

However, there exists an interesting possibility to construct eigenmodes from a subset of scalar operators living at the cut-off or very closed to cut-off which we will describe now. Consider operators of dimension $(h,\bar{h})$ located at $z=z_{+}+\epsilon e^{i\phi_{\epsilon}},\bar{z}=z_{-}+\epsilon e^{-i\phi_{\epsilon}}$. Then (\ref{action on local op}) implies
\begin{align}
    &[\hat{\mathcal{L}}_{0}^{\epsilon},\mathcal{O}(z,\bar{z})]|_{z=z_{+}+\epsilon e^{i\phi_{\epsilon}},\bar{z}=z_{-}+\epsilon e^{-i\phi_{\epsilon}}} \nonumber \\
 &   = \frac{\Lambda}{\pi}[(i\sqrt{d}+2\beta\epsilon e^{i\phi_{\epsilon}})h+(-i\sqrt{d}+2\beta\epsilon e^{-i\phi_{\epsilon}})\bar{h}]\mathcal{O}(z,\bar{z})|_{z=z_{+}+\epsilon e^{i\phi_{\epsilon}},\bar{z}=z_{-}+\epsilon e^{-i\phi_{\epsilon}}}  \\
 & \; \; \; \; \;\;\;\;\;\;+\frac{i\epsilon\sqrt{d}\Lambda}{\pi}\left[e^{i\phi_{\epsilon}}\partial_{z}+e^{-i\phi_{\epsilon}}\partial_{\bar{z}}\right]\mathcal{O}(z,\bar{z})|_{z=z_{+}+\epsilon e^{i\phi_{\epsilon}},\bar{z}=z_{-}+\epsilon e^{-i\phi_{\epsilon}}}
\end{align}
For scalar operators $(h=\bar{h})$, one end up with
\begin{align}\label{primary eigenmodes}
  &[\hat{\mathcal{L}}_{0}^{\epsilon},\mathcal{O}(z,\bar{z})]|_{z=z_{+}+\epsilon e^{i\phi_{\epsilon}},\bar{z}=z_{-}+\epsilon e^{-i\phi_{\epsilon}}} \nonumber \\
 &  =   \left(\frac{8\beta\epsilon\Lambda\cos\phi_{\epsilon}}{\pi}h+ \frac{i\epsilon\sqrt{d}\Lambda}{\pi}\left[e^{i\phi_{\epsilon}}\partial_{z}+e^{-i\phi_{\epsilon}}\partial_{\bar{z}}\right]\right)\mathcal{O}(z,\bar{z})|_{z=z_{+}+\epsilon e^{i\phi_{\epsilon}},\bar{z}=z_{-}+\epsilon e^{-i\phi_{\epsilon}}} \nonumber \\
 & \approx \frac{8\beta\epsilon\Lambda\cos\phi_{\epsilon}}{\pi}h \mathcal{O}(z_{+}+\epsilon e^{i\phi_{\epsilon}},z_{-}+\epsilon e^{-i\phi_{\epsilon}}) \; \; \text{when} \; h\sim \mathcal{O}(\frac{1}{\epsilon})
\end{align}
This is quite unusual feature, since the eigenmodes depend on the cut-off location! It also crucially depends on the CFT in which such heavy scalar operators ($h\sim \mathcal{O}(\frac{1}{\epsilon})$) exists\footnote{Non-scalar operators would have imaginary eigenvalue and hence can not be part of algebra of observables.}. This subset of operators are non vanishing in taking $\epsilon \rightarrow 0$ limit with $h\epsilon$ fixed. 

However one may still construct a set of \textit{non-local} eigenmodes constructed out of primaries. A similar type of modular eigenmodes were constructed earlier in \cite{Das:2019iit} in Lorentzian set-up without cut-off with a potential connection to OPE blocks\footnote{See also other works on such connection: \cite{Czech:2016xec}-\cite{Nath:2024aqh}.}. Following \cite{Ishibashi:2016bey}, we use a systematic procedure to obtain a mode expansion of primary fields. We consider the following scalar modes $\mathcal{O}_{k,k'}^{\epsilon}$ defined as:
\begin{align}\label{primary modes}
    \mathcal{O}_{k,k'}^{\epsilon} = \frac{\beta^{h-1}}{(2\pi i)^{2}}\int_{C^{\epsilon}}dz\int_{\bar{C}^{\epsilon}}d\bar{z} (z-z_{+})^{h-\frac{ik}{\sqrt{d}}-1}(z-z_{-})^{h+\frac{ik}{\sqrt{d}}-1}(\bar{z}-z_{+})^{h-\frac{ik'}{\sqrt{d}}-1}(\bar{z}-z_{-})^{h+\frac{ik'}{\sqrt{d}}-1}\mathcal{O}(z,\bar{z})
\end{align}
Using $T\mathcal{O}$ OPE, we find the following action of Hamiltonian on these modes:
\begin{align}
   & [\mathcal{L}_{0}^{\epsilon}+\bar{\mathcal{L}}_{0}^{\epsilon},\mathcal{O}_{k,k'}^{\epsilon}] \nonumber \\
    &= \frac{\beta^{h}}{(2\pi i)^{2}}\int_{\bar{C}^{\epsilon}}d\bar{z} (\bar{z}-z_{+})^{h-\frac{ik'}{\sqrt{d}}-1}(\bar{z}-z_{-})^{h+\frac{ik'}{\sqrt{d}}-1}\int_{C^{\epsilon}}dz(z-z_{+})^{h-\frac{ik}{\sqrt{d}}-1}(z-z_{-})^{h+\frac{ik}{\sqrt{d}}-1} \nonumber \\
    &\;\; \; \; \;\;\;\;\;\;\;\;\;\; \times\oint_{z} dz' (z'-z_{+})(z'-z_{-})\left[ \frac{h\mathcal{O}(z,\bar{z})}{(z'-z)^{2}}+\frac{\partial_{z}\mathcal{O}(z,\bar{z})}{z'-z}\right] \nonumber \\
    & + \frac{\beta^{h}}{(2\pi i)^{2}}\int_{C^{\epsilon}}dz(z-z_{+})^{h-\frac{ik}{\sqrt{d}}-1}(z-z_{-})^{h+\frac{ik}{\sqrt{d}}-1}\int_{\bar{C}^{\epsilon}}d\bar{z} (\bar{z}-z_{+})^{h-\frac{ik'}{\sqrt{d}}-1}(\bar{z}-z_{-})^{h+\frac{ik'}{\sqrt{d}}-1} \nonumber \\
    &\;\; \; \; \;\;\;\;\;\;\;\;\;\; \times\oint_{\bar{z}} d\bar{z}' (\bar{z}'-z_{+})(\bar{z}'-z_{-})\left[ \frac{h\mathcal{O}(z,\bar{z})}{(\bar{z}'-\bar{z})^{2}}+\frac{\partial_{\bar{z}}\mathcal{O}(z,\bar{z})}{\bar{z}'-\bar{z}}\right]
\end{align}
After some straightforward algebraic steps, we will end up with
\begin{align}
   [\mathcal{L}_{0}^{\epsilon}+\bar{\mathcal{L}}_{0}^{\epsilon},\mathcal{O}_{k,k'}^{\epsilon}] =  -(k+k')\mathcal{O}_{k,k'}^{\epsilon} + \text{`boundary term'}
\end{align}
Here the boundary term has the following expression:
\begin{align}
    &\;\;\;\;\;\;\;\;\;\;\;\;\;\;\;\;\;\;\;\;\;\;\;\;\;\;\;\;\;\;\;\;\;\;\;\;\;\;\;\;\;\;\;\;\;\;\;\;\;\;\;\text{`boundary term'} \nonumber \\
    &= \epsilon^{h}(i\sqrt{d})^{h}e^{\frac{k\phi_{\epsilon}}{\sqrt{d}}}\int_{\bar{C}^{\epsilon}}d\bar{z} A_{k'}(\bar{z})\left[ e^{\frac{ik}{\sqrt{d}}\log\left(\frac{i\sqrt{d}}{\beta\epsilon}\right)+i\phi_{\epsilon} h}\mathcal{O}(z_{+}+\epsilon e^{i\phi_{\epsilon}},\bar{z})-(-1)^{h-\frac{ik}{\sqrt{d}}}e^{-\frac{ik}{\sqrt{d}}\log\left(\frac{i\sqrt{d}}{\beta\epsilon}\right)-i\phi_{\epsilon} h}\mathcal{O}(z_{-}+\epsilon e^{-i\phi_{\epsilon}},\bar{z})\right] \nonumber \\
    &+ \epsilon^{h}(i\sqrt{d})^{h}e^{\frac{k'\phi_{\epsilon}}{\sqrt{d}}}\int_{C^{\epsilon}}dz A_{k}(z)\left[ e^{\frac{ik'}{\sqrt{d}}\log\left(\frac{i\sqrt{d}}{\beta\epsilon}\right)+i\phi_{\epsilon} h}\mathcal{O}(z, z_{+}+\epsilon e^{i\phi_{\epsilon}})-(-1)^{h-\frac{ik'}{\sqrt{d}}}e^{-\frac{ik'}{\sqrt{d}}\log\left(\frac{i\sqrt{d}}{\beta\epsilon}\right)-i\phi_{\epsilon} h}\mathcal{O}(z, z_{-}+\epsilon e^{-i\phi_{\epsilon}})\right] \nonumber \\
    &= \epsilon^{h}(i\sqrt{d})^{h}e^{\frac{k\phi_{\epsilon}}{\sqrt{d}}}\left[ e^{\frac{ik}{\sqrt{d}}\log\left(\frac{i\sqrt{d}}{\beta\epsilon}\right)+i\phi_{\epsilon} h}-(-1)^{h-\frac{ik}{\sqrt{d}}}e^{-\frac{ik}{\sqrt{d}}\log\left(\frac{i\sqrt{d}}{\beta\epsilon}\right)-i\phi_{\epsilon} h}\right]\mathcal{O}_{k'} \nonumber \\
    &+ \epsilon^{h}(i\sqrt{d})^{h}e^{\frac{k'\phi_{\epsilon}}{\sqrt{d}}}\left[ e^{\frac{ik'}{\sqrt{d}}\log\left(\frac{i\sqrt{d}}{\beta\epsilon}\right)+i\phi_{\epsilon} h}-(-1)^{h-\frac{ik'}{\sqrt{d}}}e^{-\frac{ik'}{\sqrt{d}}\log\left(\frac{i\sqrt{d}}{\beta\epsilon}\right)-i\phi_{\epsilon} h}\right]\mathcal{O}_{k}
\end{align}
Here $A_{k}(z)= (z-z_{+})^{h-\frac{ik}{\sqrt{d}}-1}(z-z_{-})^{h+\frac{ik}{\sqrt{d}}-1}$ and similar expression holds for $A_{k'}(\bar{z})$. In the last step, we have used a consistent definition of holomorphic and anti-holomorphic modes:
\begin{align}
    \mathcal{O}_{k}\equiv\int_{C^{\epsilon}}dz A_{k}(z)\mathcal{O}(z,\bar{z}), \; \mathcal{O}_{k'} \equiv \int_{\bar{C}^{\epsilon}}d\bar{z} A_{k'}(\bar{z})\mathcal{O}(z,\bar{z}).
\end{align}
Hence, the boundary term will be of $\mathcal{O}(\epsilon^{h})$. In $\epsilon \rightarrow 0$ limit, this will vanish as $h>0$ for any Unitary CFT\footnote{Here we refer unitarity with respect to radial quantization.}. At finite $\epsilon$, we can still ignore the effect of boundary term consistently as long as $h\geq 1$. This further implies for scalars with $h\geq1$, we have
\begin{align}
    [\hat{\mathcal{L}}^{\epsilon}_{0},\mathcal{O}_{k,k'}] = -\frac{\Lambda}{\pi}(k+k')\mathcal{O}_{k,k'}
\end{align}
Here we have assumed $k$ to be discrete parameter. However we can argue that $k$ must have the same quantized expression as in (\ref{k quant}) and it becomes continuous in $\epsilon \rightarrow 0$ limit. To see this, we use the following mode expansion of primary $\mathcal{O}(z,\bar{z})$ in terms of $\mathcal{O}_{k,k'}$:
\begin{align}\label{primary mode exp}
    \mathcal{O}(z,\bar{z}) =  \frac{\pi^{2}\beta^{-h}}{\Lambda^{2}} \sum_{k,k'} (z-z_{+})^{\frac{ik}{\sqrt{d}}-h}(z-z_{-})^{-\frac{ik}{\sqrt{d}}-h}(\bar{z}-z_{+})^{\frac{ik'}{\sqrt{d}}-h}(\bar{z}-z_{-})^{-\frac{ik'}{\sqrt{d}}-h}\mathcal{O}_{k,k'}
\end{align}
If we put the above mode expansion of scalar primary in (\ref{primary modes}), we obtain:
\begin{align}\label{mode check}
   & \;\;\;\;\;\;\;\;\;\;\;\;\;\;\;\;\;\;\;\;\;\;\;\;\;\;\;\;\;\;\;\;\;\;\;\;\;\;\;\;\;\;\;\;\;\;\;\;\;\;\;\;\;\;\;\;\;\;\;\;\;\;\;\;\;\;\;\;\;\;\;\;\;\mathcal{O}_{k,k'} \nonumber \\
    &= \frac{\beta^{-1}}{(2 i\Lambda)^{2}}\sum_{m,m'}\int_{C^{\epsilon}}dz (z-z_{+})^{-\frac{i(k-m)}{\sqrt{d}}-1}(z-z_{-})^{\frac{i(k-m)}{\sqrt{d}}-1}\int_{\bar{C}^{\epsilon}}d\bar{z} (\bar{z}-z_{+})^{-\frac{i(k'-m')}{\sqrt{d}}-1}(\bar{z}-z_{-})^{\frac{i(k'-m')}{\sqrt{d}}-1}\mathcal{O}_{m,m'}
\end{align}
Using (\ref{map}), we map $(z,\bar{z})\rightarrow(\omega,\bar{\omega})$ and get the following relations:
\begin{align}
    dz= \frac{d}{4\beta}\csc^{2}\left(\frac{\sqrt{d}}{2}\omega\right)d\omega, \; (z-z_{+})(z-z_{-})=\frac{d}{4\beta^{2}}\csc^{2}\left(\frac{\sqrt{d}}{2}\omega\right).
\end{align}
By using this and (\ref{no cutoff curve}), we simplify (\ref{mode check}) to be
\begin{align}
 \mathcal{O}_{k,k'} =   \frac{1}{(2\Lambda)^{2}}\sum_{m,m'}e^{(k-m+k'-m')s}\int^{\Lambda}_{-\Lambda}d\theta\; e^{i(k-m)\theta} \int^{\Lambda}_{-\Lambda}d\theta \; e^{-i(k'-m')\theta}\mathcal{O}_{m,m'}
\end{align}
To solve the above equation, we must have $k=k(n)=\frac{n\pi}{\Lambda}$. Hence one can easily see that taking $\epsilon \rightarrow 0$ limit, the modes become continuous and one should replace sum with integral. Thus in this limit, one should use the following mode expansion for primary fields
\begin{align}
     \mathcal{O}(z,\bar{z}) =  \pi^{2}\beta^{-h}\int^{\infty}_{0}dk\int^{\infty}_{0}dk' (z-z_{+})^{\frac{ik}{\sqrt{d}}-h}(z-z_{-})^{-\frac{ik}{\sqrt{d}}-h}(\bar{z}-z_{+})^{\frac{ik'}{\sqrt{d}}-h}(\bar{z}-z_{-})^{-\frac{ik'}{\sqrt{d}}-h}\mathcal{O}_{k,k'}
\end{align}
We note that the above primary mode expansion is strictly valid for primaries not located at fixed points. All the above discussion on mode expansions will not be valid for primaries at fixed points in $\epsilon \rightarrow 0$ limit. We will see in later section that, this fact will become crucial in constructing a \textit{meaningful} Hilbert space structure in the same limit out of \textit{edge states}.
\subsection{Highest weight representation and GNS Hilbert space}
The goal of this present section is to construct a Hilbert space at finite $\epsilon$ cut-off, on which the local algebra acts. From the previous section, we have obtained a single copy of \textit{modular Virasoro algebra} (\ref{Final form of mod Vir algebra}) and the following kind of eigenmodes of $\hat{\mathcal{L}}^{\epsilon}_{0}$:
\begin{align}\label{all eigenmodes}
 &[ \hat{\mathcal{L}}^{\epsilon}_{0}, \hat{\mathcal{L}^{\epsilon}_{n}}]=-n\hat{\mathcal{L}}^{\epsilon}_{n} \\
& [ \hat{\mathcal{L}}^{\epsilon}_{0}, \mathcal{O}_{n,n'}]=-(n+n')\mathcal{O}_{n,n'}\\
& [ \hat{\mathcal{L}}^{\epsilon}_{0}, \hat{\mathcal{L}}^{\epsilon}_{m}\mathcal{O}_{n,n'}]=-(m+n+n')\hat{\mathcal{L}}^{\epsilon}_{m}\mathcal{O}_{n,n'} \\
& [\hat{\mathcal{L}}^{\epsilon}_{0}, \mathcal{O}(z_{+}+\epsilon e^{i\phi_{\epsilon}}, z_{-}+\epsilon e^{-i\phi_{\epsilon}})] = [\hat{\mathcal{L}}^{\epsilon}_{0}, \mathcal{O}(z_{-}+\epsilon e^{-i\phi_{\epsilon}},z_{+}+\epsilon e^{i\phi_{\epsilon}})] = \frac{8h\beta\epsilon\Lambda\cos\phi_{\epsilon}}{\pi}\mathcal{O}, \; h\sim \mathcal{O}(\frac{1}{\epsilon})
\end{align}
Using these eigenmodes, we will now construct a Hilbert space step by step.

\textbf{\textit{Hermitian conjugate:}} In \cite{Das:2024mlx}, the Hermitian conjugates of primaries and stress tensor in modular quantization have been discussed in details. Here we summarize the main results. To get an Unitary Lorentzian dynamics, governed by a Hermitian Hamiltonian $H=H^{\dagger}$, one need to define a conjugation operator in Euclidean time consistently. In particular, by demanding operators are Hermitian at time(t)=0\footnote{This is true for scalar operators by definition.}, would automatically imply the time evolution is Unitary. In Euclidean time($t_{E}$), this would require $\mathcal{O}^{\dagger}(t_{E})=\mathcal{O}(-t_{E})$. In the context of modular quantization, we require $\mathcal{O}^{\dagger}(\omega,\bar{\omega})=\mathcal{O}(-\bar{\omega},-\omega)$. In complex plane $(z,\bar{z})$ this would translate to\cite{Das:2024mlx}:
\begin{align}\label{conjugate}
    \mathcal{O}^{\dagger}(z,\bar{z}) = \mathcal{O}(z_{+}+z_{-}-\bar{z},z_{+}+z_{-}-z)
\end{align}
To check the consistency, one can consider scalars at fixed point $z=z_{+}$. Being zero modes of the Hamiltonian, scalars at fixed point would not flow with time and hence they will not change under Hermitian conjugation. This is also directly shown from (\ref{conjugate}) as
\begin{align}\label{fixed point Hermitian}
    \mathcal{O}^{\dagger}(z_{+},z_{-}) =  \mathcal{O}(z_{+},z_{-})
\end{align}
The same definition also holds with cut-off as one can easily check using the map($z'\rightarrow\omega$) with cut-off as in (\ref{map}). We will now derive how the scalar modes will transform under this conjugation. From (\ref{primary mode exp}), we have
\begin{align}
    \mathcal{O}^{\dagger}(z,\bar{z})= \frac{\pi^{2}\beta^{-h}}{\Lambda^{2}} \sum_{k,k'} (\bar{z}-z_{-})^{-\frac{ik}{\sqrt{d}}-h}(\bar{z}-z_{+})^{\frac{ik}{\sqrt{d}}-h}(z-z_{-})^{-\frac{ik'}{\sqrt{d}}-h}(z-z_{+})^{\frac{ik'}{\sqrt{d}}-h}\mathcal{O}^{\dagger}_{k,k'}
\end{align}
Also,
\begin{align}
  \mathcal{O}(z_{+}+z_{-}-\bar{z},z_{+}+z_{-}-z) =   \frac{\pi^{2}\beta^{-h}}{\Lambda^{2}} \sum_{k,k'} (z_{-}-\bar{z})^{\frac{ik}{\sqrt{d}}-h}(z_{+}-\bar{z})^{-\frac{ik}{\sqrt{d}}-h}(z_{-}-z)^{\frac{ik'}{\sqrt{d}}-h}(z_{+}-z)^{-\frac{ik'}{\sqrt{d}}-h}\mathcal{O}_{k,k'}
\end{align}
Hence, (\ref{conjugate}) implies\begin{align}
    \mathcal{O}^{\dagger}_{k,k'}=\mathcal{O}_{-k,-k'}
\end{align}
In a similar manner, one can also define Hermitian conjugation for $T(z), \bar{T}(\bar{z})$ and it's mode expansion as \cite{Das:2024mlx}
\begin{align}\label{unitarity 1}
    (\hat{\mathcal{L}}_{n}^{\epsilon})^{\dagger} = \hat{\mathcal{L}}^{\epsilon}_{-n}
\end{align}

\textbf{\textit{Construction of vacuum:}} The most vital part of constructing a Hilbert space is the choice of vacuum state. In our set-up, it must satisfy
\begin{align}
    \hat{\mathcal{L}}^{\epsilon}_{0}|0\rangle_{\epsilon} = 0
\end{align}
Apriori, there will be no reason to identify this vacuum $|0\rangle_{\epsilon}$ with the conformal vacuum $|0\rangle$ of radial quantization. Also, $\hat{\mathcal{L}}^{\epsilon}_{\pm 1}$ are not the usual global symmetry generators. Again, the uniqueness of $|0\rangle_{\epsilon}$ is not guaranteed. However, that is not an issue to construct a GNS Hilbert space which can be built from a cyclic state. Hence to find a cyclic vacuum, we will now follow a standard prescription motivated from constructing ground state in Euclidean path integral.

Under Euclidean time($t_{E}$) evolution, one can construct a state and then project it to the ground state of the Hamiltonian by sending $t_{E}\rightarrow -\infty$. Similarly, in Euclidean QFT, one can construct a wave-functional $\psi[\phi_{f}]=\langle \phi_{f}|\psi\rangle$ using transition amplitude $|\psi\rangle=U(t_{E}^{f}-t_{E}^{i})|\phi_{i}\rangle$ and projecting it to ground state by taking $t_{E}^{i}\rightarrow-\infty$ with $t_{E}^{f}$ fixed. Using path integral, one can construct ground state wave-functional by inserting identity operator $\mathcal{I}$ at $t^{i}_{E}=-\infty$.
\begin{align}
    \langle\phi_{f}|0\rangle =\lim_{t_{E}^{i}\rightarrow -\infty} \langle\phi_{f}|e^{-(t_{E}^{f}-t_{E}^{i})H}|\phi_{i}\rangle = \int^{\phi(t_{E}=t_{E}^{f})=\phi_{f}}_{\phi(t_{E}=-\infty)=\mathcal{I}} D\phi e^{-S[\phi]}
\end{align}
This definition is also consistent with conformal vacuum in radial quantization. Here we would like to adopt this definition to define vacuum in modular quantization. However, unlike radial quantization, here the Euclidean time $s$ is periodic. Naively, this looks like an obvious obstruction to define the vacuum. However, there is a special section of the curve, where taking $s\rightarrow-\infty$ still makes sense\footnote{I thank Bobby Ezhuthachan for noticing and emphasizing this special curve in related discussion.}.

As we already mentioned earlier, the contractible time$(s)$ circles are shrunk to zero size at $\theta=\pm \infty$, while they become of infinite size at $\theta=0$. The time curve equation (\ref{no cutoff curve}) at fixed $\theta$ has the following form \cite{Das:2024mlx}:
\begin{align}
    (x+\frac{\alpha}{2\beta})^{2}+(y-\frac{\sqrt{d}}{2\beta}\coth{\sqrt{d} \theta})^{2}=\frac{d}{4\beta^{2}\sinh^{2}(\sqrt{d}\theta)}, \; \text{where}\; x=Re(z),y=Im(z)
\end{align}
Hence, it is clear that $\theta=0$ correspond to the infinite time circle. Thus, at this point($\theta=0$), it makes sense to unwrap the $s$ circle: $s:(-\infty,\infty)$\footnote{In \cite{Das:2024mlx}, I have used a different argument to unwrap the time circle near the cut-off. That looks naive in the present context.}. For this, we do not need to use any wick rotation or Lorentzian analysis. We can now use this special point to define a vacuum, which is more convenient from a path integral point of view. In path integral language, we will insert identity operator at $(\theta=0,s=-\infty)$.

This choice of vacuum will therefore constrain the modes as in radial quantization. In more details, we will now use the fact that $T(z)|0\rangle_{\epsilon}$ and $\phi(z,\bar{z})|0\rangle_{\epsilon}$ are regular at $s=-\infty$. Using mode expansion for $T(z)$, which is similar to that for primaries (\ref{primary mode exp}) with $h=2$,
\begin{align}
    T(z)|0\rangle_{\epsilon} = \frac{\pi^{2}}{\Lambda^{2}\beta^{2}} \sum_{n}(z-z_{+})^{\frac{ik(n)}{\sqrt{d}}-2}(z-z_{-})^{-\frac{ik(n)}{\sqrt{d}}-2}\tilde{\mathcal{L}}_{n}^{\epsilon}|0\rangle_{\epsilon}
\end{align}
Here $k(n)=\frac{n\pi}{\Lambda}$. Using $\frac{z-z_{+}}{z-z_{-}}=e^{i\frac{\sqrt{d}\Lambda}{\pi}\omega}$ and $(z-z_{+})(z-z_{-})= \frac{d}{4\beta^{2}}\csc^{2}\left(\frac{\sqrt{d}\Lambda}{2\pi}\omega\right)$, we obtain
\begin{align}
  \lim_{\theta=0,s\rightarrow-\infty}T(z)|0\rangle_{\epsilon} = \lim_{\theta=0,s\rightarrow-\infty}\sum_{n}\left(\frac{4\beta^{2}}{d}\right)^{2}e^{-ns}\sinh^{2}\left(\frac{\sqrt{d}\Lambda}{2\pi}s\right)\tilde{\mathcal{L}}^{\epsilon}_{n}|0\rangle_{\epsilon}  
\end{align}
Thus regularity of $T(z)$ and $\bar{T}(\bar{z})$ acting on vacuum implies the following:
\begin{align}\label{vacuum cond 1}
    \tilde{\mathcal{L}}^{\epsilon}_{n}|0\rangle_{\epsilon} = \tilde{\bar{\mathcal{L}}}^{\epsilon}_{n}|0\rangle_{\epsilon} =\hat{\mathcal{L}}^{\epsilon}_{n}|0\rangle_{\epsilon}=  0, \; \forall\; n\geq0
\end{align}
Similarly finiteness of $\mathcal{O}(z,\bar{z})|0\rangle_{\epsilon}$ at the same point would imply:
\begin{align}
     \lim_{\theta=0,s\rightarrow-\infty}\mathcal{O}(z,\bar{z})|0\rangle_{\epsilon} = \lim_{\theta=0,s\rightarrow-\infty}\sum_{n,n'}\left(\frac{4\beta^{2}}{d}\right)^{2h}e^{-(n+n')s}\sinh^{4h}\left(\frac{\sqrt{d}\Lambda}{2\pi}s\right)\mathcal{O}_{n,n'}|0\rangle_{\epsilon}  = \text{finite}
\end{align}
From this, we have
\begin{align}\label{vacuum cond 2}
    \mathcal{O}_{n,n'}|0\rangle_{\epsilon}=0, \; \forall \;(n+n')>0
\end{align}
In a similar way, one can also show that sending $s\rightarrow \infty$, one would obtain the dual vector $_{\epsilon}\langle 0|$. Hence we can consistently develop an inner product with respect to vacuum. Using this, one have $_{\epsilon}\langle0|\hat{\mathcal{L}_{n}}=0$ for all $n\leq0$. From this we have 
\begin{align}
    _{\epsilon}\langle0|T(z)|0\rangle_{\epsilon}=_{\epsilon}\langle0|\bar{T}(\bar{z})|0\rangle_{\epsilon}=0,\; _{\epsilon}\langle0|\mathcal{O}(z,\bar{z})|0\rangle_{\epsilon} = \frac{\pi^{2}\beta^{-h}}{\Lambda^{2}}|z-z_{+}|^{2}|z-z_{-}|^{2}\;_{\epsilon}\langle 0|\mathcal{O}_{0,0}|0\rangle_{\epsilon}.
\end{align}
Note that, from $[\hat{\mathcal{L}}^{\epsilon}_{0},\mathcal{O}_{0,0}]=0$, we have either $\mathcal{O}_{0,0}|0\rangle_{\epsilon}=0$ or, $\mathcal{O}_{0,0}|0\rangle_{\epsilon}\propto |0\rangle_{\epsilon}$. This choice of vacuum will enable us to define a Unitary Hilbert space. By definition, here the primary bilocal modes on the vacuum $|0\rangle_{\epsilon}$ acts as a highest weight state.

\textbf{\textit{States in the Hilbert space and Unitarity:}} Using the Hermiticity of the cut-off Hamiltonian $(\hat{\mathcal{L}}^{\epsilon}_{0})^{\dagger}=\hat{\mathcal{L}}^{\epsilon}_{0}$ and it's eigenmodes (\ref{all eigenmodes}), together with the vacuum $|0\rangle_{\epsilon}$, we can always construct an orthonormal basis which span the Hilbert space. For instance, at level 1(eigenvalue 1), one can have degenerate eigenstates $\hat{\mathcal{L}}^{\epsilon}_{-1}|0\rangle_{\epsilon},\mathcal{O}_{-1,0}|0\rangle_{\epsilon},\mathcal{O}_{0,-1}|0\rangle_{\epsilon}$\footnote{From the modular Virasoro algebra, one can get $\hat{\mathcal{L}}^{\epsilon}_{1}\hat{\mathcal{L}}^{\epsilon}_{-1}|0\rangle_{\epsilon} = 0.$ Usually one would assume that this implies $\hat{\mathcal{L}}^{\epsilon}_{-1}|0\rangle_{\epsilon} =0$. We have not assumed this in our construction.}. From this, using standard Gram-Schmidt procedure, one can find an orthonormal basis. At level 2, similarly one have: 
\begin{align}
\text{level 2}: \;\hat{\mathcal{L}}^{\epsilon}_{-2}|0\rangle_{\epsilon}, (\hat{\mathcal{L}}^{\epsilon}_{-1})^{2}|0\rangle_{\epsilon, }\mathcal{O}_{-2,0}|0\rangle_{\epsilon}, \mathcal{O}_{0,-2}|0\rangle_{\epsilon}, \mathcal{O}_{-1,-1}|0\rangle_{\epsilon}, \hat{\mathcal{L}}^{\epsilon}_{-1}\mathcal{O}_{-1,0}|0\rangle_{\epsilon}, \hat{\mathcal{L}}^{\epsilon}_{-1}\mathcal{O}_{0,-1}|0\rangle_{\epsilon}
\end{align}
However, unlike radial quantization, this construction is not purely kinematical. This is primarily due to lack of a state-operator correspondence. For instance, to orthonormalize $\mathcal{O}_{-n,-n'}|0\rangle_{\epsilon}$, one need to know about CFT data like OPE coefficient. However, for the present note, the details of the Hilbert space is not needed. 

On the other hand, we still have some Unitarity constraint, which is already manifested in radial quantization. For instance, using
(\ref{unitarity 1}) and (\ref{vacuum cond 1}) we obtain the following from positivity of the norm:
\begin{align}
    _{\epsilon}\langle0|\hat{\mathcal{L}}^{\epsilon}_{n}\hat{\mathcal{L}}^{\epsilon}_{-n}|0\rangle_{\epsilon}\geq 0 \; \implies c\geq 0
\end{align}
We expect interesting dynamical constraints may come from the positivity of norm of states like $\mathcal{O}_{-n,-n'}|0\rangle_{\epsilon}$. For this, one need to understand and solve `Ward identities' of modular quantization involving two point function of primaries. At present, we do not have anything more to say on these.

\textbf{\textit{Cyclicity of vacuum and local operator $\implies$ states:}} Even though, we presently lack a detailed structure of the Hilbert space, the constructibility of a well defined Hilbert space should be guaranteed. We can use the power of GNS Hilbert space construction out of algebras of bounded operators with respect to a cyclic state. In our context, we need to show the vacuum $|0\rangle_{\epsilon}$ is a cyclic state. 

In CFTs(in the absence of a global charge), all local operators can be built out of primaries and Virasoro descendants. From the Hermitian conjugate of primaries and stress tensor (\ref{conjugate}), we can see that in vacuum $|0\rangle_{\epsilon}$, the norm of primaries $\mathcal{O}(z,\bar{z})$ and \textit{modular Virasoro descendants} $\hat{\mathcal{L}}^{\epsilon}_{-n_{1}}\dots\hat{\mathcal{L}}^{\epsilon}_{-n_{k}}\mathcal{O}(z,\bar{z})$ will be finite. The short distance divergences will be absent from the definition (\ref{conjugate}), within the cut-off picture\footnote{The coincident point limit is taking $Re(z)=-\frac{\alpha}{\beta},Im(z)=0$, which lies in the branch-cut axes of quantization contour as in fig(). Thus the algebra by definition does not contain any operators lying on the branch-cut.}. Hence all $\mathcal{O},T\mathcal{O},TT\mathcal{O},\dots$ will have finite norm and hence they are bounded local operators with respect to vacuum. Now since, any other states in the constructed Hilbert space are locally look like vacuum \cite{Witten:2018zxz}, the absence of short distance or coincident point divergence will be guaranteed. 

Also, using operator product expansion\footnote{Even though, OPE is derived from state-operator correspondence in radial quantization, we can still use it as a pure operator statement of CFT in complex plane. This does not depend on states of any Hilbert space.}, we can define a local algebra of operators $\mathcal{A}$ from local operators in the CFT. Therefore we can construct an irreducible representation of bounded operators $\pi_{|0\rangle_{\epsilon}}(A)$ acting on $\mathcal{H}_{|0\rangle_{\epsilon}}$- the Hilbert space constructed out of $|0\rangle_{\epsilon}$ and the eigenbasis of $\hat{\mathcal{L}}^{\epsilon}_{0}$. Here $A\in \mathcal{A}$. Since all local operators do have mode expansion in terms of eigenmodes of $\hat{\mathcal{L}}^{\epsilon}_{0}$ as we described earlier, one can always write such operators acting on $|0\rangle_{\epsilon}$ as linear combination of basis states in the Hilbert space $\mathcal{H}_{|0\rangle_{\epsilon}}$. This tells
\begin{align}
    \textit{(Local operators)}|0\rangle_{\epsilon} \implies \textit{States in}\; \mathcal{H}_{|0\rangle_{\epsilon}} 
\end{align}
This will show the cyclicity of the vacuum and \textit{one-sided} operator $\implies$ state correspondence. However, we still do not have state $\implies$ local operator correspondence. 

\textbf{\textit{Null states:}} As in radial quantization, one can also in principle construct null states $|v\rangle_{\epsilon}$, such that $_{\epsilon}\langle v^{\dagger}v\rangle_{\epsilon} = 0$ and $_{\epsilon}\langle v^{\dagger}\psi\rangle_{\epsilon} =0$. Here $|\psi\rangle_{\epsilon}$ is any other state in the Hilbert space $\mathcal{H}_{|0\rangle_{\epsilon}}$. To construct this, we again need to know the details of the Hilbert space. However, the existence of such null states can never be ignored. To construct the physical Hilbert space, we must project out the null states. 

\textbf{\textit{GNS Hilbert space and type-I Von-Neuman algebra:}}
Using all these ingredients, we can now proceed towards constructing a GNS Hilbert space. For a review, see appendix A of \cite{Magan:2020iac}. The usefulness of GNS construction relies on the fact that in spite of not knowing the detailed dynamics of a system(or knowing the Hamiltonian governs any dynamics), one can still construct a Hilbert space $\mathcal{H}_{\omega}$ out of an abstract $C^{*}$ algebra $\mathcal{A}$ and a given linear functional or a state $\omega$ which maps $\mathcal{A}\rightarrow\mathbb{C}$. Then one find representation of algebra $\pi_{\omega}(\mathcal{A})$ acting on $\mathcal{H}_{\omega}$. Here following\cite{Magan:2020iac}, we provide a step by step procedure to construct a GNS Hilbert space from modular quantization.
\begin{itemize}
    \item In a $C^*$ algebra, there exists a map $A \rightarrow A^{*}$ $\forall A\in \mathcal{A}$, such that for $B\in \mathcal{A}$
    \begin{align}
        (AB)^{*}=B^{*}A^{*},\; (aA)^{*} = \bar{a} A^{*}, \; \forall a\in \mathbb{C}.
    \end{align}

In modular quantization, we have seen the conjugate operation $\dagger$ plays the same role of $*$, satisfying all of the above condition.
\item $\omega$ is a positive and normalized linear functional which maps $\mathcal{A}\rightarrow\mathbb{C}$. $\omega$ satisfies the following properties:
\begin{align}
    \omega(A^{*}A)\geq 0, \; \omega(\mathcal{I})=1
\end{align}
In modular quantization, we have constructed $|0\rangle_{\epsilon}$, which plays the role of $\omega$. Positivity of the norm as well as the definition of vacuum would guarantee the above conditions.
\item To construct the Hilbert space $\mathcal{H}_{\omega}$, one need to define an inner product structure which can be written in $\omega$ as
\begin{align}
    \langle A|B\rangle = \omega(A^{*}B), \; \forall A.B\in \mathcal{A}.
\end{align}
In our set-up, we construct a pure state $|0\rangle_{\epsilon}$ and define it's conjugate $_{\epsilon}\langle 0|$, which also define an inner product structure satisfying the above. Hence we can write $\omega$ as pure state density operator $\rho_{0}=|0\rangle_{\epsilon}\langle 0|$. Hence by definition this is a projection operator. Here $\omega(A^{\dagger}B)$ is replaced by $Tr(\rho_{0}A^{\dagger}B)$.
\item There may exists non zero operators $W$, for which $\omega(W^{*}W)=0$. The set of such operators $\{W\}$ form a left Ideal $I$, such that for any element $A\in \mathcal{A}$, $AW\in I$. GNS Hilbert space can be constructed by quotienting the ideal subspace. Hence $\mathcal{H}_{\omega}\equiv \mathcal{A}/I$ acting on $\omega$. In a similar fashion, we have already mentioned how null states could exist in the modular quantized Hilbert space and we should quotient out those states. Hence we should have states $|[\psi]\rangle_{\epsilon}\in \mathcal{H}_{|0\rangle_{\epsilon}}$, which is made out of equivalent class of operators $\mathcal{A}+I$. 
\item One can define a GNS induced representation $\pi_{\omega}(\mathcal{A})$ as
\begin{align}
 \pi_{\omega}:\mathcal{A}\rightarrow\mathcal{B}(\mathcal{H}_{\omega}) \; \text{such that}, \; \pi_{\omega}(A)|[B]\rangle_{\omega} = [|AB]\rangle_{\omega}   
\end{align}
Here $\mathcal{B}(\mathcal{H}_{\omega})$ is the set of bounded operators in the Hilbert space. In our set-up, existence of OPE and bounded local operators $\implies$ states justify the above representation $\pi_{|0\rangle_{\epsilon}}(A)$. Since $\omega$ is pure, this is irreducible representation. 

\item To summarize, we have obtained a GNS Hilbert space(isomorphic to the full Hilbert space constructed out of highest weight representation) out of local(bounded) operators in the CFT with regulated boundaries and a cyclic vacuum $|0\rangle_{\epsilon}$. In this sense, we have a \textit{state-operator correspondence}, albeit different from the radial quantization.

\item We will now call this Hilbert space $\mathcal{H}_{|0\rangle_{\epsilon}}$ as the \textit{bulk Hilbert space}\footnote{Since our original set-up is actually a BCFT due to conformal cut-off boundaries, there exists two different notion of Hilbert spaces. Here by \textit{bulk} Hilbert space, we mean operators living away from the boundary and have a bulk OPE channel.} or sometimes \textit{asymptotic Hilbert space}\footnote{This is purely motivated from the fact that we construct the vacuum at $\theta=0$, faraway from the cut-off region $\theta=\pm \Lambda$. This is \textit{asymptotic} region with respect to cut-off.}.
\end{itemize}
Now we can move to the original starting problem where we have a CFT on cylinder with Hamiltonian $H$ and the existence of fixed points would force us to divide the spatial region into two sub-parts. The quantization contour at fixed time-slice would describe either of two sided of the spatial circle divided by two fixed points. To quantize the one-sided Hamiltonian we have to find spectrum. We have seen, to find a sensible Hilbert space made out of algebras of operators of each sides, we must use cut-offs near two fixed points. In other words, by using cut-offs we have a \textit{emergent modular Virasoro algebra} and hence describe a BCFT, different from the usual radial quantization. In the presence of cut-off, we have two independent Hilbert spaces $\mathcal{H}^{L}_{\epsilon}$ and $\mathcal{H}^{R}_{\epsilon}$ constructed out of local algebras in two subregions. Here the total Hilbert space is constructed out of quantization of $H^{total}_{\epsilon}=H^{R}_{\epsilon}\otimes\mathcal{I}_{L}-\mathcal{I}_{R}\otimes H^{L} _{\epsilon}$ \cite{Das:2024vqe}. Even though we start with $C^*$ algebra, we have
\begin{align}
    \left(\mathcal{A}(\mathcal{B}(\mathcal{H}_{\epsilon}^{L})\otimes\mathcal{I}^{R})\right)^{''} = \mathcal{A}(\mathcal{I}^{L}\otimes\mathcal{B}(\mathcal{H}_{\epsilon}^{R}))
\end{align}
Here $\mathcal{A}^{'}$ defines the commutant algebra of $\mathcal{A}$. This is guaranteed in our cut-off set-up at a constant time-slice. By bi-commutant theorem, the algebras $\mathcal{A}^{\epsilon}_{L,R}$ are Von-Neuman algebras. Hence by split property, we have a factorization of Hilbert space:
\begin{align}
    \mathcal{H}^{tot}_{\epsilon}=\mathcal{H}^{L}_{\epsilon}\otimes\mathcal{H}^{R}_{\epsilon}
\end{align}
From all of the above discussion, the algebras of bounded operators in one-sided modular quantized space form a type-I Von-Neuman factor\footnote{Factor is quite obvious from the existence of cut-off dividing the Hilbert spaces. There is no common spatial region exists for the two algebras.}. The discreteness of the one-sided modular spectrum also manifests this.

\subsection{Emergent type-III$_{1}$ algebra in $\epsilon \rightarrow 0$ limit}
We have so far only discussed the \textit{bulk Hilbert space} in our BCFT set-up. However, as we know there also exist a \textit{boundary Hilbert space} with respect to a dual channel in BCFT literature. This is constructed out of a localized(normalized) boundary state\footnote{In principle, the boundary state is not normalized. Using certain cut-off one can make it normalizable \cite{Calabrese:2016xau}.} and local excitations on it. The local excitations on the cut-off boundaries can be treated as boundary condition changing(bcc) operators. Hence, we expect there should be another OPE, which is sometimes called the boundary OPE, where a single bulk operator can be expanded in terms of \textit{localized} boundary operators located at that cut-off boundaries\footnote{In other words, this should be justified from the fact that one point functions of bulk operators(in vacuum $|0\rangle_{\epsilon}$) are nonzero as we have seen earlier.}. The consistency in BCFT hence requires equivalence of two different channels of OPE. In other language, this is clearly visible from \textit{`open-closed string' duality}, where `open string' partition function can be written as transition amplitude of `closed strings' propagating between two boundary states. This duality can also be understood from modular invariant partition function on annulus. We will use this to understand the boundary Hilbert space and the shrinking limit $\epsilon \rightarrow 0$.

\textbf{\textit{Shrinking limit and regularity at fixed points:}}
The conformal boundary condition we have used at the cut-off has the following form:
\begin{align}
    \left(T(\omega)-\bar{T}(\bar{\omega})\right)|_{\theta=\pm \Lambda} = 0
\end{align}
By inverting (\ref{modes in omega}), we can write $T(\omega),\bar{T}(\bar{\omega})$ in terms of $\mathcal{L}_{n},\bar{\mathcal{L}}_{n}$ as the following:
\begin{align}
    T(\omega)=\frac{\pi}{\Lambda}\sum_{m}e^{-\frac{m\pi}{\Lambda}\omega}\left(\mathcal{L}_{m}^{\epsilon}+\frac{c\Lambda d}{24\pi}\delta_{m,0}\right), \; \bar{T}(\bar{\omega}) = \frac{\pi}{\Lambda}\sum_{m}e^{-\frac{m\pi}{\Lambda}\bar{\omega}}\left(\mathcal{L}_{m}^{\epsilon}+\frac{c\Lambda d}{24\pi}\delta_{m,0}\right)
\end{align}
One can easily check that the boundary condition now translates to
\begin{align}
    \mathcal{L}^{\epsilon}_{n}-\bar{\mathcal{L}}^{\epsilon}_{-n} = 0,\; \forall \; n\in \mathbb{Z}.
\end{align}
We are interested to take it's $\epsilon \rightarrow 0$ limit, for which the condition will reduce to $T(\omega)=\bar{T}(\bar{\omega})=\frac{c}{24}d$ at $\theta=\pm \infty$. Also at the fixed points, the same limit must satisfy (\ref{fixed point bc}) as we discussed previously. Hence a consistent regularity condition at fixed point requires\footnote{Since we demand $T(\omega)|f\rangle, \bar{T}(\bar{\omega})|f\rangle$ to be regular for $\theta\rightarrow \pm \infty$. This is a consistent demand to get finite $T(\omega),\bar{T}(\bar{\omega})$ at fixed points by conformal transformation as we argued before, where by definition $T(z_{\pm})$ should be regular.} 
\begin{align}\label{bc in limit}
    \lim_{\epsilon\rightarrow 0}\mathcal{L}^{\epsilon}_{n}|f\rangle \; \text{and} \; \lim_{\epsilon\rightarrow 0}\bar{\mathcal{L}}^{\epsilon}_{n}|f\rangle\; \text{to be regular}
\end{align}
Here $|f\rangle$ is a state which satisfies the above regularity condition at fixed points. Since we argued at $\epsilon \rightarrow 0$, the spectrum becomes continuous, we use (\ref{Vir modes}) $\mathcal{L}_{k}$ in terms of $k$, as the representation for $\lim_{\epsilon\rightarrow 0}\mathcal{L}^{\epsilon}_{n}$. In writing this, we can use radial mode expansion of $T(z)$ around $z=z_{+}$ as $T(z)=\sum_{m\in \mathbb{Z}}(z-z_{+})^{-m-2}L_{m}$. By substituting this mode expansion in (\ref{Vir modes}), we obtain:
\begin{align}
    \mathcal{L}_{k} = \frac{\beta}{2\pi i}\sum_{m\in \mathbb{Z}}\int_{z_{-}}^{z_{+}} dz (z-z_{+})^{-\frac{ik}{\sqrt{d}}-m-1}(z-z_{-})^{\frac{ik}{\sqrt{d}}+1}L_{m}
\end{align}
By using change of variable $t=\frac{z-z_{-}}{z_{+}-z}$, we can rewrite the above as
\begin{align}
    \mathcal{L}_{k} &= \frac{\beta}{2\pi} \sum_{m\in \mathbb{Z}} \left(\frac{i\sqrt{d}}{\beta}\right)^{1-m}\int^{\infty}_{0}dt \; t^{\frac{ik}{\sqrt{d}}+1}(1+t)^{m-2} L_{m} \nonumber \\
    & = \frac{\beta}{2\pi} \sum_{m\in \mathbb{Z}} \left(\frac{i\sqrt{d}}{\beta}\right)^{1-m} B\left(\frac{ik}{\sqrt{d}}+2,-m-\frac{ik}{\sqrt{d}}\right)L_{m}
\end{align}
The analytic Beta function $B(z_{1},z_{2})$ with complex variables ($z_{1},z_{2}$) is defined for $Re(z_{1})>0, Re(z_{2})>0$. Since we want the action of $\mathcal{L}_{k}$ to be well-defined at fixed point, we must want $B\left(\frac{ik}{\sqrt{d}}+2,-m-\frac{ik}{\sqrt{d}}\right)$ to be analytic. Hence regularity of $\mathcal{L}_{k}$ acting on $|f\rangle$ demands $L_{m}|f\rangle=0$ for all $m>0$. We know one such solution is primary state as being the highest weight state in radial quantization. In a similar manner, one can also get $\bar{\mathcal{L}}_{k}|f\rangle$ is regular $\implies \bar{L}_{n}|f\rangle = 0$ for all $n>0$. This is again satisfied by primaries at fixed points. This would satisfy the regularity condition in (\ref{bc in limit}). 

Independently, it has also been observed in (\ref{primary eigenmodes}), that scalar primaries located at fixed points are eigenmodes of one-sided Hamiltonian $\hat{\mathcal{L}}^{\epsilon}_{0}$ in the $\epsilon \rightarrow 0$ limit. They are zero modes for $h<<\mathcal{O}(\frac{1}{\epsilon})$ and become non-zero modes when $h\geq\mathcal{O}(\frac{1}{\epsilon})$ with $h\epsilon$ fixed. However they are zero modes for the full or two-sided Hamiltonian $H$ by definition. In \cite{Das:2024mlx}, it has been shown that such zero modes acting on conformal vacuum\footnote{SL(2,$\mathbb{C}$) invariant vacuum.} are degenerate vacuum states of the full two-sided Hamiltonian. However, the non-trivial point is that \textit{this conformal vacuum $|0\rangle$ is not a part of the one-sided Hilbert space $\mathcal{H}_{|0\rangle_{\epsilon}}$ we constructed in $\epsilon \rightarrow 0$ limit.}. Hence the natural question is, how such conformal primary states could emerge in taking $\epsilon \rightarrow 0$ limit. To understand this, we first study the localized boundary state satisfying the boundary condition at cut-offs.

\textbf{\textit{Open-closed string duality:}} We will now show that these primary states will naturally appear in closed string channel after shrinking the boundary states to the fixed point in $\epsilon \rightarrow 0$. To define the closed string channel, we must work in ($\omega,\bar{\omega}$) or proper conformal frame of quantization. The topology of the $\omega$ frame is described by a \textit{finite thermal cylinder} where time is periodic. To understand it better, one can think of the map $z\rightarrow \omega$  (\ref{map}) as the following:
\begin{align}
    z\rightarrow \zeta=\frac{z-z_{+}}{z-z_{-}}=e^{i\sqrt{d}\omega}
\end{align}
Here $\zeta\rightarrow \omega$ is a mapping to thermal cylinder, where at $\zeta=0,\infty$, the cylinder smoothly caps off to the fixed points $\theta=\pm\infty$\footnote{More correctly, this geometry is similar to an American football.}. However, since we remove a disk of radius $\epsilon$ from the complex plane around $z=z_{\pm}$, upon mapping to $\omega$ plane, the cylinder has finite length of $2\Lambda$ where $\theta:(-\Lambda,\Lambda)$. We have imposed conformal boundary conditions at two edges of the cylinder. The Hamiltonian in $\omega$ frame can be written as (\ref{modes in omega}):
\begin{align}
  H_{\omega}=  \frac{1}{2\pi}\int^{\Lambda}_{-\Lambda}d\theta (T(\omega)+\bar{T}(\bar{\omega})) = \mathcal{L}^{\epsilon}_{0}+\bar{\mathcal{L}}^{\epsilon}_{0}+\frac{c\Lambda d}{12\pi}
\end{align}
Since the Hamiltonian $H_{\omega}$ generates time translation $s\sim s+\frac{2\pi}{\sqrt{d}}$, the thermal partition function on the cylinder can be written as:
\begin{align}
    Z_{\epsilon} = Tr_{\mathcal{H}_{\epsilon}}e^{-\frac{2\pi}{\sqrt{d}}H_{\omega}}=Tr_{\mathcal{H}_{\epsilon}}e^{-\frac{2\pi}{\sqrt{d}}\left(\mathcal{L}^{\epsilon}_{0}+\bar{\mathcal{L}}^{\epsilon}_{0}+\frac{c\Lambda d}{12\pi}\right)} = Tr_{\mathcal{H}_{\epsilon}}e^{-\frac{2\pi^{2}}{\Lambda\sqrt{d}}\left(\tilde{\mathcal{L}}^{\epsilon}_{0}+\tilde{\bar{\mathcal{L}}}^{\epsilon}_{0}+\frac{c\Lambda^{2} d}{12\pi^{2}}\right)} = Tr_{\mathcal{H}_{\epsilon}}e^{-\frac{2\pi^{2}}{\Lambda\sqrt{d}}\left(\hat{\mathcal{L}}^{\epsilon}_{0}+\frac{c\Lambda^{2} d}{12\pi^{2}}\right)}
\end{align}
We recall that we have absorbed $\hat{\mathcal{L}}_{0}^{\epsilon} \rightarrow \hat{\mathcal{L}}_{0}^{\epsilon} +\frac{c\Lambda^{2} d}{12\pi^{2}}$ \footnote{We neglect the 1 factor in (\ref{shift}) as $\Lambda^{2}>>1$.}. Thus effectively $Z_{\epsilon} = Tr_{\mathcal{H}_{\epsilon}}e^{-\frac{2\pi^{2}}{\Lambda\sqrt{d}}H_{\epsilon}}$. Note that, we can write a modular parameter\footnote{This is related to the modular parameter of the annulus as finite cylinder is related to it conformally.} $q=e^{2\pi i\tau}$, where $\tau=\frac{i\pi}{\Lambda\sqrt{d}}$ and $|\tau|<1$. In the dual closed string channel, the SL(2,$\mathbb{Z}$) invariant partition function can be written as a transition amplitude of closed strings between two states \cite{DiFrancesco:1997nk}, where time and space reverses the role. The $\theta$ direction is the time evolution in the closed string channel and the corresponding Hamiltonian can be written by mapping $\omega \rightarrow \zeta$ plane,\footnote{Note that, $\zeta$ is the annulus frame or a sphere with two cuts at two poles.} where $\zeta=e^{i\sqrt{d}\omega}$. We can write it as,
\begin{align}
    H_{\zeta} = \sqrt{d}\left(L_{0}^{\zeta}+\bar{L}_{0}^{\bar{\zeta}}-\frac{c}{12}\right)
\end{align}
Hence the transition amplitude should be written as $_{\epsilon}\langle b|e^{-2\Lambda H_{\zeta}}|a\rangle_{\epsilon}$, where $|b\rangle_{\epsilon}$ and $|a\rangle_{\epsilon}$ are two boundary states correspondingly. To check that this transition amplitude is related to modular S-transformation $\tau \rightarrow -\frac{1}{\tau}$, we define $\tilde{q}=e^{-\frac{2\pi i}{\tau}}$.
\begin{align}
    \tilde{q}^{\left(L_{0}^{\zeta}+\bar{L}_{0}^{\bar{\zeta}}-\frac{c}{12}\right)} = e^{-2\Lambda\sqrt{d}\left(L_{0}^{\zeta}+\bar{L}_{0}^{\bar{\zeta}}-\frac{c}{12}\right)} = e^{-2\Lambda H_{\zeta}}
\end{align}
Thus by open-closed string duality we obtain:
\begin{align}\label{open closed part}
    Z_{\epsilon}^{ab} = Tr_{\mathcal{H}_{\epsilon}^{\{ab\}}}e^{-\frac{2\pi^{2}}{\Lambda\sqrt{d}}H_{\epsilon}} =\; _{\epsilon}\langle b|e^{-2\Lambda\sqrt{d}\left(L_{0}^{\zeta}+\bar{L}_{0}^{\bar{\zeta}}-\frac{c}{12}\right)}|a\rangle_{\epsilon}
\end{align}
Here $\mathcal{H}_{\epsilon}^{\{ab\}}$ refers to the Hilbert space subject to conformal boundary conditions $|a\rangle_{\epsilon}, |b\rangle_{\epsilon}$. In stead of computing the partition function in the $\epsilon \rightarrow 0$ limit, we will now interested in the shrinking limit of the boundary states itself. Computing partition function will only dominates the vacuum saddle as we will see next section. However, here we are purely interested to understand the shrinking limit of the states, which will be there beyond dominant thermodynamic saddles. By definition, those do not contribute to the `open string' partition function at leading order.

Since the Hilbert space of the closed string channel is generated by $H_{\zeta}$, we can use states of radial quantization to describe it. The boundary condition at $\theta=\pm \Lambda$ can be mapped to $T(\zeta)\zeta^{2}=\bar{T}(\bar{\zeta})\bar{\zeta}^{2}$. By using mode expansion we get 
\begin{align}
    \sum_{n}L^{\zeta}_{n}\zeta^{-n} = \sum_{m}\bar{L}^{\zeta}_{m}\bar{\zeta}^{-m}
\end{align}
After writing $\zeta=e^{i\sqrt{d}\omega}$ and using the conformal boundary condition at $\theta=\pm\Lambda$, one would end up with
\begin{align}\label{bc in zeta}
    (L_{n}^{\zeta'}-\bar{L}_{-n}^{\zeta'})|a\rangle_{\epsilon} = (L_{n}^{\zeta''}-\bar{L}_{-n}^{\zeta''})|b\rangle_{\epsilon} = 0
\end{align}
Here $\zeta'=e^{\Lambda\sqrt{d}}\zeta$ and $\zeta''=e^{-\Lambda\sqrt{d}}\zeta$. This is the standard Cardy boundary condition with $L^{\zeta'}_{n}$ and $L_{n}^{\zeta''}$, defined at two cut-offs along the contour of radius of the order $\epsilon$ and $\frac{1}{\epsilon}$ respectively. This is because the cut-offs in $z$ plane are mapped to $\zeta^{c}_{\pm} = (-\frac{i\beta\epsilon e^{i\phi_{\epsilon}}}{\sqrt{d}},-\frac{i\sqrt{d}e^{i\phi_{\epsilon}}}{\beta\epsilon})$.

\textbf{\textit{Shrinking limit and comment to asymptotic-state correspondence in `angular quantization':}}
The exact same set-up of $\zeta\rightarrow\omega$ has been discussed in the context of `angular quantization' introduced in \cite{Agia:2022srj}. The $\zeta$ plane is thought of as a sphere with two excised holes cut near two fixed points $\zeta_{\pm}=0,\infty$ of sizes $|z|=\epsilon$ and $|z|=\frac{1}{\epsilon}$ respectively. The authors prescribed a natural shrinking limit $\epsilon \rightarrow 0$ in the closed string channel. To do this, the authors provides a path integral argument where it was \textit{assumed} that a physical boundary state wave functional could be expanded in terms of local operator wave-functional basis located at the fixed points. The resulting shrinking limit obtained is proportional to $\epsilon^{\Delta_{\mathcal{O}_{min}}}$, where $\mathcal{O}_{min}$ is the lowest-lying primary located at fixed point. Hence the authors argue that most of the conformal boundary condition shrinks to identity from the entropic dominance. While this is true for our case too, here we present a different and direct approach to find \textit{the shrinking limit of the states itself, rather than finding it's contribution to dominant saddle of the partition function.}

The solution of (\ref{bc in zeta}) is well known as Cardy boundary state, which is a linear combination of Ishibashi state $|h\rangle\rangle$:
\begin{align}
    |a\rangle_{\epsilon} = \sum_{h}a_{h}|h\rangle\rangle, \; \text{where}\; |h\rangle\rangle = \sum_{N=0}^{\infty}\sum_{j=1}^{d(N)}|h,N,j\rangle\otimes\overline{|h,N,j\rangle}
\end{align}
Here $|h,N,j\rangle$ are orthonormal basis states made out of primaries $|h\rangle$ and their descendants $(L_{-1}^{\zeta'})^{N_{1}}(L_{-2}^{\zeta'})^{N_{2}}\dots|h\rangle$ of level $N$. $d(N)$ is the degeneracy of level N. We can see that $L_{-N}^{\zeta'}$ can be written in terms of $L_{-N}^{\zeta}$ in the following way:
\begin{align}
    L_{-n}^{\zeta'} = \oint d\zeta' \frac{T(\zeta')}{(\zeta')^{n-1}} = \oint_{0}d\zeta \; e^{-n\Lambda\sqrt{d}}\frac{T(\zeta)}{\zeta^{n-1}}
\end{align}
Hence one obtain
\begin{align}
    L_{-n}^{\zeta'}|h\rangle = L_{-n}^{\zeta'}\mathcal{O}_{h,\bar{h}}(0,0)|0\rangle = \oint_{0}\frac{d\zeta}{\zeta^{n-1}}e^{-n\Lambda\sqrt{d}}\left[\frac{\partial_{\zeta}\mathcal{O}(0,0)}{\zeta}+\frac{h\phi(0,0)}{\zeta^{2}}\right]
\end{align}
Taking $\Lambda \rightarrow \infty$, we can see all $L_{-n}^{\zeta'}|h\rangle$ vanishes for all $n>0$. Hence the only non-vanishing contribution comes from $n=0$. Thus we have
\begin{align}
    \lim_{\epsilon\rightarrow 0}L_{-n}^{\zeta'}|h\rangle = |h\rangle
\end{align}
Since the Ishibashi state is created with descendants of the form $L_{-n}^{\zeta'}$, after taking the shrinking limit we would end with a single primary state. As we know in Ishibashi state, the dimension $h=\bar{h}$, we will only end up with scalar operators. Hence finally we have,
\begin{align}
    \lim_{\epsilon\rightarrow 0}|a\rangle_{\epsilon} = \sum_{h}a_{h}\mathcal{O}_{h,h}(0,0)|0\rangle.
\end{align}
In a similar way, we can also obtain,
\begin{align}
    \lim_{\epsilon\rightarrow 0}\; _{\epsilon}\langle b| = \lim_{\zeta\rightarrow\infty}\sum_{h}b_{h}\zeta^{2h}\langle0|\mathcal{O}_{h,h}(\zeta,\zeta)
\end{align}
Hence for a given conformal boundary condition at fixed point cut-off, we end up with scalar primaries located at fixed points after taking $\epsilon\rightarrow 0$. The $a_{h}$ and $b_{h}$ depend on physical boundary condition, which satisfy open-closed string duality.

\textbf{\textit{Type-III$_{1}$ algebra in $\epsilon \rightarrow 0$:}} 
We will now take the $\epsilon\rightarrow 0$ limit of one-sided algebra $\mathcal{A}(\mathcal{B}(\mathcal{H}_{\epsilon}))$. We have already discussed in the previous section that in this limit, the eigenmodes $\mathcal{O}_{k,k'},\mathcal{L}_{k}$ becomes continuous and unbounded. Since we constructed the GNS Hilbert space by acting local operators on $|0\rangle_{\epsilon}$ and local operators can be expanded in those modular eigenmodes, those states become unbounded and not well-defined. This already indicates a type-III algebra. To understand the exact class of type-III, one need to study intersection spectrum of $\Delta_{\psi}\equiv e^{-H^{tot}_{\psi}}$\cite{Witten:2018zxz}. Here $H^{tot}_{\psi}$ is identified with two-sided or total modular Hamiltonian $H_{R}-H_{L}$ with respect to some normal state $\psi$. The intersection is called Averson spectrum \cite{Furuya:2023fei} having the property:
\begin{align}
    S(\mathcal{A}) \equiv \cap_{\psi} \text{Spec}(\Delta_{\psi}) = \mathbb{R}_{+} \; \text{for} \; \text{Type-III}_{1}
\end{align}
As argued by Witten \cite{Witten:2018zxz}, it is enough to check Spec$(\Delta_{0})$ for a cyclic and separating vacuum $|0\rangle$ to see whether the algebra is type-III$_{1}$. Here if we choose $|0\rangle_{\epsilon}$ to be the cyclic vacuum, then from the continuous spectrum it is guaranteed that $S(\mathcal{A}) = \mathbb{R}_{+}$, since spectrum of $H^{tot}_{|0\rangle_{\epsilon}}$ contains all continuous numbers in $\epsilon \rightarrow 0$.

However, the interesting point is to see whether the algebra has a trivial center or not in $\epsilon \rightarrow 0$. 
To be more precise, one can think of a sequence of scalar operators $\{\mathcal{O}(z_{\pm}+\epsilon),\mathcal{O}(z_{\pm}+\epsilon^{2}),\mathcal{O}(z_{\pm}+\epsilon^{3}),\dots \mathcal{O}(z_{\pm}+\epsilon^{N})\}$ while taking $\epsilon \rightarrow 0$ limit. In $N \rightarrow \infty$ limit, the sequence reduces to $\mathcal{O}(z_{\pm})$. In other words, for such sequence, taking $N \rightarrow \infty$ limit is same as taking $\epsilon \rightarrow 0$ limit for a sequence of operators of the above. This limiting procedure can be best understood from closed string channel, where the boundary states also have a $\epsilon \rightarrow 0$ limit as we have discussed. Hence as in \cite{Gesteau:2025obm}, we can explicitly see by denoting the sequence $S_{\epsilon}$, that
\begin{align}
 \lim_{\epsilon\rightarrow 0}   \langle b_{\epsilon}|S_{\epsilon}|a_{\epsilon}\rangle = \langle b|S|a\rangle ,
\end{align}
where $|b\rangle, |a\rangle$ are the limiting primaries at the fixed points, whereas the $S$ denotes the limiting fixed point scalars.

However, as we showed in Unitarity argument, these are strictly unbounded with infinite norm in modular quantization. We can still construct bounded operators from those fixed point scalars by smearing them appropriately \cite{Witten:2018zxz}. After forming the bounded sector of those fixed point scalars, we can include them to construct a larger algebra in a similar spirit to crossed product construction:
\begin{align}
\mathcal{A}_{new} \equiv  \mathcal{A}(\mathcal{B}(\mathcal{H}_{\epsilon\rightarrow 0}))\otimes (\oplus_{h}\mathcal{A}_{f}(h))  
\end{align}
 Here $\mathcal{A}_{f}$ denotes the affiliated algebra of fixed point scalars which are included in the closed limit. In Euclidean, it can be trivially shown that scalar at the fixed points corresponding to each superselection sector $|h\rangle$, would commute with any other local operators located outside the fixed points. Since we construct the algebra $\mathcal{A}(\mathcal{B}(\mathcal{H}_{\epsilon}))$ out of local operators(which are bounded as we shown), they commute with any fixed point scalar operators by Euclidean commutivity. However, it may not seem obvious after analytic continuation to Lorentzian signature, where microcausality manifests only in strict spacelike separated operators. Since the scalars at fixed points $\mathcal{O}_{f}$ are zero and non zero modes of the one-sided Hamiltonian $H_{R}$(after taking $\epsilon \rightarrow 0$ limit), we can consider the following time-like separated commutator in Lorentzian time $t=is$
\begin{align}
    [\mathcal{O}_{f}(x_{f},0),\mathcal{O}'(x,t)]= \mathcal{O}_{f}(x_{f},0)e^{iH_{R}t}\mathcal{O}'(x,0)e^{-iH_{R}t}-e^{iH_{R}t}\mathcal{O}'(x,0)e^{-iH_{R}t}\mathcal{O}_{f}(x_{f},0) 
\end{align}
Since $\mathcal{O'}(x,t)$ are localized operator in one-sided ($R$) algebra, this will commute with $H_{L}$, the Hamiltonian of the other side or compliment region\footnote{Since $H_{L}$ is made out of local stress tensor integrated over compliment region, one can explicitly check that they will commute with any other local operator of the other side region $R$. We can also show $[H_{L},H_{R}]=0$ but strictly in a limiting sense of $\epsilon \rightarrow 0$.}. Hence we can insert the identity $e^{-iH_{L}t}e^{iH_{L}t}$ inside the commutator and get:
\begin{align}
 &[\mathcal{O}_{f}(x_{f},0),\mathcal{O}'(x,t)] \nonumber \\
 &= \mathcal{O}_{f}(x_{f},0)e^{iH_{R}t}e^{-iH_{L}t}e^{iH_{L}t}\mathcal{O}'(x,0)e^{-iH_{R}t}-e^{iH_{R}t}e^{-iH_{L}t}e^{iH_{L}t}\mathcal{O}'(x,0)e^{-iH_{R}t}\mathcal{O}_{f}(x_{f},0)  \nonumber \\
 &= \mathcal{O}_{f}(x_{f},0)e^{iH^{tot}t}\mathcal{O}'(x,0)e^{-iH^{tot}t}-e^{iH^{tot}t}\mathcal{O}'(x,0)e^{-iH^{tot}t}\mathcal{O}_{f}(x_{f},0)  \nonumber \\
 &=e^{iH^{tot}t}[\mathcal{O}_{f}(x_{f},0),\mathcal{O'}(x,0)]e^{-iH^{tot}t}=0, \; \text{since} \; (x_{f},x) \; \text{spacelike separated}
\end{align}
Here we have used the fact that two-sided modular Hamiltonian $H^{tot}$ commutes with all fixed point scaalrs $\mathcal{O}_{f}$. Thus any local operators will also commute with fixed point scalar at a fixed superselection sector in Lorentzian modular time. However, the center is \textit{not} a trivial one due to the presence of scalar operators rather than c-number scalars. Hence from all of these observations, we may say that the emergent algebra $\mathcal{A}_{new}$ in $\epsilon\rightarrow 0$ has been changed from type-III$_{1}$ factor $\mathcal{A}$ to an algebra with \textit{non-trivial center}\footnote{It is known that algebra of observables for a local QFT in a fixed subregion is type III$_{1}$ factor\cite{Witten:2018zxz}. However as we described, in the strict limiting procedure, we can construct a new algebra with an emergent non-trivial center.}. The type of the algebra depends on which Hilbert space it will act on. If we still consider $\mathcal{H}_{\epsilon \rightarrow 0}$ as the resulting Hilbert space, the algebra remains type-III$_{1}$.

\subsection{`Edge Hilbert space' and `interior' Hilbert space}

As we have shown from the limiting procedure, the emergent fixed point primaries and their linear combinations acting on $|0\rangle$ itself form a Hilbert space in the `closed string' channel. State- operator correspondence as well as the highest weight representation of Virasoro algebra in radial quantization is self-sufficient to construct it. Hence we do not bother about boundedness of primaries at fixed points\footnote{Note that, in radial quantization, the norm of a primary $\mathcal{O}(z,\bar{z})$ is infinite when $|z|^{2}=1$.  We can not exclude those points in usual sense. Still we can construct a Hilbert space purely based on state-operator correspondence in radial quantization.}. Hence, we readily construct an emergent Hilbert space of fixed point scalars acting on $|0\rangle$ appears in the shrinking limit of closed string channel. We call this \textit{interior Hilbert space $\mathcal{H}_{int}$}\footnote{The reason behind this name will be clear from the Lorentzian discussion in next section.}:
\begin{align}
    \mathcal{H}_{int}:=\oplus_{f}\{|0\rangle,\mathcal{O}_{f}|0\rangle\}
\end{align}
Any linear combination of those states (e.g: $|a\rangle,|b\rangle$) is also a part of the $\mathcal{H}_{int}$ by construction. Since the total modular Hamiltonian or the two-sided Hamiltonian is made out of SL(2,$\mathbb{C}$) generators $H^{tot}=\alpha L_{0}+\beta(L_{1}+L_{-1})+h.c$ , it will satisfy \cite{Das:2024mlx}\footnote{Note that, this is purely a statement of the global symmetry group, not a statement of the Hilbert space itself. Hence we can not treat $|0\rangle$ as a ground state of the modular Hamiltonian. We should rather think this as a statement of constraint in the same spirit of Witten's background independent observer algebra.}:
\begin{align}
    H^{tot}|0\rangle=0, \; H^{tot}\mathcal{O}_{f}|0\rangle = 0.
\end{align}
Hence, the states of $\mathcal{H}_{int}$ are also \textit{degenrate vacuum of two-sided total modular Hamiltonian $H^{tot}$}.

On the other hand, the inclusion of center by \textit{bounded function of fixed point scalars} $\mathcal{O}_{f}$ after taking closure of the algebra in $\epsilon \rightarrow 0$ limit, would act on the one-sided Hilbert space $\mathcal{H}_{\epsilon\rightarrow 0}$. Since after taking the $\epsilon \rightarrow 0$ limit, the Hilbert space becomes unbounded, the action of the center on states made out of local operators on $|0\rangle_{\epsilon}$ is not well-defined except on the identity or $|0\rangle_{\epsilon}$ itself. We have constructed $|0\rangle_{\epsilon}$ by identity insertion at $\theta = 0$ with $s \rightarrow -\infty$. The effect of the boundary at $\theta = \pm \Lambda$ is there under such construction. However after taking $\epsilon \rightarrow 0$ or $\Lambda \rightarrow \infty$ limit, the effect of the boundary will be negligible on the state. Hence we may naturally think $|0\rangle_{\epsilon \rightarrow 0} \sim |0\rangle$, albeit from a different construction of vacuum. Precisely, this will define a form of \textit{`complimentarity'} between two Hilbert spaces as we will discuss now. Hence we can restrict to the states $\{\mathcal{O}_{f}\}|0\rangle_{\epsilon \rightarrow 0}$, where $\{\mathcal{O}_{f}\}$ refers to the bounded smeared operators made out of fixed point scalars. The one-sided Hamiltonian $H_{R}$ has the following property after taking $\epsilon \rightarrow 0$:
\begin{align}
    H_{R}|0\rangle_{\epsilon} = 0, \; [H_{R},\mathcal{O}^{\lambda}_{f}]=\lambda\mathcal{O}_{f}^{\lambda}
\end{align}
Here $\lambda$ denotes the eigenvalue proportional to $ h\epsilon$ of the fixed point eigen-modes(both zero and non-zero) in $\epsilon \rightarrow 0$. From this, we can see $\{\mathcal{O}_{f}\}|0\rangle_{\epsilon}$ forms eigenbases of the Hamiltonian. This would further tell that after inclusion of the center in $\epsilon \rightarrow 0$, the center acting on vacuum of modular quantization would form a Hilbert space. This is strictly emergent in $\epsilon \rightarrow 0$ limit. We call this Hilbert space as \textit{edge Hilbert space $\mathcal{H}_{edge}$}:
\begin{align}
\mathcal{H}_{edge}:=\oplus_{f}\{|0\rangle_{\epsilon\rightarrow 0},\{\mathcal{O}_{f}\}|0\rangle_{\epsilon \rightarrow 0}\}    
\end{align}
Again any linear combination of those states will also become part of the $\mathcal{H}_{edge}$. 

From the open-closed string duality as well as the same operator content on the center as fixed point scalars, would suggest a natural isomorphism between $\mathcal{H}_{int}$ and $\mathcal{H}_{edge}$. 
\begin{align}
    \mathcal{H}_{int} \cong \mathcal{H}_{edge}
\end{align}
The isomorphism should be thought of as a demand of self-consistent open-closed string duality after the inclusion of the algebraic center. In particular, the new Hilbert space becomes:
\begin{align}
    \mathcal{H}_{new} =\mathcal{H}_{edge} \;.
\end{align}
Correspondingly, the closed string channel should also include $\mathcal{H}_{int}$. This will modify the duality as the following:
\begin{align}
    Tr_{\mathcal{H}_{edge}}\left[e^{-\frac{2\pi^{2}}{\Lambda\sqrt{d}}H} \mathcal{O}_{1}\mathcal{O}_{2}\dots\right] =\; \sum_{a,b}\langle b|\mathcal{O}_{1}\mathcal{O}_{2}\dots|a\rangle
\end{align}
In particular, we should identify $\mathcal{H}_{\epsilon \rightarrow 0} \rightarrow \mathcal{H}_{edge}$ after the inclusion of the center.  In other words, the `lost states' of $\mathcal{H}_{\epsilon \rightarrow 0}$ has some \textit{non-isometric embedding} in $\mathcal{H}_{edge}$\footnote{It is strongly indicated that such identification should come from a different bulk-boundary OPE channel due to presence of cut-off boundaries at finite $\epsilon$. Hence taking $\epsilon\rightarrow 0$ limit, the algebra of bcc operators should reduce to the algebra of fixed point scalars acting on $|0\rangle_{\epsilon\rightarrow 0}$.}. 
Clearly, after constructing this new algebra acting on $\mathcal{H}_{new}$, the existence of pure states indicates that the algebra \textit{will not be of type-III anymore and more of a type-I}. 

\subsection{Comment on Lorentzian modular quantization on causal diamond}
Modular Hamiltonian of a CFT vacuum on a causal diamond can be written explicitly after (global) conformal transformation from Rindler wedge\cite{Casini:2011kv}, \cite{Kabat:2017mun}. For a line segment $A:\{X\in(-1,1)\}$ in Lorentzian time $T=0$, the expression for vacuum modular Hamiltonian in lightcone coordinates $Z_{\pm}=(X\mp T),$ is:
\begin{align}
    H_A = 2\pi\int^{u_{+}}_{v_{+}}dZ_{+} \frac{(u_{+}-Z_{+})(Z_{+}-v_{+})}{u_{+}-v_{+}}T_{++}(Z_{+}) +2\pi\int^{v_{-}}_{u_{-}}dZ_{-} \frac{(v_{-}-Z_{-})(Z_{-}-u_{-})}{v_{-}-u_{-}}T_{--}(Z_{-})
\end{align}
Here upper and lower tips are located at $(v_{+},v_{-})=(-1,1)$ and $(u_{+},u_{-})=(1,-1)$ respectively. Hence we have
\begin{align}
   H_{A}= \pi \int^{1}_{-1}dZ_{+}(1-Z_{+}^{2})T_{++}(Z_{+}) + \pi \int^{1}_{-1}dZ_{-}(1-Z_{-}^{2})T_{--}(Z_{-})
\end{align}
Upon analytic continuation to complex plane by wick rotating $(Z_{+},Z_{-})\rightarrow(-iz,-i\bar{z})$, we end up with
\begin{align}
    H_{A} = \frac{\pi}{i}\int^{i}_{-i}dz(z+i)(z-i)T(z) + \frac{\pi}{i}\int^{i}_{-i}d\bar{z}(\bar{z}+i)(\bar{z}-i)\bar{T}(\bar{z}).
\end{align}
Note that, this is exactly the same expression for the one-sided Hamiltonian(with $d=1,\beta=\frac{1}{2}$) of Euclidean modular quantization upto an overall constant factor. Similarly the proper Rindler-like coordinate can also be obtained by transforming \cite{Kabat:2017mun}:
\begin{align}
    e^{\eta_{+}}=\frac{Z_{+}-1}{Z_{+}+1}, \; e^{\eta_{-}} = \frac{Z_{-}-1}{Z_{-}+1}; \; \eta_{\pm}=\theta\mp t
\end{align}
Again by analytic continuation to $(\eta_{+},\eta_{-}) \rightarrow(-i\omega,-i\bar{\omega})$, one obtain the exact same transformation
(\ref{no cutoff curve}) with $d=1$. Hence, the local Rindler time (or Minkowski boost) $t$ is the wick-rotated version of Euclidean modular time $s$. By implementing the standard rule of quantizing CFT, we begin with CFT in Euclidean to construct the Unitary Hilbert space\footnote{This must satisfy Lorentzian Unitarity upon analytic continuation, as we discussed in defining Hermitian conjugate.} and then wick rotate back to Lorentzian after computing any observables. Similarly, modular quantization can be thought of as the Euclidean version of quantizing modular Hamiltonian in a causal diamond. The cut-offs at finite $\epsilon$ are translated to \textit{stretched horizons} outside the causal horizons \cite{Banks:2025ttj}. Using Lorentzian version of $TT$ light-cone OPE \cite{Kabat:2017mun}, one can also  obtain the same modular Virasoro algebra within the diamond bounded by stretched horizons. Note that, here we have explicitly used the expression for modular Hamiltonian on a single line interval at $T=0$, which is precisely the Lorentzian version of one-sided contour in Euclidean modular quantization. Hence the other part of the modular quantization in complex plane has no direct analogue in the causal diamond picture as the other side is not a analytic continuation of the rest of the real line interval. However, once moving into the Rindler-like coordinate $\eta_{\pm}$, it will have direct analogue to the analytic continuation of $(\omega,\bar{\omega})$ frame. Here, instead we have two Rindler wedges to be the analytic continuation of two sides of the proper modular time frame\footnote{This is also true in radial quantization in 2d CFT. We use complex plane as a natural place to study quantization. However the original proper-time frame is the Euclidean spatial cylinder, which can be wick-rotated back to Lorentzian cylinder with compactified spatial circle.}. One fixed point maps to the bifurcating horizon, whereas the other one is mapped back to infinities.

\textit{More correctly}, this should be identified with two hyperbolic patches \cite{Parikh:2012kg}. In another way, in an alternative dual AdS$_{3}$ foliation as described in \cite{Das:2022pez}, AdS$_{2}$ black hole patch has been obtained on a AdS$_{3}$ foliation of dual CFT in heating phase. It has the following metric:
\begin{align}
    ds^{2} = \frac{1}{\sin^{2}\phi}\left(d\phi^{2}+\frac{-dt^{2}+d\theta^{2}}{\sinh^{2}2\theta}\right); \; \phi:(0,\pi),t:(-\infty,\infty),\theta:(0,\infty)
\end{align}
Here $\phi$ is the bulk coordinate. Within such `brane' embedding, the boundary metric at $\phi =0$, is the AdS$_{2}$ black hole patch. This is precisely obtained by embedding the bulk diffeomorphism of heating phase Hamiltonian in the AdS$_{3}$ Poincare metric, for which $\theta$ is restricted to $(0,\infty)$. In \cite{Das:2025cuq}, such AdS$_{2}$ black hole is also obtained by considering the same heating phase Hamiltonian on strip. Hence, in a direct approach, one can not extend $\theta$ from $(0,\infty)$ to $(-\infty,\infty)$.
However, one can do the same by using a Weyl rescaling:
\begin{align}
    ds^{2}|_{bdy} \rightarrow \tanh^{2}(2\theta)\; ds^{2}|_{bdy}= \frac{-dt^{2}+d\theta^{2}}{\cosh^{2}2\theta}
\end{align}
In this metric, we can extend $\theta:(-\infty,\infty)$. This is precisely the dS$_{2}$ metric in static patch\footnote{It is obtained by taking Nariai limit of SdS black hole.}. Hence under this Weyl rescaling, the CFT Hilbert space would not change. Thus we may conclude,\textit{the same modular quantization also describes a CFT quantization in dS$_{2}$ static patch}. Note that  in \cite{Das:2025eez}, it is explained why a consistent (B)CFT can not be defined in the Milne patch: the analytic continuation of the static patch metric(cosmological horizon) in past or future wedge. Hence a consistent CFT description of 2d static patch can be obtained by modular quantization as we explained. However, for the rest of the note, we will denote modular quantization as the Euclidean version of \textit{`Rindler-like' quantization}.


In \cite{Das:2020goe}, a certain version of modular eigenmodes, made out of primaries and stress tensor of total or two-sided modular Hamiltonian have been studied purely in Lorentzian. In particular, \cite{Das:2020goe} observed two copies of Virasoro algebra with familiar central extension from modular eigenmodes. In stead of putting a cut-off the contour was deformed to get those algebras. The modes were defined as:
\begin{align}
   &\mathbb{L}_{n} = \pi \int^{\infty}_{-\infty} dZ_{+} (1-Z_{+})^{-n+1}(1+Z_{+})^{n+1}T_{++}(Z_{+}) , \nonumber \\
  & \bar{\mathbb{L}}_{n} = \pi \int^{\infty}_{-\infty} dZ_{-} (1-Z_{+})^{-n+1}(1+Z_{-})^{n+1}T_{--}(Z_{-}); \; \forall n\in \mathbb{Z}
\end{align}
Here the contour is deformed such a way that it does not pass through $Z_{\pm}=1,-1$ \footnote{A similar deformation will be discussed in the context of Euclidean quantization in the appendix (\ref{app1}), where the central extension term will be vanishing.}. Using Lorentzian version of TT ope, it was shown that \cite{Das:2020goe}
\begin{align}
    [\tilde{\mathbb{L}}_{m},\tilde{\mathbb{L}}_{n}]= (m-n)\tilde{\mathbb{L}}_{m+n}+\frac{c}{12}n(n^{2}-1)\delta_{m+n,0}
\end{align}
Similar is true for $\tilde{\bar{\mathbb{L}}}_{n}$s. Here we redefined 
\begin{align}
    \tilde{\mathbb{L}}_{n} \equiv \frac{1}{2\pi i}\mathbb{L}_{n}
\end{align}
Even though it seems to generate a similar highest weight representation of Virasoro algebra, it will be manifestly \textit{non-unitary} with respect to the total modular Hamiltonian $H=\tilde{\mathbb{L}}_{0}+\bar{\tilde{\mathbb{L}}}_{0}$. Hence, the resulting Hilbert space will not have any physical relevance in our prior context of studying outside wedge observers.

However, the `Rindler-like quantization'  will be perfectly well-defined upon analytic continuation of modular quantization once transformed back to $(\omega,\bar{\omega})$ as we argued. Correspondingly, the primaries associated with one of the fixed points will be located at the (stretched) bifurcating horizons after the inclusion of center into the algebra. 

\subsection{Observers}
We conclude this section by presenting the observer perspective, synthesizing the insights developed in the previous two sections.

In summary, we consider a CFT on a cylinder governed by certain class of SL(2,$\mathbb{R}$)-deformed Hamiltonians. The quantization is rigorously defined in the presence of a conformal cutoff: a boundary of radius $\epsilon$ surrounding the fixed points. In other way round, one may describe an one-sided BCFT observer characterized by a Hamiltonian $H_{\epsilon}$, without explicit reference to any particular state. The conformal boundary conditions imposed at the cutoff introduce a single copy of the modular Virasoro algebra. The associated highest-weight state acting on vacuum $|0\rangle_{\epsilon}$, generates a GNS Hilbert space and, correspondingly, a type-I Von Neumann algebra. This construction yields a factorization of the Hilbert space into complementary parts, each defined by the same set of local operators on either side of the cutoff.

If we identify the CFT Hamiltonian with the modular Hamiltonian of a subregion bounded by the fixed points, defined with respect to the KMS conformal vacuum $|0\rangle$, we must take the limit $\epsilon \to 0$. This thermodynamic limit will naturally reproduce type-III$_{1}$ factor. In this limit, we also observe a notion of `non-trivial centralizer' made out of fixed point scalar operators. Adjoining this into the algebra, one can define an edge Hilbert space $\mathcal{H}_{edge}$, built from fixed-point scalar operators acting on $|0\rangle_{\epsilon\rightarrow 0}$. From `open-closed string' duality, this Hilbert space provides a dual description of $\mathcal{H}_{int}$, which is constructed from field theory limit of `closed strings': fixed-point scalar primaries acting on $|0\rangle$. The center of the algebra, $Z(\mathcal{A} \cap \mathcal{A}')$, characterizes these dual Hilbert spaces relative to the two conformal vacua $|0\rangle$ and $|0\rangle_{\epsilon \rightarrow 0}$. Thus, in the limit $\epsilon \to 0$, when a BCFT observer associated with a type-I algebra $\mathcal{A}(\mathcal{B}(\mathcal{H}_{\epsilon}))$ and Hamiltonian $H_{\epsilon}$, restricted to a given subregion with a sharp horizon and a two-sided vacuum $|0\rangle$, we can equivalently describe the system in terms of the new Hilbert spaces $\mathcal{H}_{edge}, \mathcal{H}_{int}$ appeared after inclusion of the center to the algebra. In $\mathcal{H}_{int}$, the conformal vacuum $|0\rangle$ resides \textit{naturally}, which is also a vacuum of total (two-sided) modular Hamiltonian.

In the Lorentzian picture, the one-sided \textit{``open-string" channel} corresponds to a Rindler BCFT observer following a worldline confined within a single wedge bounded by stretched horizon. 
In $\epsilon \rightarrow 0$ limit, the `Rindler' observer after the inclusion of center, could detect any of the states in $\mathcal{H}_{edge}$-the Hilbert space generated by scalar operators localized on the (stretched)horizon acting on the vacuum accessible to the Rindler observer. Apart from vacuum or identity sector, the (stretched) horizon contains all those microstructures as experienced by the external observer. At the same time, \textit{if} we associate the infalling observer by `closed string' channel, lived in the corresponding dual state in dual $\mathcal{H}_{int}$, the infalling experience within smooth boundary can also be captured by $\mathcal{H}_{edge}$ due to isomorphism between $\mathcal{H}_{edge}\cong \mathcal{H}_{int}$. Except the vacuum, all the other states could be thought of `smooth', horizonless `Rindler geometry', as the bifurcating horizon is replaced by fixed point scalar or bcc operator in the correponding states with respect to `Rindler observer'. We may think of this picture as \textit{`Rindler' complementarity} or more appropiately, \textit{`Fuzzy Rindler complimentarity'}. 
However, one might wonder whether another \textit{natural} possibility\footnote{Since those states are made out of primary excitations on $|0\rangle$, the natural vacuum for two-sided $H_{mod}$.} exists to interpret $\mathcal{H}_{int}$ in terms of \textit{purely} interior or `wormhole' sector. However, the same state \textit{can not} have two different descriptions within a single geometry! While the former description of $\mathcal{H}_{int}$ in terms of infalling observer within boundary surface is precisely concrete, the later one is, perhaps, nothing more than \textit{an illusion.} Hence, there is \textit{no independent interior construction} in this framework!

\section{BTZ black hole in AdS/CFT}\label{sec6}
Being the simplest black hole solution of Einstein gravity, BTZ black hole serves a perfect laboratory for studying non-perturbative aspects of quantum black holes, where some form of information loss puzzle do exist. Since the metric is locally AdS$_{3}$, one can describe it using large diffeomorphism of asymptotic AdS$_{3}$. In particular, the most general solution of vacuum Einstein equation in three dimension with negative cosmological constant is known as Banados geometry \cite{Banados:1998gg}. In the Fefferman-Graham gauge, the metric can be written as the following \cite{Roberts:2012aq}:
\begin{align}\label{FG}
    ds^{2} = \frac{dy^{2}+d\omega d\bar{\omega}}{y^{2}}+ \frac{L(\omega)}{2}d\omega^{2}+\frac{\bar{L}(\bar{\omega})}{2}d\bar{\omega}^{2}+\frac{y^{2}L(\omega)\bar{L}(\bar{\omega})}{4}d\omega d\bar{\omega}
\end{align} 
Here $L(\omega)$ and $\bar{L}(\bar{\omega})$ are the boundary stress tensors for the corresponding boundary-valued diffeomorphism or conformal transformation for $z\rightarrow f(\omega)$. 
\begin{align}\label{Banadox sch}
    L(\omega) = -\frac{12}{c}T(\omega) \equiv -\frac{f'''(\omega)f'(\omega)-\frac{3}{2}(f''(\omega))^{2}}{(f'(\omega))^{2}}
\end{align}
The anti-holomorphic part is defined for $\bar{L}(\bar{\omega})$. $(z,\bar{z})$ should be thought of as boundary coordinates for AdS$_{3}$ metric in Poincare coordinate \cite{Das:2024vqe}. For the boundary-valued diffeomorphism:
\begin{align}\label{boundary diff}
    z= -\frac{\alpha}{2\beta}-\frac{\sqrt{d}}{2\beta}\cot\left(\frac{\Lambda\sqrt{d}}{2\pi}\omega\right), \; \bar{z}= -\frac{\alpha}{2\beta}-\frac{\sqrt{d}}{2\beta}\cot\left(\frac{\Lambda\sqrt{d}}{2\pi}\bar{\omega}\right),
\end{align}
we have $L(\omega)=\bar{L}(\bar{\omega})=-\frac{\Lambda^{2}d}{2\pi^{2}}$. For this, (\ref{FG}) reduces to
\begin{align}
 ds^{2} = \frac{dy^{2}+d\omega d\bar{\omega}}{y^{2}}- \frac{\Lambda^{2}d}{4\pi^{2}}d\omega^{2}-\frac{\Lambda^{2}d}{4\pi^{2}}d\bar{\omega}^{2}+\frac{y^{2}\Lambda^{4}d^{2}}{16\pi^{4}}d\omega d\bar{\omega}   
\end{align}
Using $\omega=s+i\theta, \bar{\omega}=s-i\theta$ and the change of variable $r^{2}=\frac{(\frac{\Lambda^{2}d}{\pi^{2}}y^{2}+4)^{2}}{16y^{2}}$, we obtain:
\begin{align}
     ds^{2} = (r^{2}-\frac{\Lambda^{2}d}{\pi^{2}})ds^{2}+\frac{dr^{2}}{r^{2}-\frac{\Lambda^{2}d}{\pi^{2}}}+r^{2}d\theta^{2}
\end{align}
Here $r:[\frac{\Lambda\sqrt{d}}{\pi},\infty)$. To remove conical singularity of the near horizon Rindler geometry(in $s,r$ plane), the Euclidean time should be periodically identified $s\sim s+\frac{2\pi^{2}}{\Lambda\sqrt{d}}$ \cite{Townsend:1997ku}. Classocally, this is a planar BTZ or AdS-Rindler type geometry with $\theta:(-\infty,\infty)$. To obtain cylindrical BTZ black hole, we need to identify $\theta\sim \theta+2\pi$. The boundary geometry will correspondingly $S^{1}\times S^{1}$ or a Euclidean torus. After analytic continuation to real time $s\rightarrow it$, one would end up with a two-sided eternal BTZ black hole with two boundaries in the standard Kruskal extension. The Bekenstein-Hawking entropy of this BTZ geometry is $S_{BTZ}=\frac{\pi r_{h}}{2G_{N}}=\frac{\Lambda\sqrt{d}}{2G_{N}}$.

Although the classical identification of $\theta$ may appear straightforward, the introduction of modular quantization renders the situation considerably more subtle. More precisely, the boundary diffeomorphism~(\ref{boundary diff}) defines the map (\ref{map}) relevant for modular quantization, as discussed in the previous section. The presence of fixed points of the Hamiltonian flow at $\theta=\pm \infty$ naturally induces a quantization on either side of the $\theta$ contour. 

To formulate the boundary quantization rigorously, one must introduce conformal cutoffs $\theta=\pm \Lambda$ around these fixed points\footnote{The same $\Lambda$ also appeared in the (\ref{boundary diff}) as a consequence of getting the modular Virasoro algebra as we described in the last section.}. Consequently, a single copy of the modular Virasoro algebra emerges in each side of the $\theta$ contour, from which we constructed a GNS Hilbert space representation. As argued earlier, however, there exists a limit, $\Lambda \to \infty$ or $\epsilon \to 0$, in which these cutoffs collapse onto the fixed points. This raises the question of how the identification of $\theta$ remains well-defined in the corresponding limiting procedure. 

A closer examination reveals that two equivalent descriptions are possible, albeit with distinct physical interpretations. The essential feature underlying this pictorial equivalence is \textit{the existence of a branch cut along the quantization contour, bounded by two branch points located precisely at the fixed points, $z=z_{\pm}$}.

\begin{itemize}
    \item In the conventional description of the complex plane in the presence of a branch cut, the limiting procedure implies that the $\theta$ contours on either side terminate at the branch points and cannot be analytically continued across them. Consequently, the two sides remain disconnected. The corresponding bulk dual is described by the two-sided AdS-Rindler geometry, equivalently the planar BTZ black hole with a boundary cutoff as well as a stretched horizon \cite{Das:2024vqe}. From our earlier detailed analysis of modular quantization, avoiding the crossing of the cut is equivalent to refraining from taking the strict limit $\epsilon \to 0$. Correspondingly, the stretched horizons extend to and terminate at the boundary cut-offs near the fixed points. The apparent disconnectivity of the two sides admits a geometric interpretation in terms of the ``entanglement'' between the two AdS-Rindler wedges with stretched horizons\footnote{Any entangled state is well-defined only with respect to a factorization of the Hilbert space. By invoking the split property, such a factorization can be realized in the presence of a boundary cutoff, or equivalently, a stretched-horizon cutoff in the bulk.}. Correspondingly, a canonical TFD state can be built from the eigenstates of modular quantization at finite $\epsilon$ cut-offs. 

    \item Alternatively, the analytic crossing of the branch cut connecting the two branch points may always be geometrically interpreted in terms of two Riemann sheets joined along the branch cut to form a single 2-sheeted Riemann surface. In this description, the two sides of the $\theta$ contour reside on distinct sheets and therefore remain independent CFTs prior to crossing the cut. 

Within the framework of modular quantization on a single sheet, this issue has already been addressed through the emergent center of the operator algebra generated by fixed-point scalar operators. These operators act as boundary-condition-changing (bcc) operators, analogous to the twist operators appearing in discussions of entanglement entropy~\cite{Calabrese:2004eu}. 

Furthermore, we will explicitly demonstrate in the following section that, in the limit $\epsilon \to 0$,  the entropy of the large BTZ black hole can be recovered from one-sided modular quantization through the mechanism of `open-closed string' duality. While computing the partition function in shrinking limit, the KMS vacuum $|0\rangle$ dominates.

This correspondence strongly suggests a non-standard (non-perturbative) bulk description of BTZ black hole in terms of a stretched horizon in AdS-Rindler geometry with a boundary cutoff. Following the analysis of~\cite{Burman:2023kko}, we will show explicitly that the microstates as well as the entangled TFD state in this dual description reproduce the exact \textit{ Hartle-Hawking correlators} in the limit of stretched horizon approaches the event horizon as well as compactifying the boundary $\theta$ coordinates. 
\end{itemize}
From the second perspective, the emergence of two-sided bulk smoothness may be interpreted as the geometric manifestation of connectivity in the two-sheeted boundary description. This is clearly visible only in the $\epsilon \rightarrow 0$ limit of the `entanglement' or the canonical TFD picture, where the identity or vacuum sector dominates in the thermal partition function. 
\subsection{Microstate counting of BTZ from modular quantization}

The major success of the torus description in radial quantization relies on Cardy entropy counting of BTZ microstates from dual CFT. 
The key step in deriving the Cardy entropy is the thermodynamic limit $\beta\rightarrow 0$ while keeping the length of spatial circle $L$ finite on a torus\footnote{See \cite{Banerjee:st4}, \cite{Carlip:2000nv} for a review.}. This corresponds exactly to the limit in which the thermal circle decompactifies, causing the torus to approach a spatial circle, where radial quantization becomes manifest.
The same limit can also be obtained by taking $\beta$ finite and $L\rightarrow \infty$\footnote{I thank Sanchari Pal for asking this question in a discussion \cite{Pal:2022uoo}, long time ago.}. The final answer will apparently be the same from modular invariance of the torus partition function in Cardy analysis. However, In a torus, taking $\beta$ finite and $L \rightarrow \infty$ is precisely a \textit{thermal cylinder limit} of the torus. In the same limit, we \textit{can not} write the torus partition function as the same thermal partition function, since the trace of the Hilbert space should be taken over $\mathcal{H}_{\mathbb{R}^{1}}$ or over states in a thermal cylinder. This is not carried out by CFT in radial quantization. Using modular quantization, we will show how we could recover the BTZ entropy of large AdS. 


\textbf{\textit{Recovering Cardy density using `open-closed string' duality in modular quantization:}}

We now consider thermal partition function of CFT in modular quantization and use open-closed string duality as in (\ref{open closed part}). 
\begin{align}
   Z^{\{ab\}}_{\tau} =  Tr_{\mathcal{H}_{\epsilon}^{\{ab\}}}e^{2\pi i \tau\hat{\mathcal{L}}^{\epsilon}_{0}} = \langle b|e^{-\frac{2\pi i}{\tau}\left(L_{0}^{\zeta}+\bar{L}_{0}^{\bar{\zeta}}-\frac{c}{12}\right)}|a\rangle
\end{align}
Here we have used $H_{\epsilon}=\hat{\mathcal{L}}^{\epsilon}_{0}$, where we have shifter $\hat{\mathcal{L}}^{\epsilon}_{0}+\frac{c\Lambda^{2} d}{12\pi^{2}}\rightarrow \hat{\mathcal{L}}^{\epsilon}_{0}$, to get the usual Virasoro algebra with SL(2,$\mathbb{R}$) subalgebra. Also we have one modular parameter $\tau = \frac{i\pi}{\Lambda\sqrt{d}}$. We can again treat $\epsilon \rightarrow 0$ or $\Lambda \rightarrow \infty$ as the `thermodynamic limit', where the spectrum of $\hat{\mathcal{L}}^{\epsilon\rightarrow 0}_{0}$ is approximated by continuous label $E$, as we described earlier. Correspondingly, the partition function can be written in terms of the continuous density of states $\rho_{m}(E)$ in modular quantization:
\begin{align}\label{annulus duality}
    \lim_{\epsilon \rightarrow 0}Z^{\{ab\}}_{\tau} = \int dE_{\epsilon} \rho_{m}(E_{\epsilon})e^{2\pi i\tau E_{\epsilon} } = \lim_{\epsilon \rightarrow 0}\sum_{\Delta_{\zeta}}\langle b|\Delta_{\zeta}\rangle \langle\Delta_{\zeta}| e^{-\frac{2\pi i}{\tau}\left(\Delta_{\zeta}-\frac{c}{12}\right)}|a\rangle 
\end{align}
In $\epsilon \rightarrow 0$, the vacuum contribution from $\Delta_{\zeta}$ dominates i.e: $|\Delta_{\zeta}\rangle\approx|0\rangle$. Hence we have
\begin{align}\label{vacuum dominance}
\lim_{\epsilon \rightarrow 0}\sum_{\Delta_{\zeta}}\langle b|\Delta_{\zeta}\rangle \langle\Delta_{\zeta}| e^{-\frac{2\pi i}{\tau}\left(\Delta_{\zeta}-\frac{c}{12}\right)}|a\rangle = e^{\frac{2\pi i}{\tau}\frac{c}{12}}\lim_{\epsilon\rightarrow 0}\langle b|0\rangle\langle 0|a\rangle \approx   e^{\frac{2\pi i}{\tau}\frac{c}{12}+\lim_{\epsilon \rightarrow 0}(\log g_{a}+\log g_{b})}  
\end{align}
Here $\log g_{a,b}$ are the boundary entropies\cite{Cardy:2016fqc}. In $\epsilon \rightarrow 0$, these are $\mathcal{O}(1)$ number. Thus the leading contribution in $\epsilon \rightarrow 0$ can be written as:
\begin{align}\label{annulus limit}
  \lim_{\epsilon \rightarrow 0}Z^{ab}_{\tau} \approx  e^{\frac{2\pi i}{\tau}\frac{c}{12}} 
\end{align}
Hence combining (\ref{annulus duality}) and (\ref{annulus limit}) we get
\begin{align}
    \int dE_{\epsilon} \rho_{m}(E_{\epsilon}) e^{2\pi i\tau E_{\epsilon}} = e^{\frac{2\pi i}{\tau}\frac{c}{12}}
\end{align}
By writing $\rho_{m}(E_{\epsilon})$ as a complex integration along a contour enclosing $e^{2\pi i \tau} = 0$ in a similar way of \cite{Carlip:2000nv}, and using  (\ref{annulus limit}) we finally obtain:
\begin{align}
    \rho_{m}(E_{\epsilon}) = \int d\tau e^{-2\pi i\tau E_{\epsilon}+\frac{2\pi i}{\tau}\frac{c}{12}} 
\end{align}
The range of $\tau$ integration covers the full imaginary axis. This integral can be evaluated using saddle point approximation provided the imaginary of $\tau$ at the saddle point becomes very large. The saddle point is:
\begin{align}
    \tau^{*} = i\sqrt{\frac{b}{a}}, \; \text{where} \; a= E_{\epsilon}, \; b = \frac{c}{12}
\end{align}
This yields \cite{Carlip:2000nv}: 
\begin{align}
    \rho_{m}(E_{\epsilon}) \approx \left(\frac{b}{4a^{3}}\right)^{\frac{1}{4}}e^{4\pi\sqrt{ab}}
\end{align}
The saddle approximation demands $E_{\epsilon}>>0$. By choosing $E_{\epsilon} = \frac{c\Lambda^{2}d}{12\pi^{2}}$, the microcanonical entropy will be 
\begin{align}
    S_{micro}\approx\underbrace{\frac{c}{3}\Lambda\sqrt{d}}_{S_{E}}-\frac{3}{2}\log(S_{0})
\end{align}
Here $S_{E}$ refers to the microcanonical entropy at $E=\frac{c\Lambda^{2}d}{12\pi^{2}}$, which we have chosen. This choice is determined by ensemble equivalence in thermodynamic limit. In particular, for canonical ensemble with an effective inverse temperature $\beta_{eff} = 2\pi i\tau= \frac{2\pi^{2}}{\Lambda\sqrt{d}}$, we get
\begin{align}
    Z_{\tau\rightarrow 0}^{\{ab\}} = e^{\frac{c\pi^{2}}{3\beta_{eff}}}
\end{align}
Hence the thermal entropy at $\beta_{eff}$ is
\begin{align}
    S_{th} = \frac{\partial}{\partial\left(\frac{1}{\beta_{eff}}\right)}\left[ \frac{\log Z^{\{ab\}}_{\tau\rightarrow 0}}{\beta_{eff}}\right] = \frac{2\pi^{2}c}{3\beta_{eff}}=\frac{c}{3}\Lambda\sqrt{d}
\end{align}
One can recognize $S_{th}$ as the vacuum entanglement entropy of a CFT interval $z\in (z_{+},z_{-})$ bounded by two fixed points \cite{Calabrese:2004eu}. Here $\Lambda\sqrt{d} = \log\left(\frac{\sqrt{d}}{\beta\epsilon}\right)=\frac{1}{2}\log\left(\frac{|z_{+}-z_{-}|}{\epsilon}\right)$ . However, we emphasize once again that such an identification is meaningful only in the presence of a finite $\epsilon$ cutoff, owing to the split property. Consequently, while we relate the entropy to entanglement entropy, the limit $ \epsilon \to 0$ should not be taken literally.

From this observation of entanglement entropy as well as vacuum dominance in obtaining the thermal entropy in (\ref{vacuum dominance}), we can write thermal correlators in open string channel as vacuum correlators:
\begin{align}\label{open closed}
    \lim_{\epsilon\rightarrow 0}Tr_{\mathcal{H}_{\epsilon}^{\{ab\}}}\left[e^{-\frac{2\pi^{2}}{\Lambda\sqrt{d}}H_{\epsilon}}\mathcal{O}_{1}(s_{1},\theta_{1})\mathcal{O}_{2}(s_{2},\theta_{2})\dots \right] = \langle 0|\mathcal{O}_{1}(s_{1},\theta_{1})\mathcal{O}_{2}(s_{2},\theta_{2})\dots|0\rangle
\end{align}


We can also identify the same entropy to a large BTZ entropy when we can take $\epsilon \rightarrow 0$ limit. 
  Since $\Lambda\sqrt{d}=\log\left(\frac{\sqrt{d}}{\beta\epsilon}\right)$ is dimensionless, we may identify:
\begin{align}
    \Lambda\sqrt{d} = \frac{\pi r_{h}}{l_{AdS}} \; \text{for} \; \Lambda\sqrt{d}>>1
\end{align}
This identification can be understood in large BTZ, for which the circumference $2\pi\frac{r_{h}}{l_{AdS}} <4\Lambda\sqrt{d}$, whereas the boundary length ($|\theta|$) of each side will be $2\Lambda$. Hence BTZ entropy will be identified with $S_{micro}$.
\begin{align}
    S_{micro} = \frac{c}{3}\Lambda\sqrt{d}-\frac{3}{2}\log(S_{E}) =\underbrace{\frac{\pi r_{h}}{2G_{N}}}_{S_{bh}}-\frac{3}{2}\log(S_{bh})=S_{BTZ}, \; \text{where} \; c=\frac{3l_{AdS}}{2G_{N}}.
\end{align}
This $-\frac{3}{2}\log S_{bh}$ is different from what is obtained from radial quantization \cite{Sen:2012dw}. In literature, a similar factor appeared in loop quantum gravity framework, as obtained by Kaul-Majumdar in \cite{Kaul:2000kf}\footnote{However, to the best of our knowledge, this striking similarity seems accidental.}. 

However, to establish the BTZ or torus connection explicitly, we have to consider the following rule while computing correlators in modular quantization.

To lift the thermal correlators of modular quantization to that of the torus, we must use the compactification of $\theta\sim \theta+2\pi$ and then taking $\epsilon \rightarrow 0$ limit. This will lift correlators of $R^{1}\times S^{1}$ to $S^{1}\times S^{1}$. The usual strategy is to implement sum of images \cite{Furuya:2023fei}. Using (\ref{open closed}) we have: 
\begin{align}\label{torus corr equiv}
    \lim_{\beta\rightarrow 0} \; _{torus}\langle \mathcal{O}_{1}(s_{1},\theta_{1})\mathcal{O}_{2}(s_{2},\theta_{2})\rangle_{torus} &=\sum_{m,n=-\infty}^{\infty} \lim_{\epsilon\rightarrow 0}Tr_{\mathcal{H}_{\epsilon}^{\{ab\}}}\left[e^{-\frac{2\pi^{2}}{\Lambda\sqrt{d}}H_{\epsilon}}\mathcal{O}_{1}(s_{1},\theta_{1}+2\pi m)\mathcal{O}_{2}(s_{2},\theta_{2}+2\pi n) \right] \\ 
    &= \sum_{m,n=-\infty}^{\infty}\langle 0|\mathcal{O}_{1}(s_{1},\theta_{1}+2\pi m)\mathcal{O}_{2}(s_{2},\theta_{2}+2\pi n)|0\rangle
\end{align}
However, we emphasize that compactifying $\theta$ is not equivalent to identifying the  $\theta$-contour with that of the opposite side. As discussed earlier, the $\theta$ orbits are intrinsically circular. However, the presence of fixed points requires the quantization contour to terminate at either side of the fixed-points set when quantizing the Hamiltonian\footnote{I thank Bobby Ezhuthachan, for discussion on this point.}. Rather, the two sides may be compactified independently using image sum prior taking the $\epsilon \to 0$ limit. 

In the next section, we will employ modular quantization in $\epsilon \rightarrow 0$ limit to obtain an exact dual description of the BTZ geometry in certain limit from the perspective of modular quantization.

\subsection{Bulk description of modular quantization in $\epsilon \rightarrow 0$ limit}
Within the framework of AdS/CFT, viewed broadly as an instance of gauge/gravity duality beyond supersymmetric cases, the holographic CFT in radial quantization admits a bulk description in terms of AdS spacetime coupled to light matter fields in the limit ($G_{N}\to 0$), where Einstein gravity provides the dominant contribution. In this description, (small) finite $G_{N}$ effects can, in principle, be systematically incorporated within perturbation theory through an $(\mathcal{O}(G_{N}))$ expansion by solving the Einstein equations and accounting for matter backreaction together with higher-derivative gravitational corrections order by order. Nevertheless, such a perturbative framework is inherently incapable of capturing genuinely non-perturbative contributions, such as effects of order $\mathcal{O}(e^{-l_{AdS}/G_{N}})$. 

Beyond specific examples with a known exact duality, AdS/CFT can be viewed more broadly as a correspondence between quantum gravity in asymptotically AdS (AAdS) spacetimes and a dual holographic CFT. The AdS$_3$/CFT$_2$ case, with a holographic large-c CFT characterized by $c=\frac{3l_{AdS}}{2G_{N}}$, provides a notable example, having passed many nontrivial consistency checks even without an explicit construction of the dual CFT.

In the framework of holographic CFT formulated through modular quantization, viewed as an observer motivated alternative to radial quantization, we may propose a non-standard bulk perspective capable of probing finite-\(c\) features of the black hole Hilbert space \footnote{This picture is closely analogous to the `fixed area state' in gravitational path integral \cite{Dong:2018seb}, which serves a realization of `holography as quantum error correcting code' \cite{Almheiri:2014lwa}.}. Here is a smoother and more formal rephrasing:

 As discussed at the beginning of this section, the large diffeomorphism that generates a BTZ black hole can be naturally incorporated within the framework of modular quantization. The key idea is to identify the quantity dual to the BTZ mass, or equivalently the ADM Hamiltonian, with the modular Hamiltonian of the holographic CFT. In this picture, the ADM clock associated with an asymptotic observer determines the (Euclidean) modular time (s) of the dual geometry. The primary motivation for this background-independent construction is to investigate the Hilbert space of the BTZ black hole in a non-perturbative setting. 
 In our view, this constitutes one of the essential aspects of Witten's background-independent algebraic framework.

In particular, within the framework of dual boundary modular quantization, the presence of a cut-off is naturally incorporated in the bulk via a stretched horizon in AdS-Rindler geometry with boundary cut-off \cite{Das:2024vqe}, whereas taking the removal of the cut-off limit with a boundary compactification corresponds more closely to the conventional bulk description of a BTZ black hole in the \(G_{N}\rightarrow 0\) limit, as discussed earlier in the context of entropic identification to BTZ.  However, such a geometry is \emph{not} an exact solution of Einstein's equations. This observation suggests that the finite cut-off configuration may admit a bulk interpretation inaccessible within the standard large-\(c\) or \(G_{N}\rightarrow 0\) formulation of AdS/CFT. 

Motivated by this possibility, we investigate an alternative dual description of modular quantization, inspired by t'hooft's brick-wall type constructions \cite{tHooft:1984kcu}- \cite{Iizuka:2013kma}. Following closely the proposal of \cite{Burman:2023kko}, we consider a plausible finite-\(G_{N}\) bulk description, constructed purely from a bottom-up perspective, consisting of free massless scalar fields propagating in a AdS-Rindler background equipped with a Dirichlet stretched horizon\footnote{See also \cite{Banerjee:2024dpl}-\cite{Caceres:2025qlh}, \cite{Krishnan:2026mpa}, \cite{Gayari:2026jwn} for related developments.}. In the absence of an underlying top-down derivation, we assume this stretched horizon to be static, spacelike, and endowed with Dirichlet boundary conditions for free massless scalar fields. Since the construction is intrinsically bottom-up, several parameters remain undetermined, including the precise location of the stretched horizon. To constrain these parameters, we adopt the philosophy of \cite{tHooft:1984kcu} \footnote{However, here we took slightly different perspective from \cite{Burman:2023kko}, where the Planckian stretched horizon  is introduced as an effective cutoff for low-energy physics below the Planck scale. In contrast, here we do not treat the stretched horizon as an EFT cutoff or UV regulator. Instead, we regard it as a feature of an alternative genuine non-perturbative finite-$G_{N}$ feature, arising from the UV-complete dual CFT description within modular quantization. An arbitrary cut-off would obey any formal boundary condition. However, modular quantization with a conformal cut-off should select admissible boundary conditions in the bulk. Here we simply choose Dirichlet boundary condition.} and fix them using low-energy semiclassical input in the \(G_{N}\rightarrow 0\) limit. One important input arises from black hole thermodynamics, in particular the Bekenstein-Hawking entropy. Our approach is therefore to construct a microscopic statistical model consistent with a prescribed thermodynamic description\footnote{This reflects a remarkable aspect of black hole thermodynamics: the entropy is known even in the absence of a complete microscopic understanding. A similar feature appears in the conventional Euclidean gravitational path integral, where the Bekenstein-Hawking entropy emerges from semiclassical saddle points without requiring an explicit knowledge of microstates.}.

\textit{\textbf{An alternative finite-\(G_{N}\) bulk description from a bottom-up perspective in modular quantization:}}

In the present work, we explicitly use `observer-motivated' AdS/CFT framework, in which a natural UV-complete CFT observer resides at the asymptotic boundary and is equipped with a Hamiltonian that determines the notion of time evolution in the dual bulk description. 

A stretched-horizon model of the BTZ geometry with Dirichlet boundary conditions was studied in detail in \cite{Das:2022evy}-\cite{Krishnan:2023jqn}, including explicit calculations of microstate correlators in \cite{Burman:2023kko}, \cite{Burman:2024egy}. Here we summarize some of its essential features and emphasize it's relation to the framework of boundary modular quantization. Following \cite{Burman:2023kko}, we consider a free massless scalar field \(\phi(r,t,\theta)\) \footnote{In principle, there should exist many such fields, along with additional matter fields. However, in the absence of a precise top-down theory, we restrict ourselves to this minimal phenomenological framework, which is sufficient to capture the essential physics and address the question of interest.} propagating in a one-sided BTZ background with coordinates \((r,t,\theta)\), together with a stretched horizon located at \footnote{From now, we will take $l_{AdS}=1$ to avoid cluttering.}
\begin{align}
 r=\frac{\Lambda\sqrt{d}}{\pi}+\epsilon_{b} \;.   
\end{align}
 From the boundary cut-off picture of modular quantization, the stretched horizon ends on the boundary cut-offs $
\theta =\pm\Lambda$. The parameter \(\epsilon\), appearing in the boundary cut-off relation $
\Lambda=\log\left(\frac{\sqrt{d}}{\beta\epsilon}\right)$, is treated independently from the stretched-horizon regulator \(\epsilon_{b}\) in the dual large AdS black hole in order to maintain a self-consistent dual description. This construction provides a precisely \emph{kinematical} dual realization of modular quantization with a cut-off for the one-sided contour. As discussed earlier, the usual BTZ picture from modular quantization is recovered by first taking the \(\Lambda\rightarrow\infty\) limit and subsequently compactifying \(\theta\sim \theta+2\pi\). Correspondingly in the bulk, this is associated with the \textit{removing the stretched horizon} limit \(\epsilon_{b} \to 0\) and correspondingly the geometry becomes a large BTZ black hole.

Consider a scalar field admits the mode expansion
\begin{align}
    \phi(r,t,\theta)
    =
    \frac{1}{\sqrt{r}}
    \sum_{\omega}
    \sum_{J=-J_{\mathrm{cut}}}^{J_{\mathrm{cut}}}dJ\;
    e^{-i\omega t}
    e^{iJ\theta}
    \phi_{\omega,J}(r).
\end{align}
The factor \(e^{-i\omega t}\) follows directly from the time-translation invariance of the background geometry. In contrast, the boundary cut-off \(\Lambda\) explicitly breaks translational invariance along the \(\theta\)-direction. We incorporate this breaking effectively through a cut-off in angular momentum, denoted by \(J_{\mathrm{cut}}\). Nevertheless, in the limit \(\epsilon\rightarrow 0\), consistency requires
\[
J_{\mathrm{cut}}\rightarrow\infty,
\]
so that translational symmetry along \(\theta\) is restored for the one-sided BTZ black hole geometry. Consequently, the factor \(e^{iJ\theta}\) may still be employed without loss of generality in this limit.

We impose Dirichlet boundary conditions for the scalar field at the stretched horizon:
\begin{align}
    \phi(r=\frac{\Lambda\sqrt{d}}{\pi}+\epsilon_{b},t,\theta)=0.
\end{align}
As a consequence, the frequencies \(\omega\) become discretized. In addition, one requires the field to satisfy normalizability conditions as \(r\rightarrow\infty\). Solving these conditions yields the normal mode spectrum \cite{Krishnan:2023jqn}
\begin{align}\label{normal mode}
    \omega_{n}
    \approx
    \frac{2n\Lambda\sqrt{d}}
    {\log\left(\frac{\Lambda\sqrt{d}}{2\pi\epsilon_{b}}\right)}, \; n\in \mathbb{Z}_{+} \; .
\end{align}
The dependence of \(\omega\) on the angular momentum quantum number \(J\) can be neglected because of the large stretched-horizon logarithmic factor $
\log\left(\frac{\Lambda\sqrt{d}}{2\pi\epsilon_{b}}\right)$,
together with the assumption of finite \(J_{\mathrm{cut}}\), as argued in \cite{Burman:2023kko}. It is worth noting that an identical form of the normal mode spectrum also emerges when the fields are restricted to constant-\(\theta\) slices, corresponding effectively to an AdS\(_2\) black hole geometry \cite{Soni:2023fke}. Since both the boundary conditions and the normalizability constraints on \(\phi\) are independent of \(\theta\), this may indeed provide a valid approximation for determining the spectrum of \(\omega_{n}\).

Proceeding in the same spirit as \cite{Burman:2023kko}, we compute the microcanonical entropy of the scalar field configuration at energy$
E=M_{BTZ}
=
\frac{\Lambda^{2}d}{8G_{N}\pi^{2}}$, where the radius of the full BTZ black hole geometry (after identification of the \(\theta\)-direction) is given by $r_{h}
=
\frac{\Lambda\sqrt{d}}{\pi}$.
\begin{align}
    S(E=M_{BTZ}) = 2\pi \sqrt{\frac{M_{BTZ}}{2G_{N}}\left[\frac{2G_{N}\log\left(\frac{\Lambda\sqrt{d}}{2\pi\epsilon_{b}}\right)J_{cut}}{6\Lambda\sqrt{d}}\right]} 
\end{align}
To promote the `kinematical' dual to a `dynamical' dual description, we will fix 
\begin{align}
   S(E=M_{BTZ})=S_{BTZ}=\frac{\pi r_{h}}{2G_{N}}=\frac{\Lambda\sqrt{d}}{2G_{N}}. 
\end{align}
This precisely uplifts the modular quantization to bulk massless scalar field quantization in the fix microcanonical ensemble of stretched horizon background and making the set-up more than a `kinematical' equivalence of Bekenstein-Hawking entropy\footnote{In other words, we associate (Bekenstein-Hawking) entropic degrees of freedom with a dynamical set-up that enables exact microstate counting. This is in the same spirit of background independent setting, where the gravity dof is dictated by matter dof after using the gravitational constraint.}. This identification fixes the parameters as the following:
\begin{align}\label{fixing}
    \frac{1}{6}\log\left(\frac{\Lambda\sqrt{d}}{2\pi\epsilon_{b}}\right)J_{cut} = \frac{\Lambda \sqrt{d}}{2G_{N}}
\end{align}
As discussed earlier, taking the boundary cut-off removal limit \(\Lambda\rightarrow\infty\), equivalently \(\epsilon\rightarrow 0\), requires removal of the stretched horizon limit \(\epsilon_{b}\rightarrow 0\) and an angular compactification of boundary coordinate requires \(J_{\mathrm{cut}}\rightarrow\infty\). According to (\ref{fixing}), we can interpret this as
\begin{align}\label{fixing identific}
    \epsilon_{b}=\frac{a\Lambda\sqrt{d}}{2\pi}e^{-\frac{1}{G_{N}}}, \; J_{cut} = \frac{6\Lambda\sqrt{d}}{a}, \; \text{where} \; a=\text{arbitrary constant.}
\end{align}
Even though, this is not the unique choice of fixing, this is purely physically motivated as the following. Since, $J_{cut} \rightarrow \infty$ is precisely obtained while taking the removal of boundary cut-off $\Lambda \rightarrow \infty$. On the other hand, stretched horizon should be thought of as a purely non-perturbative effect in $G_{N}$. Hence similar to prior discussion in the earlier section, $\epsilon_{b} \rightarrow 0$ should be thought of as a \textit{semiclassical limit or $G_{N} \rightarrow 0$}\footnote{This is a rather striking feature of the construction. From the boundary perspective, the parameters \(c\) and \(\Lambda\) are treated independently, so that taking \(\Lambda\rightarrow\infty\) does not by itself imply \(c\rightarrow\infty\). However, once the system is interpreted holographically through the BTZ bulk description, we must need both $\Lambda \rightarrow \infty$ and $c \rightarrow \infty$ simultaneously. In our holographic description, this implies taking $\epsilon_{b} \rightarrow 0$ and $J_{cut}\rightarrow \infty$ simultaneously. This parallels the standard Cardy entropy analysis in radial quantization, where the Cardy formula remains valid for arbitrary \(c\), and the high-temperature limit \(\beta\rightarrow 0\) does not independently require \(c\rightarrow\infty\). To interpret the Cardy entropy as the BTZ black hole entropy, one must additionally invoke the AdS/CFT correspondence and identify the semiclassical limit \(c\rightarrow\infty\), or equivalently \(G_{N}\rightarrow 0\), in which the BTZ saddle dominates the gravitational path integral.}. In the framework of modular quantization, the identification (\ref{fixing identific}) furnishes a bulk analogue of the corresponding boundary limiting procedure, with the boundary central charge $c$ implicitly encoded in the parameter of the massless scalar field  $\phi$ through normalization and boundary condition on stretched horizon \footnote{In this way, we treat our construction slightly different from \cite{Burman:2023kko}, where extra input like Planckian stretched horizon as effective field theory cut-off is not needed by construction. See also related discussion on $J_{cut}$ \cite{Gayari:2026jwn}.}.

As in \cite{Burman:2023kko}, the saddle $\beta^{*}$ is given by
\begin{align}
    \beta^{*}(E) = \sqrt{\log\left(\frac{\Lambda\sqrt{d}}{2\pi\epsilon_{b}}\right)\frac{\pi^{2} J_{cut}}{6\Lambda\sqrt{d}E}}
\end{align}
Using $E=M_{BTZ}$ and (\ref{fixing}), we obtain:
\begin{align}
    \beta^{*} = \frac{2\pi^{2}}{\Lambda\sqrt{d}} =\beta_{BTZ}
\end{align}
Hence, the microcanonical constraint also fixes the Hawking temperature of the BTZ black hole, \(\beta_{\mathrm{BTZ}}\). Similarly, the energy eigenstates in the microcanonical energy window of the massless free scalar field on the one-sided stretched-horizon background provide a corresponding ``bulk'' description of the microstates.
\begin{align}\label{energy level}
    E=M_{BTZ}= \sum_{n,J}\omega_{n}N_{n}
\end{align}
Here $N_{n}$ is the occupation number in $n$-th level eigenstates defined as:
\begin{align}\label{microstate}
   |N\rangle \equiv \prod_{n,J}\frac{1}{\sqrt{N_{n}!}}(a^{\dagger}_{n,J})^{N_{n}}|0\rangle_{SH}; \; \text{where} \; a_{n,J}|0\rangle_{SH} = 0 
\end{align}
$a_{n,J},a^{\dagger}_{n,J}$ are annihilation and creation operator defined via the mode expansion of $\phi$. Since $E$ is large($E>>1$), one can treat it as thermodynamic limit and replace $N_{n}$ by it's mean value $\langle N_{n}\rangle$ during certain correlator calculation as in \cite{Burman:2023kko}. This is justified if the variance is sufficiently suppressed in the same limit. Denoting the integer-partitioned level $\sum_{n}nN_{n}=N_{lev}$, one can use standard statistical mechanics result of Bose-Einstein mean:
\begin{align}\label{BE}
    \langle N_{n} \rangle = \frac{1}{e^{\frac{\pi n}{\sqrt{6N_{lev}}}}-1}
\end{align}
Using (\ref{normal mode}) and (\ref{energy level}), we have\footnote{Here $2J_{cut}>>1$ as $G_{N}$ is small but finite.}
\begin{align}
    M_{BTZ} = \frac{2\Lambda\sqrt{d} }{\log\left(\frac{\Lambda\sqrt{d}}{2\pi\epsilon_{b}}\right)}2J_{cut}N_{lev}
\end{align}
Using the expression for $M_{BTZ}$, we will end up with 
\begin{align}
    N_{lev} = \frac{\left(\log\frac{\Lambda\sqrt{d}}{2\pi\epsilon_{b}}\right)^{2}}{96\pi^{2}}
\end{align}
Consequently, 
\begin{align}
  \exp\left({\frac{n\pi}{\sqrt{6N_{lev}}}}\right) = \exp\left({\frac{4n\pi^{2}}{\log\left(\frac{\Lambda\sqrt{d}}{2\pi\epsilon_{b}}\right)}}\right) . 
\end{align}
 Note that,
\begin{align}
    \beta_{BTZ}\omega_{n} = \frac{2\pi^{2}}{\Lambda\sqrt{d}}\frac{2 n\Lambda\sqrt{d}}{\log\left(\frac{\Lambda\sqrt{d}}{2\pi\epsilon_{b}}\right)} = \frac{4\pi^{2}n}{\log\left(\frac{\Lambda\sqrt{d}}{2\pi\epsilon_{b}}\right)}
\end{align}
Thus we can see the Bose-Einstein mean is converted to usual Boltzman factor in equilibrium temperature $\beta_{BTZ}$:
\begin{align}\label{occupation}
    \langle N_{n}\rangle = \langle N_{n}\rangle_{\beta_{BTZ}} = \frac{1}{e^{\beta_{BTZ}\omega_{n}}-1}
\end{align}
This factor played a crucial role in deriving the microstate two-point correlator in \cite{Burman:2023kko}, which reproduces the thermal Hartle–Hawking correlator in the boundary limit. 

\subsection{Bulk-boundary matching of microstate correlators}

In \cite{Burman:2023kko}, the boundary limit of microstate two-point correlator in position space was computed explicitly by incorporating the overall normalization of the field in a Dirichlet stretched-horizon background. To obtain a well-defined field normalization using the Klein-Gordon inner product in the cut-off background, the boundary condition plays a crucial role. It is expected that any conformal boundary condition would serve a similar purpose \footnote{In other words, this suggests that a well-defined quantization in the stretched-horizon background emerges in the presence of conformal boundary conditions.}. One of the most remarkable features of this computation is the existence of a smooth $\epsilon_{b}\rightarrow 0$ limit for the correlators. Rather than becoming divergent or vanishing, the correlators reduce to a form exactly identical to the `one-sided Hartle-Hawking correlator' in the BTZ background. More specifically, following the notation and detailed derivation of \cite{Burman:2023kko}, our set-up gives
\begin{align}\label{pure state corr}
    G_{c}^{+} &\equiv \lim_{r,r' \rightarrow\infty} \langle N|\phi(r,t,\theta)\phi(r',t',\theta')|N\rangle \nonumber \\
    &\approx\frac{\Lambda\sqrt{d}}{\pi\log\left(\frac{\Lambda\sqrt{d}}{2\pi\epsilon_{b}}\right)}\sum_{n,J}\left[(N_{n}+1)\left(e^{-i\omega_{n}(t-t')}e^{iJ(\theta-\theta')}+e^{i\omega_{n}(t-t')}e^{-iJ(\theta-\theta')}\right)+e^{-i\omega_{n}(t-t')}e^{iJ(\theta-\theta')}\right]\tilde{f}(\omega_{n},J) \nonumber \\
   & = \frac{\Lambda\sqrt{d}}{\pi\log\left(\frac{\Lambda\sqrt{d}}{2\pi\epsilon_{b}}\right)}
   \sum_{n=-\infty}^{\infty} 
   \sum_{J=-J_{cut}}^{J_{cut}}
   \frac{e^{\beta_{BTZ}\omega_{n}}}{e^{\beta_{BTZ}\omega_{n}}-1}
   \tilde{f}(\omega_{n},J)
   e^{-i\omega_{n}(t-t')}
   e^{iJ(\theta-\theta')} + \text{variance}
\end{align}
Using typical state construction, it can be explicitly checked that the variance part will be exponentially suppressed in entropy\cite{Burman:2023kko}\footnote{Since the boundary limit of two point function in $|N\rangle$ will be same as that of the typical state constructed in the microcanonical energy silver, we can alternatively use the typical state construction to make it more rigorous.}. Hence, we can safely use the mean of occupation number as being Boltzman factor of occupation number. Here $\tilde{f}(\omega_{n},J)$ is defined by
\begin{align}\label{QNM}
 \tilde{f}(\omega_{n},J) 
 \equiv 
 \frac{
 \left(\frac{\omega_{n}^{2}-J^{2}}{4}\right)
 \sinh\left[\frac{\pi^{2}\omega_{n}}{\Lambda\sqrt{d}}\right]
 }{
 \sinh\left[\frac{\pi^{2} (\omega_{n}+J)}{2\Lambda\sqrt{d}}\right]
 \sinh\left[\frac{\pi^{2} (\omega_{n}-J)}{2\Lambda\sqrt{d}}\right]
 }.
\end{align}


In the $\epsilon_{b}\rightarrow 0$ limit, the spectrum becomes continuous, $\omega_{n}\rightarrow\omega$, and the sums over $n$ can be replaced by integrals of $\omega$ \textit{smoothly}. This is explicitly seen by considering (\ref{fixing identific}) by which we obtain: $\omega_{n+1}-\omega_{n}\sim 2\Lambda\sqrt{d}G_{N}$. In $G_{N} \rightarrow 0$, we have $(\omega_{n+1}-\omega_{n})\rightarrow d\omega$. Further, taking the compactification limit $J_{cut} \rightarrow \infty$, we obtain \cite{Burman:2023kko}
\begin{align}\label{continuum}
  \lim_{\epsilon_{b}\rightarrow 0}G_{c}^{+} 
  =
  \int^{\infty}_{-\infty}\frac{d\omega}{2\pi}
  \sum^{\infty}_{J=-\infty}\,
   \frac{e^{\beta_{BTZ}\omega}}{e^{\beta_{BTZ}\omega}-1}
   \tilde{f}(\omega,J)
   e^{-i\omega(t-t')}
   e^{iJ(\theta-\theta')}
\end{align}
\begin{align}
   \equiv
   \int^{\infty}_{-\infty}\frac{d\omega}{2\pi}
   \sum^{\infty}_{J=-\infty}\;
   G(\omega,J)
   e^{-i\omega(t-t')}
   e^{iJ(\theta-\theta')}.
\end{align}

Using the property $\tilde{f}(\omega,J)=-\tilde{f}(-\omega,J)$,
one immediately finds
\begin{align}
   G(-\omega,J)=e^{-\beta_{BTZ}\omega}G(\omega,J).
\end{align}
This is precisely the KMS condition for the momentum-space correlator. In position space, a thermal correlator must satisfy the Euclidean time periodicity condition on a strip of length $\beta$:
\begin{align}
    \langle\mathcal{O}_{1}(t-i\beta)\mathcal{O}_{2}(0)\rangle_{\beta}
    =
    \langle \mathcal{O}_{2}(0)\mathcal{O}_{1}(t)\rangle_{\beta}.
\end{align}
One can immediately identify (\ref{continuum}) as the boundary limit of the \textit{one-sided HHI correlator} as in \cite{Burman:2023kko}. Conversely, in the strict semiclassical limit $\epsilon_{b}\to 0$, the microstates can be directly identified with the HHI state, since the exponentially suppressed variance vanishes in this limit.

Consequently, the left-right ($LR$) correlator can be computed using the standard $J=CRT$ operator prescription familiar from the two-sided eternal black hole construction. This identification is further supported by the emergence of a type-III algebra in the strict semiclassical limit, where the spectrum becomes continuous as $\omega_n \to \omega$. Only within a type-III algebra can one consistently define both the modular operator and modular conjugation \cite{Leutheusser:2021frk}.

This feature is also evident from the definition of the state (\ref{microstate}) $|N\rangle$, where the operators $a_{n,J}$ and $a_{n,J}^{\dagger}$ become continuous, and their commutator approaches a Dirac delta function in the strict limit. Equivalently, the states acquire unbounded norm and therefore become ill-defined in the semiclassical limit, which is precisely the hallmark of a type-III algebra. See related discussion in \cite{Furuya:2023fei}.

Hence, we obtain
\begin{align}
\langle N|\mathcal{O}_{L}(t_{1},\theta_{1})\mathcal{O}_{R}(t_{2},\theta_{2})|N\rangle
=
\langle N|\mathcal{O}_{L}(t_{1},\theta_{1})
\mathcal{O}_{L}\left(t_{2}-\frac{i\beta_{BTZ}}{2},\theta_{2}\right)|N\rangle .
\end{align}
As can be checked explicitly, this does not alter the temporal behavior of the correlator. Moreover,
\begin{align}
\langle N|\mathcal{O}|N\rangle =0 .
\end{align}
Therefore, the connected $RL$ correlator should be identified with the \textit{exact} HHI correlator, since the disconnected contribution vanishes.\footnote{I thank Chethan Krishnan for clarifying this point.} See \cite{Burman:2024egy} for an explicit bulk construction of the two-sided HHI state correlator from the microstate correlator in the same limit. We may therefore conclude that
\begin{align}
\langle N| \mathcal{O}_{1}\mathcal{O}_{2}|N\rangle = \langle HHI|\mathcal{O}_{1}\mathcal{O}_{2}|HHI\rangle
\qquad \text{for} \qquad
\epsilon_{b}\to 0,\quad J_{\rm cut}\to \infty .
\end{align}

In this exact semiclassical limit, the $\mathcal{O}(e^{-S/2})$ suppression of the variance can be consistently neglected. This is precisely the regime in which the Maldacena's information loss puzzle becomes sharp \cite{Maldacena:2001kr}, namely when the $G_N\to 0$ limit is taken prior to the $t\to\infty$ limit in the boundary correlator. A notable aspect of the stretched-horizon bulk description is that it reproduces the smooth semiclassical horizon physics exactly, despite originating from a finite-$G_N$ framework in which bulk smoothness is not manifest from the outset.

However, this identification suggests a similar construction of the bulk TFD state $|TFD^{b}\rangle_{\epsilon_{b}}$ at finite $\epsilon_{b}$ stretched horizon cut-off and then take the $\epsilon_{b} \rightarrow 0$ limit.

\begin{align}
    |TFD^{b}\rangle_{\epsilon_{b}} = \frac{1}{\sqrt{Z_{\beta_{BTZ}}}}\sum_{E_{\epsilon_{b}}}e^{-\frac{\beta_{BTZ} E_{\epsilon_{b}}}{2}}|E_{\epsilon_{b}}\rangle_{L}|E_{\epsilon_{b}}\rangle_{R}
\end{align}
We will be first interested in computing single-sided correlator in the prescribed semiclassical limit $\epsilon_{b} \rightarrow 0$ and $J_{cut} \rightarrow \infty$ related to $G_{N}$ by (\ref{fixing identific}). For single-sided correlators, this precisely describes the semiclassical thermal correlators at temperature $\beta_{BTZ}$ with thermal average over all states in the Hilbert space. We will not explicitly show this due to complicated thermal averages, however we will provide strong justification behind this claim.


From the bulk side, we want to compute
\begin{align}
 &\lim_{\epsilon_{b}\rightarrow 0,\; J_{cut}\rightarrow \infty} \;\;\lim_{r,r'\rightarrow\infty}  \langle \phi_{L}(r,t,\theta)\phi_{L}(r',t',\theta')\rangle_{\beta_{BTZ}} \nonumber \\
 &=\lim_{\epsilon_{b}\rightarrow 0,\; J_{cut}\rightarrow \infty} \;\;\lim_{r,r'\rightarrow\infty}Tr_{\mathcal{H}_{\epsilon_{b}}}\left[e^{-\beta_{BTZ}H_{SH}}\phi_{L}(r,t,\theta)\phi_{L}(r',t',\theta') \right]
\end{align}
Here $H_{SH}$ defines the stretched horizon Hamiltonian for massless free scalars. Since this is just a thermal average, we can simply add sum of all energy eigenstate correlators with thermal exponential weight factor and then take the limit. For the microstate correlators at energy $E=M_{BTZ}$, we have already obtained exact Hartle-Hawking type correltor in a suitable bulk limit $G_{N}\rightarrow 0$. On the other hand, if any of the eigenstates shows quasi-periodic behavior, that will be enough to stop exact late-time thermal decay of microstate correlators. Consider the following eigenstate correlator:

 \begin{align}\label{eigenstate corr}
 &\lim_{r,r' \rightarrow\infty} \langle N|\phi(r,t,\theta)\phi(r',t',\theta')|N\rangle \nonumber \\
    &\approx\frac{\Lambda\sqrt{d}}{\pi\log\left(\frac{\Lambda\sqrt{d}}{2\pi\epsilon_{b}}\right)}\sum_{n,J}\left[(N_{n}+1)\left(e^{-i\omega_{n}(t-t')}e^{iJ(\theta-\theta')}+e^{i\omega_{n}(t-t')}e^{-iJ(\theta-\theta')}\right)+e^{-i\omega_{n}(t-t')}e^{iJ(\theta-\theta')}\right]\tilde{f}(\omega_{n},J) \nonumber \\
\end{align}
Here, if we assume the energy of the eigenstates $E\sim \mathcal{O}(M_{BTZ})=\sum_{n,J}\omega_{n}N_{n}$, the thermal decay will be manifested as we have shown previously. Even though, one may not exactly obtain $\langle N_{n}\rangle=\langle N\rangle_{\beta_{BTZ}}$, the order of energy level $N_{lev}$ is enough to convert the Bose-Einstein mean to a standard Boltzman factor with temperature of $\mathcal{O}(\beta_{BTZ})$. This is straightword from our previous analysis as one can explicitly check. The exact thermal decay will be governed by the quasi normal modes(QNM) which will contain explicit $\beta_{BTZ}$ factor\footnote{As observed in \cite{Burman:2023kko}, this factor has exact same structure of BTZ quasinormal modes in frequency space. See also \cite{Banerjee:2024dpl}, for related observation.}. Hence, we will only obtain thermal decay from such correlators with $e^{-\beta_{BTZ}t}$ dependence like HHI correlator. However, if we consider $1<<E<<M_{BTZ}$, we can still use the Bose-Einstein mean, but fail to rewrite it as a function of $\omega_{n}$. 
As we argued previously, rewriting the sum into integral it is essential to write $\langle N_{n}\rangle \sim g(\omega_{n})$. The smooth emergence of thermal decay is purely a consequence of converting the sum into integrals. If we can not convert the discrete sum into the integral, we might expect to have quasi-periodic in time behavior. However, in such cases, the prefactor $\frac{\Lambda\sqrt{d}}{\pi \log\left(\frac{\Lambda\sqrt{d}}{2\pi \epsilon_{b}}\right)} \sim \Lambda\sqrt{d}G_{N}$ has not been absorbed in a \textit{smooth integral}. Hence if we take $G_{N} \rightarrow 0$ limit, the prefactor vanishes. Hence, those correlators will further be vanishing. When $N_{n} \sim \mathcal{O}(1)$ the sum can be converted to integral due to exact $\omega_{n}$ dependence of the correlator and it will show thermal decay of $e^{-\beta_{BTZ}t}$. 
Thus, even though, we can not exactly do the sum, we have given strong evidences why it will have only thermal decay of $\mathcal{O}(e^{-\beta_{BTZ}t})$.  

If we now consider two-sided correlator of the form:
\begin{align}
    \lim_{r,r'\rightarrow\infty}  \langle \phi_{L}(r,t,\theta)\phi_{R}(r',t',\theta')\rangle_{\beta_{BTZ}} = \sum_{n,m}e^{-\frac{\beta_{BTZ}}{2}(E^{n}_{\epsilon_{b}}+E^{m}_{\epsilon_{b}})}\; _{L}\langle E^{n}_{\epsilon_{b}}|\phi_{L}|E_{\epsilon_{b}}^{m}\rangle_{L} \; _{R}\langle E^{n}_{\epsilon_{b}}|\phi_{R}|E_{\epsilon_{b}}^{m}\rangle_{R}
\end{align}
Since these are sum of one point function of cross-correlators, these will not vanish generically. For instance, the only non-vanishing cross-correlators will be of the particular combination $|E^{n}_{\epsilon_{b}}\rangle_{L,R} = |N\rangle$, $|E^{m}_{\epsilon_{b}}\rangle_{L,R} = |N+1\rangle$ and the vice versa. We find using a straightforward calculation:
\begin{align}
  \lim_{r,r'\rightarrow\infty}   \langle N|\phi_{L}|N+1\rangle\langle N|\phi_{R}|N+1\rangle  \approx\frac{\Lambda\sqrt{d}}{\pi\log\left(\frac{\Lambda\sqrt{d}}{2\pi\epsilon_{b}}\right)}\sum_{n,J}\left[(N_{n}+1)e^{-i\omega_{n}(t-t')}e^{iJ(\theta-\theta')}\right]\tilde{f}(\omega_{n},J)
\end{align}
Using a similar logic as previous equal-side correlator, we can again see that generically this cross correlator will also be decaying exponentially. Thus, the opposite side connected $LR$ correlator\footnote{The disconnected part will eventually be zero as easily can be argued.} will also be non-vanishing and gives the thermal temporal decay. Without explicitly computing, one can also directly argue this by using KMS condition to change $RL$ to $LL$ correlators as we have shown before. 

This simply illustrate why only thermal decaying nature of $\mathcal{O}(e^{-\beta_{BTZ}t})$ survives in the $|TFD^{b}\rangle_{\epsilon_{b}}$ after taking $\epsilon_{b} \rightarrow 0$ limit. To summarize, whenever we can write the individual correlators as some function of $\omega_{n}$, we can immediately convert it to integral in the semiclassical limit and we will eventually obtain a thermal decay with temperature $\beta_{BTZ}$. On the other hand, when we can not write them as a function of $\omega_{n}$, the prefactor ensures the vanishing of those correlators. Hence, it is expected that after the explicit thermal average, one would finally obtain the same HHI correlator, since by definition the HHI state is a TFD state made out of bulk EFT Fock space\footnote{This is analogous to Rindler eigenstates, whch in the bulk will correspond to excited states on Boulware vacuum $|0\rangle_{B}$. $|0\rangle_{SH}$ reduces to $|0\rangle_{B}$ in $\epsilon_{b} \rightarrow 0$ limit as argued in \cite{Mukohyama:1998rf}.} in each wedges.

From this argument, we may finally conclude that\footnote{Here we are not distinguishing HHI and Unruh state, featuring same correlation function in the exact semiclassical limit for (eternal)BTZ.}
\begin{align}
 \langle N|\mathcal{O}_{1}\mathcal{O}_{2} |N\rangle= \langle HHI|\mathcal{O}_{1}\mathcal{O}_{2}|HHI\rangle = \lim_{\epsilon_{b} \rightarrow 0}\; _{\epsilon_{b}} \langle TFD^{b}|\mathcal{O}_{1}\mathcal{O}_{2}|TFD^{b}\rangle_{\epsilon_{b}} 
\end{align}

\textbf{\textit{Matching with boundary correlators:}}
To connect with boundary observer story, we also need to identify this as the reference KMS state in $\epsilon \rightarrow 0$, which is the conformal vacuum $|0\rangle$. As in our boundary prescription to connect correlators of modular quantization to BTZ correlators, we compute two point correlator in the conformal vacuum $|0\rangle$ and compactify the $\theta$ coordinate. From (\ref{open closed}) we have
\begin{align}
    \lim_{\epsilon\rightarrow 0}Tr_{\mathcal{H}_{\epsilon}}\left[e^{-\frac{2\pi^{2}}{\Lambda\sqrt{d}}H_{\epsilon}}\mathcal{O}_{1}(s_{1},\theta_{1})\mathcal{O}_{2}(s_{2},\theta_{2}) \right] = \langle 0|\mathcal{O}_{1}(s_{1},\theta_{1})\mathcal{O}_{2}(s_{2},\theta_{2})|0\rangle
\end{align}
Using this connection, we can explicitly compute one-sided correlator as follows:
\begin{align}
    \sum_{m=-\infty}^{\infty}\langle 0|\mathcal{O}(s,\theta+2\pi m)\mathcal{O}(0,0)|0\rangle =\sum_{m=-\infty}^{\infty} \prod_{i=1}^{2}\left(\frac{\partial z'_{i}}{\partial \omega_{i}}\right)^{h_{\mathcal{O}}}\left(\frac{\partial \bar{z}'_{i}}{\partial \bar{\omega}_{i}}\right)^{h_{\mathcal{O}}}\langle 0|\mathcal{O}(z'_{1},\bar{z}'_{1})\mathcal{O}(z'_{2},\bar{z}'_{2})|0\rangle
\end{align}
Here, we need to use (\ref{boundary diff}) $z'=-\frac{\alpha}{2\beta}-\frac{\sqrt{d}}{2\beta}\cot\left(\frac{\Lambda\sqrt{d}}{2\pi}\omega\right)$. Using the conformal transformation 

\begin{align}
 \sum_{m=-\infty}^{\infty}\langle 0|\mathcal{O}(t,\theta+2\pi m)\mathcal{O}(0,0)|0\rangle    \approx \sum_{m=-\infty}^{\infty} \left[\cosh\left(\frac{2\pi}{\beta_{BTZ}}(\theta+2\pi m)\right)-\cosh\left(\frac{2\pi t}{\beta_{BTZ}}\right) \right]^{-2h_{\mathcal{O}}}
\end{align}

Since the conformal vacuum $|0\rangle$ is the KMS vacuum of $H_{tot}=H_{R}-H_{L}$ due to SL(2,$\mathbb{C}$) symmetry, $H|0\rangle = 0$. As observed in \cite{Das:2024vqe}, the algebra will be type-III$_{1}$ factor. Hence by KMS property, we can again obtain the $RL$ correlator in $|0\rangle$ by $\mathcal{O}_{R}(t_{R},\theta_{R})=\mathcal{O}_{L}(t_{L}-\frac{i\beta_{BTZ}}{2},\theta_{R})$. From this we can obtain again the same HHI correlator.  Also, by definition $\langle 0|\mathcal{O}|0\rangle = 0$ due to conformal transformation. Hence we exactly reproduce the connected HHI $RL$ correlator from boundary conformal vacuum. This is also consistent with the `open-closed string' duality as in (\ref{open closed}). In other words, we have explicitly made the following identification:
\begin{align}
 & \sum_{m=-\infty}^{\infty}   \lim_{\epsilon\rightarrow 0}Tr_{\mathcal{H}_{\epsilon}^{ab}}\left[e^{-\frac{2\pi}{\Lambda\sqrt{d}}H}\mathcal{O}_{L}(s,\theta)\mathcal{O}_{L}(0,0)\right] = \sum_{m=-\infty}^{\infty} \langle 0|\mathcal{O}_{L}(s,\theta)\mathcal{O}_{L}(0,0)|0\rangle \\ \nonumber
  &= \lim_{\epsilon_{b} \rightarrow 0}\langle N|\mathcal{O}_{L}(s,\theta)\mathcal{O}_{L}(0,0)|N\rangle = \langle HHI|\mathcal{O}_{L}(s,\theta)\mathcal{O}_{L}(0,0)|HHI\rangle
\end{align}
This is precisely the setting discussed in the previous section, where the type-III$_{1}$ algebra arises from the continuous spectrum of $H_{\epsilon}$ in the $\epsilon \rightarrow 0$ limit. This provides the algebraic counterpart of the equivalence between bulk QFT in a fixed BTZ background and the boundary CFT in modular quantization, prior to incorporating the center into the algebra.

A similar story is also straightforward in microcanonical ensemble. From the bulk picture, the constructed individual microstates have reproduced exact HHI behavior of correlators. Hence the previous discussion will be true for a microcanonical TFD \cite{Chandrasekaran:2022eqq}, constructed at finite $\epsilon_{b}$. It will eventually reproduce same connected $LR$, as well as $LL$ correlators as we have justified. On the other hand, from the boundary modular quantization, we again have shown exact ensemble equivalence in obtaining the BTZ entropy. Thus, immediately we can identify the KMS vacuum $|0\rangle$ to be the microcanonical TFD:
\begin{align}
  |0\rangle \sim \lim_{\epsilon\rightarrow 0}
    \sum_{j}
    e^{-\frac{\beta_{BTZ}}{2}\left(E_{j}-\frac{c\Lambda^{2}d}{12\pi^{2}}\right)}
    f\!\left(E_{j}-\frac{c\Lambda^{2}d}{12\pi^{2}}\right)
    |E_{j}\rangle_{L}|E_{j}\rangle_{R}.  
\end{align}
As in \cite{Chandrasekaran:2022eqq}, $f$ may be taken to be a Gaussian function sharply peaked around the microcanonical energy silver of $M_{BTZ}$: $E_{j}=\frac{c\Lambda^{2}d}{12\pi^{2}}$. Correspondingly, it will reproduce HHI correlators as in the previous discussion. 

As observed in \cite{Das:2024vqe}, the modular Hamiltonian lies in the ``\textit{heating phase}'' in the context of Floquet CFTs\footnote{Such dynamical phases have also been interpreted in terms of a first-order phase transition from certain bulk descriptions in AdS$_3$ \cite{Das:2022pez}.}. In \cite{Das:2024vqe}, this phase was related to an analogue of the Hawking-Page transition occurring above $T>T_{HP}$. Consequently, within modular quantization there is no need to distinguish separate descriptions of the TFD state below and above the Hawking-Page temperature. This is furthermore consistent with the bulk description, where $\lim_{\epsilon_{b}\rightarrow 0}|TFD^{b}\rangle_{\epsilon_{b}}$ has been argued to reproduce an HHI-type thermal decay.

\subsection{Comments on adding the center}
To complete the discussion of the change of algebra in modular quantization, however, one must understand the `closed-string' channel of the dual description. In particular, we first analyze the shrinking limit of the localized Dirichlet boundary states at the stretched horizon. Motivated by the previous discussion regarding the exact equivalence between bulk and boundary HHI correlators in the thermodynamic and semiclassical limits respectively, one may anticipate a similar mechanism governing the identity sector in the $\epsilon_{b} \rightarrow 0$ limit of the entropic dominance. Since from the boundary modular quantization framework, we have seen all conformal boundary conditions shrink to the vacuum $|0\rangle$ due to the vacuum dominance in the `closed string sector'. Also we have identified $|0\rangle$ with the HHI state in the dual description within modular quantization in canonical ensemble. This is also justified with the Dirichlet stretched horizon description of BTZ, in which the microstates reduces to HHI state. However, as we stated before, the main issue in the semiclassical limit is the type-III$_{1}$ transition of the algebra, in which the finite $\epsilon_{b}$ microstates are apparently `lost'. This is an algebraic version of `information loss problem' as first formulated in \cite{Leutheusser:2021frk} within standard AdS/CFT with eternal black hole.

In the conventional formulation of AdS/CFT, the change in the type of algebra is understood through the inclusion of gravity, or equivalently the ADM Hamiltonian, on the boundary via a suitable crossed-product construction \cite{Witten:2021unn}. The resulting extended algebra acts on an enlarged Hilbert space and admits a notion of trace, thereby allowing one to define a classical entropy and, subsequently, the generalized area law. Even without incorporating $1/N$ corrections, this change of algebra can still be understood in a KMS reference state through the microcanonical TFD construction \cite{Chandrasekaran:2022eqq}. One may expect an analogous mechanism to arise here, since the boundary modular Hamiltonian is treated as the dual of the ADM mass for the BTZ geometry. Moreover, the starting algebra is already a type-III$_1$ factor in the semiclassical limit, as discussed earlier. However, within this algebraic lifting, the pure states do not exist, but one can compute entropy as in classical entropy.

The principal novelty of modular quantization, however, is that it offers an alternative realization of this change of algebra directly within the strict thermodynamic limit $\epsilon\rightarrow 0$. We have shown that it is possible to introduce an `emergent center,' constructed from bounded functions of fixed-point scalar primaries as affiliated operators of the algebra. The inclusion of this center transforms the original factor into a non-factor endowed, in principle, with an infinite-dimensional center. Perhaps the most remarkable consequence is the emergence of $\mathcal{H}_{edge}$ as a dual counterpart of $\mathcal{H}_{int}$, generated by the action of the center on two conformal vacua corresponding to two different channels. Upon incorporating $\mathcal{H}_{edge}$ into the original Hilbert space, one definitely expect the algebraic structure to undergo a substantial modification. The main novelty is the existence of pure states in the constructed new Hilbert space.

However this discussion includes sum over all boundary conditions in the description of full or new Hilbert space. On the other hand, in our large BTZ description through streteched horizon framework, we only consider the Dirichlet boundary condition on the probe scalar fields. In fact, if we use all other (non)conformal boundary conditions on the bulk stretched horizon, we may understand the emergence of $\mathcal{H}_{edge}$. 

Hence, influenced by Witten’s crossed product construction, modular quantization indicates that gravity can also be incorporated by treating the “emergent” eigenmodes of the dual modular Hamiltonian as an algebraic center, thereby allowing states to be naturally defined in terms of $\mathcal{H}_{edge}$ or $\mathcal{H}_{int}$. Correspondingly, to understand the consequence of adding the center in semiclassical physics, one should compute correlators in the state of $\mathcal{H}_{edge}$ and take the $c \rightarrow \infty$ limit.

\textbf{\textit{Comments on Finite mass BTZ:}}

The entropic vacuum dominance is only true for large eternal BTZ(infinite mass limit) black holes. To obtain a finite mass BTZ, we need to take $\Lambda\sqrt{d} =\text{finite}$. One way to take this limit correspond to taking $\frac{\sqrt{d}}{\beta} \rightarrow 0$ while $ \epsilon \rightarrow 0$. The original boundary modular Hamiltonian in such limit approaches to a phase transition in the language of floquet CFT. Using `open-closed string' duality in partition function, we have
\begin{align}
Z^{\{ab\}}_{\tau} = \int dE_{\epsilon} \rho_{m}(E_{\epsilon})e^{2\pi i\tau E_{\epsilon} } = \sum_{\Delta_{\zeta}}\langle b|\Delta_{\zeta}\rangle \langle\Delta_{\zeta}| e^{-2\Lambda\sqrt{d}\left(\Delta_{\zeta}-\frac{c}{12}\right)}|a\rangle 
\end{align}
Since $\epsilon$ is finite, we can not use the vacuum dominance, and correspondingly we have
\begin{align}
  Z^{\{ab\}}_{\tau} = e^{\frac{c\Lambda\sqrt{d}}{6}}\sum_{\Delta_{\zeta}}\langle b|\Delta_{\zeta}\rangle\langle\Delta_{\zeta}|a\rangle  
\end{align}

However, in $c\rightarrow \infty$ limit, we have the relevant contribution coming from $c$ dependent part:
\begin{align}
    \lim_{c \rightarrow \infty}Z^{\{ab\}} \approx e^{\frac{c\Lambda\sqrt{d}}{6}}
\end{align}
Hence one recovers the BTZ entropy using microcanonical counting. However, this clearly tells that one can not use the entropic vacuum dominance in the `closed string' sector and hence the boundary contribution can not be ignored. 

However, the bulk interpretation should still be a BTZ black hole in $G_{N} \rightarrow 0$ limit\footnote{This is also true for exact Cardy counting, which is only true for large BTZ black hole. See related comments on this in \cite{Krishnan:2026mpa}.}. The main point we want to emphasize is that, this is one example where \textit{nontrivial $\mathcal{H}_{edge}$} and hence \textit{non-trivial center} correspondingly  must appear in entropy counting.

\textit{\textbf{Connection to fixed area states:}}

A natural interpretation of our non-standard bulk picture may arise from the Euclidean path integral construction of fixed-area states, which admits a natural geometric analogy with quantum error-correcting codes in holography. As argued in \cite{Das:2024vqe}, the fixed-area states can be naturally identified with eigenstates of the one-sided modular Hamiltonian through the JLMS formula in $G_{N} \rightarrow 0$ limit \cite{Jafferis:2015del}. Once constructing these states, the full Hilbert space of gravity should have a natural factorization within superselection sector fixed by the value of area. Within our modular quantization framework, each of these states can be precisely realized as states in $\mathcal{H}_{edge}$, formed by the center of the algebra. For that, one needs to sum over all conformal boundary conditions as argued in \cite{Das:2024vqe}. 
\begin{align}
    \mathcal{H}_{\text{full CFT}} = \oplus_{a,b} \mathcal{H}^{\{ab\}}_{L} \otimes\mathcal{H}^{\{ab\}}_{R} \overbrace{\longrightarrow}^{\epsilon\rightarrow 0} \oplus_{f}\mathcal{H}_{L}^{f} \otimes \mathcal{H}_{R}^{f}
\end{align}
However, we should emphasize that, from thermal entropy counting of modular quantization, the strict $\epsilon \rightarrow 0$ only reproduce the vacuum sector. Thus the full $\mathcal{H}_{edge}$ and $\mathcal{H}_{int}$ do not appear in entropically dominant saddles of modular quantization in strict semiclassical or thermodynamic limit. Nevertheless, it should be there in the full Hilbert space, by construction, after adding the center or (non perturbative) gravity. 

In our bulk description, the shrinking limit consisting of edge modes should be thought of \textit{emergent} eigenmodes of ADM Hamiltonian \footnote{We should note that, in contrast to the gravitational “soft mode” proposal of \cite{Hawking:2016msc}, the zero modes of the full two-sided modular Hamiltonian constructed in this work do not correspond to charges associated with large diffeomorphisms. Rather, our characterization of the corresponding eigenstates bears a closer resemblance to the `microstate geometries' encountered in the Fuzzball paradigm.}, as the modular Hamiltonian itself describes the dual of ADM Hamiltonian. In a similar spirit to crossed product construction, one can affiliate this emergent center to the algebra of type-III$_{1}$ appeared in qft in curved background which we obtained in BTZ with Dirichlet stretched horizon picture after taking the semiclassical limit. The resulting algebra, acted on new Hilbert space $\mathcal{H}^{b}_{edge}$ should definitely be changed. The Euclidean counterpart of $\mathcal{H}^{b}_{edge}$ is very close to the construction of fixed area states of Euclidean quantum gravity in the semiclassical limit. In the Lorentzian picture, the bcc operator acting on conformal vacuum should be naturally realized by horizonless geometries. 
Most notably, \textit{in some exact supersymmetric examples of AdS/CFT, such type of microstate geometries have been realized in supergravity limit within Fuzzball paradigm \cite{Lunin:2001jy}, \cite{Bena:2007kg}-\cite{Giusto:2012yz}. We expect $\mathcal{H}_{edge}$ bears a similar version for non- supersymmetric examples in AdS/CFT. However, only a subset of $\mathcal{H}_{edge}$ should belong to ``microstate geometries".}

\textbf{\textit{Comment on Unitarity restoration after including the central algebra:}}

While our discussion has largely centered on the new Hilbert space structures, rather than a thorough analysis of the changes in the algebra, we can still explore the resulting consequences by considering the action of the centrally extended algebra on the new Hilbert space $\mathcal{H}'$:
\begin{align}
    \mathcal{A}^{L,R}_{new} \equiv  \mathcal{A}(\mathcal{B}(\mathcal{H}^{L,R}_{\epsilon\rightarrow 0}))\otimes (\oplus_{h}\mathcal{A}_{f}(h)) ; \; \mathcal{H}' = \oplus_{f}\mathcal{H}_{L}^{f} \otimes \mathcal{H}_{R}^{f}
\end{align}
Here $\mathcal{H}^{f}_{L,R}=\mathcal{H}^{L,R}_{edge}$, with $f$ being the fixed point primaries in $\mathcal{H}_{edge}$. To find it's consequence on correlators, we should compute the following redefined correlators using `open-closed string' duality in the modified Hilbert space:
\begin{align}
    Tr_{\mathcal{H}_{new}}\left[e^{-\frac{2\pi^{2}}{\Lambda\sqrt{d}}H}\mathcal{O}_{1}(s,\theta)\mathcal{O}_{2}(0,0) \right] = \sum_{f,f'}\langle \mathcal{O}_{f}|\mathcal{O}_{1}(\zeta,\bar{\zeta})\mathcal{O}_{2}(1,1)|\mathcal{O}_{f'}\rangle
\end{align}
Here $\mathcal{O}_{f,f'}$ corresponds to fixed point scalars acting on conformal KMS vacuum $|0\rangle$, which belongs to $\mathcal{H}_{int}$. Also $\zeta=e^{i\sqrt{d}\omega}$, $\bar{\zeta}=e^{-i\sqrt{d}\bar{\omega}}$. Since we have included all conformal boundary conditions, we must sum over all primaries in $\mathcal{H}_{int}$. To understand the effect in semiclassical limit, we have to take $c \rightarrow \infty$ of the correlators. In \cite{Das:2024mlx}, we have shown a remarkable consequence of modular quantization is the large $c$ HHLL blocks with $h_{H}>\frac{c}{24}$ shows atypical scar-like behavior in contrast to semiclassical typicality occurred in radial quantization. We will briefly review this in appendix-(\ref{B}). If we use this in strict large $c$ limit, we will see a quasi-periodic behavior in Lorentizan time $t$ of the correlators, coming from semiclassical $HHLL$ blocks in modular quantization. Even though, this is purely boundary computation, the exact bulk-boundary correpondence for $|0\rangle \longleftrightarrow|HHI\rangle$ shows a similar prescription should be obtained purely from bulk side once adding the central sector corresponding to the eigenmodes of boundary ADM Hamiltonian to the bulk or dual boundary type-III$_{1}$ factor.

This is reminiscent to unitarity restoration via obtaining `Page curve' in the entropy of Hawking radiation, where the Euclidean gravitational path integral acquires `replica wormhole' saddles after summing over topology, which further poses several puzzles regarding construction within a single CFT. In contrast, here we do not need any `wormholes' in the description to restore unitarity following modular quantization framework in the background independent algebra. Interestingly, $\mathcal{H}_{int}$, consisting of local primaries inserted on conformal vacuum $|0\rangle$ may also be thought of as partially entangled TFD (PETS)\cite{Antonini:2023hdh}, when we identify the vacuum $|0\rangle$ as the TFD or dual HHI via modular quantization or the corresponding bulk description. We expect this may have some direct implication in the recent discussion on baby universes in AdS, which is expected to be Lorentzian version of `wormhole' saddles in gravitational path integral. We have shown in the appendix, the thermal correlator in $\mathcal{H}_{edge}$  will restore unitarity via `open-closed string' duality, explicitly due to this sector of $\mathcal{H}_{int}$, consisting of $\mathcal{O}_{h}|0\rangle$ with $h>\frac{c}{24}$. The AS$^{2}$ construction of baby Universe also relies on heavy operator(above BTZ threshold) in PETS description. Hence we may conclude that, \textit{from the standpoint of modular quantization, `baby universes' in $\mathcal{H}_{int}$ may appear as horizonless geometry with boundary-like surfaces in $\mathcal{H}_{edge}$ due to isomorphism $\mathcal{H}_{edge}\cong \mathcal{H}_{int}$.} However, recent discussions surrounding the puzzles of constructing such baby universes  \cite{Antonini:2024mci}-\cite{Engelhardt:2026blp}, \cite{Gesteau:2025obm} have made it increasingly clear that a semiclassical description of a baby universe cannot arise within the standard formulation of AdS/CFT \footnote{Here by standard formulation, we suggest a formulation based on radial quantization, where two copies of CFT are needed to describe a black hole in a pure state.} without additional assumptions, such as ensemble averaging over CFTs or over states \cite{Magan:2025hce} \footnote{See also other discussion regarding averaging over  $N$ \cite{Liu:2025ikq}, \cite{Kudler-Flam:2026nzz}}. This is also clearly visible from modular quantization framework, where \textit{no independent description of `interior' or `wormhole' exists!} This is also supported by \cite{Witten:1999xp}, since by the no-go theorem, positive curvature boundary(dS$_{2}$ static patch) can not have any `wormhole' solution in a dual AdS description.

While states in $\mathcal{H}_{int}$ below the BTZ threshold exhibit a thermally decaying behavior in modular frame, states above the threshold halt this decay. Consequently, within modular quantization, \textit{the most interesting features} arise from the sub-threshold states: rather than displaying the exact behavior of a hard surface, the (stretched) horizon continues to exhibit thermal dissipation. A similar description has been observed in the context of a `moving membranes' or `moving mirrors' in a two-dimensional CFT, as discussed in \cite{Akal:2021foz}-\cite{Biswas:2024mlq} \footnote{See also an attempt for constructing `wormhole' like features using `moving interface' \cite{Biswas:2024xut}.}. 


From the bulk stretched horizon picture, on the other hand, one can obtain this full center by constructing non-trivial excitations on top of the stretched horizon. For instance, see related Stringy Fuzzball-inspired constructions involving nontrivial angular profiles at the stretched horizon \cite{Das:2023ulz}, which is reminiscent to boundary descriptions of 2d `moving mirrors' \cite{Akal:2021foz}-\cite{Biswas:2024mlq}. Summing over all sorts of conformal boundary conditions should naturally realize as constructing all possible admissible non-trivial stretched horizon profiles in the bulk descriptions\footnote{As noted previously, this is `analogous' to the Euclidean gravitational path integral, where summing over topologies gives rise to nontrivial Euclidean wormhole saddles that play a central role in restoring unitarity and reproducing the Page curve. However, our description does not create any parallel universe or `wormhole' to restore unitarity.}. The smooth limit of all such boundary conditions from CFT should predict a natural realization of `microstate geometries' within $\mathcal{H}_{edge}$ in the bulk semiclassical limit.

Hence, to summarize, our claim is:

\textit{Adding (non perturbative) gravity in the algebra} $\implies$ \textit{Restoration of Unitarity} $\implies$ \textit{no independent interior and (Stretched) horizon with microstructures}

\section{Discussion}\label{sec7}

In this note, we undertake a detailed analysis of modular quantization \textit{or heating phase quantization in the context of Floquet CFT} in 2d CFT, as first proposed in \cite{Das:2024mlx},\cite{Das:2024vqe}, partly motivated by the goal of understanding non-perturbative signature of black hole Hilbert space. We will summarize the key results and significance in the following:
\begin{itemize}
   
\item The black hole description arising from radial quantization in AdS/CFT is purely coming from the usual identification of $\mathcal{H}_{AAdS}\cong\mathcal{H}_{CFT}$, where the full matter plus gravity Hamiltonian in AAdS is dual to CFT Hamiltonian(under radial quantization)\footnote{I thank Bobby Ezhuthachan to clarify this point.}. In this standard basis, black hole is described via a mixed thermal state, often purified by a second copy of a CFT in a TFD state. Hence constructing a black hole Hilbert space is technically a difficult problem within it's usual construction. Usually, a subset of full CFT Hilbert space or the microstates will form the black hole Hilbert space. From the discussion on the emergent type-III algebra of single trace primaries acting on dual TFD state \cite{Leutheusser:2021frk}, the explicit realization of those microstates in the semiclassical limit is not well-understood by definition. Consequently, constructing an explicit embedding of the corresponding black hole code subspace into the full physical Hilbert space(CFT Hilbert space) remains a highly nontrivial task.

In contrast, modular quantization of the CFT provides a natural, background-independent Hilbert space description of the BTZ black hole, following Witten's notion of a background-independent operator algebra, as discussed previously. In this construction, the inclusion of gravity via emergent center of the dual algebra,\footnote{ Via JLMS relation, the center is identified with the dual of the area operator \cite{Harlow:2016vwg}, where the boundary modular Hamiltonian is dual to the area operator in leading order of $G_{N}$.}, already incorporates the relevant code subspace within well-defined $\mathcal{H}_{edge}$. Since, in this observer motivated AdS/CFT picture, we are only zooming the black hole Hilbert space sector, only the microstates of $E=M_{BTZ}=\frac{c\Lambda^{2}d}{12\pi^{2}}$ constructs the Hilbert space sector $\mathcal{H}_{\epsilon}$. Correspondingly, $\mathcal{H}_{edge}$ in $\epsilon\rightarrow 0$ is dominated by primaries $h\sim \frac{c\Lambda^{2}d}{24\pi^{2}}>\frac{c}{24}$. Although the precise nature of the non-isometric embedding\footnote{This is non-isometric, since the descendant states do not appear in $\mathcal{H}_{edge}$.} $\mathcal{H}_{\epsilon \to 0} \rightarrow \mathcal{H}_{edge}$ remains unclear, our construction nevertheless provides a well-defined and exact characterization of $\mathcal{H}_{edge}$. Nevertheless, adding the full (non-perturbative) gravity suggests to include the full center to the description and hence one need to incorporate full $\mathcal{H}_{edge}$.

\item Within the standard radial quantization framework, to the best of our knowledge, an exact derivation of the HHI correlator of BTZ black hole from either boundary CFT correlators on the torus or correlators in the TFD state above the $T_{HP}$ has not been achieved. In a broader sense, this issue appears closely related to Maldacena's formulation of the information-loss puzzle. The same limitation persists for microstate correlators constructed using the Cardy density of states on the torus. Although such correlators exhibit thermal decay through the dominance of HHLL conformal blocks\footnote{Here we are not assuming large $N$ factorization as in the usual AdS/CFT examples.}, the origin of the image sum remains unclear, despite the fact that it should be intrinsic to a description formulated on a compact torus.

By contrast, within the modular quantization framework and its associated bulk description, we have explicitly demonstrated the equivalence between the boundary HHI correlator and the vacuum correlator of the boundary theory. In this picture, the image sum admits a natural interpretation as arising from the compactification of the $\theta$ coordinate around the fixed points. This provides additional evidence in favor of the proposed dual description and offers a strong consistency check of the modular quantization framework within AdS/CFT. Consequently, this formulation of the BTZ black hole Hilbert space within AdS/CFT inherently accommodates \textit{both} background independence and the emergence of a semiclassical EFT limit.

\item To address the information-loss and unitarity paradoxes in the EFT regime of black holes, the conventional gravitational path-integral formulation of AdS/CFT incorporates replica wormhole saddles, as discussed earlier. However, in the absence of any form of ensemble averaging on the (single) CFT side, a purely CFT-based description of such wormholes has not yet been understood. Even the simplest example, namely the eternal AdS wormhole, gives rise to the factorization problem \cite{Maldacena:2004rf} within the standard AdS/CFT framework based on gravitational path integral formulation.

One of the central results of this work is to propose an alternative perspective on the resolution of the unitarity paradox. Following the notion of a background-independent operator algebra, non-perturbative gravitational degrees of freedom must be incorporated into the algebra of observables associated with the exterior region. In the observer based description of AdS/CFT, this corresponds to including the ADM Hamiltonian, or equivalently the dual CFT Hamiltonian, as an element of the algebra. Using the modular quantization framework, we have shown that this extension is naturally realized through the inclusion of the emergent center of the algebra, which in turn gives rise to the new Hilbert space $\mathcal{H}_{edge}$.

Furthermore, exploiting the `open–closed string' duality inherent in the construction, we have argued that certain heavy primary states above the BTZ threshold within $\mathcal{H}_{edge}$ restore unitarity once the full center is incorporated into the algebra. From the bulk perspective, this should admit an interpretation in terms of nontrivial boundary profiles localized on the stretched horizon that remain well-defined in the smooth limit. The scar-like behavior of the corresponding correlators further suggests that these heavy states may be associated with horizonless hard boundary like objects. On the other hand, the thermal decaying nature of the states in $\mathcal{H}_{edge}$ below the threshold, would further provide interesting horizonless exotic geometries. 

Taken together, these observations suggest that the conventional notion of a perfectly smooth black-hole horizon will not survive in a theory of quantum gravity such as AdS/CFT within this observer-based approach. Rather, our results appear more naturally aligned with proposals of the firewall or fuzzball type, in which nontrivial microscopic structure replaces the naive semiclassical horizon.  
 
\end{itemize}
\textit{\textbf{Possible future directions:}}


One of the most significant implications of modular quantization is that it naturally reproduces a horizon structure characterized by fixed point, an emergent center, and the description of both sides of the geometry within a single CFT in the Lorentzian description. Given its explicit connection to `Rindler-like' quantization\footnote{The Rindler Hamiltonian is also conformally mapped to the same set-up as we argued.}, it is natural to ask whether modular quantization may also play a role in near-horizon CFT descriptions in a theory of non-extremal black holes in quantum gravity. The explicit dS$_{2}$ static patch connection already provides an observer description in the Nariai limit of SdS black hole. Within Witten’s proposal of a background-independent algebra, if one associates a CFT observer near the horizon of other non-extremal black holes, modular quantization along the modular timeline would, by construction, yield a gravitational description through a modified Hamiltonian constraint, where $H_{tot}=H_{obs}+H_{ADM}$. This would be slightly abstract construction, where the matter dof is only incorporated through near horizon observer dof\footnote{This might be achieved via dimensional reduction of pure gravity sector in the near horizon region of non-extremal black hole \cite{Solodukhin:1998tc}.}. 
In general non-extremal black holes, such CFT observer must arise itself from a complete theory of quantum gravity rather than introducing it formally by hand. However, as in our present study within AdS/CFT, the construction of $\mathcal{H}_{edge}$ suggests a holographic description of black holes directly in terms of `horizon edge modes'. Several evidences in the literature support such near horizon holography perspective. These include membrane paradigm \cite{Price:1986yy}- \cite{Parikh:1997ma}, as well as related fluid/gravity correspondence in AdS/CFT \cite{Hubeny:2011hd}, \cite{Bhattacharyya:2007vjd}, the near horizon conformal symmetry with analogous structure of a single copy of the “near-horizon Virasoro algebra” \cite{Solodukhin:1998tc}-\cite{Halyo:2015ffa}, as noticed in \cite{Das:2024vqe}, Kerr/CFT correspondence \cite{Guica:2008mu}, Carrolian description of near horizon physics\cite{Bagchi:2023cfp}-\cite{Bagchi:2026qpi}, which may provide a natural holographic framework for non-extremal black holes, even beyond the AdS setting. However, to the best of our knowledge, no concrete description has been yet achieved.  This type of picture might provide a notion for `horizon as a holographic plate'. 

A more concrete realization of Witten’s observer-based proposal may emerge in the context of extremal or near-extremal Reissner–Nordström black holes. In such cases, the decoupling of the asymptotic region suggests the possibility of an effective CFT description in terms of an observer in the near-horizon, near-extremal limit
, in a similar spirit of recent exploration to JT gravity with boundary observer\cite{Maldacena:2016upp}. We hope to investigate this direction further in future work.

We should emphasize that a weak point of this note is not to provide an intrinsic mechanism of modular quantization, where one could directly compute correlators and other observables using some version of Ward identities etc. Here, we explicitly depend on `open-closed string' duality to uncover new structures in the algebra as well as computing correlators in the Hilbert space of modular quantization. A detailed intrinsic mechanism \textit{may be} helpful for future investigation in this framework. Another weak point is the absence of Fermionic as well as other spinning sector in the constructed physical Hilbert space of modular quantization\footnote{See also interesting observation regarding breakdown of Kerr EFT in the presence of fermions \cite{Chakraborty:2025zyb}.}. Hence a supersymmetric generalization of such quantization might be interesting.

The central point we wish to emphasize is that our AdS/CFT analysis based on modular quantization is also motivated by the search for an alternative bulk formulation that does not rely on the gravitational path integral. Although the gravitational path integral has proven to be an extraordinarily powerful tool and has played a crucial role in addressing questions of black-hole unitarity, its application has also led to a number of conceptual puzzles associated with wormhole configurations and spacetime factorization. In this work, we have explored a different perspective, guided by the fundamental spirit of the AdS/CFT correspondence, in which a single CFT, without invoking any form of ensemble averaging, is sufficient to describe a dual black-hole geometry in a unitarily consistent manner. There is no independent description of `wormhole' or interior in the presence of gravity, as we have described. While our present understanding remains incomplete, we hope that the framework developed here represents an initial step toward such a formulation.

\section{Acknowledgement}

I would like to thank Debraj Banerjee, Vaibhav Burman, Bobby Ezhuthachan, Chethan Krishnan, Parthasarathi Majumdar, Koushik Ray, Rahul Roy, Udayana Satpathy and Seiji Terashima for many insightful discussions related to this work. I am also grateful to ChatGPT for assistance during the writing process, to Claude for providing useful information relevant to this research and to Gist.Science for cooperative summary on my work. Finally, I would like to acknowledge beautiful people, family and friends from Kolkata, whose support, spirit and warmth have enriched my research in countless ways.

I am also grateful to Parthasarathi Majumdar for introducing me to Hawking radiation during my master's project at RKMVERI many years ago, an experience that partly motivated my pursuit of research in black holes. I owe a special debt of gratitude to Bobby Ezhuthachan for continually reminding me why and how 2D CFT as well as AdS$_{3}$/CFT$_{2}$ works remarkably. He also repeatedly emphasized the subtle point that, while the infinite-dimensional Virasoro algebra imposes highly nontrivial constraints on correlation functions, it simultaneously affords the \textit{freedom} to choose a conformal frame. I am equally thankful to Chethan Krishnan for rekindling my interest in black holes at IISc and for introducing me to the concept of the stretched horizon, a framework whose effectiveness I was initially skeptical about, but which ultimately proved to be far more compelling than I had expected.

I would also like to acknowledge the organisers of the school ``Introductory School on Conformal Field Theories" at IACS(18-22 Aug,2025), which helps me to clarify some basic 2D (B)CFT stuff.

Beyond academic interactions, I remain grateful to my wonderful friends, supportive family, colleagues, coffee club, stuffs and the visitors (at IACS) for their support, encouragement, and companionship. 

My research is financially supported by a DST-Inspire Faculty fellowship.

\appendix
\section{Modular Virasoro algebra with a closed contour and zero central extension}\label{app1}

As discussed in the main text, a key ingredient of the modular quantization framework is the conformal cut-off introduced around the fixed points, together with the resulting single copy of the \textit{emergent modular Virasoro algebra} on each side of the contours. All of the novel structures and implications of modular quantization arise once this algebra is established. From the bulk perspective, the stretched-horizon conformal cut-off is therefore understood as a non-perturbative effect in $G_{N}$. However, the necessity of introducing this cut-off fundamentally originates from the presence of fixed points of the Hamiltonians, which obstruct a straightforward quantization procedure. In the related context of “angular quantization” discussed in \cite{Agia:2022srj}, an analogous issue appears as a difficulty in defining the thermal trace over the Hilbert space due to these fixed points, which is of course a consequence of type-III$_{1}$ algebra. Similarly, the emergence of the modular Virasoro algebra reflects the impossibility of constructing a closed algebra of quantum charges associated with the spacetime diffeomorphism symmetries generated by the eigenmodes of the modular Hamiltonian. As a result, the contours corresponding to these charges naturally acquire a branch cut bounded by two branch points located at the fixed points. The significance of the cut is inaccessibility of quantum fields to interact between two sides under modular quantization.

As mentioned earlier, the primary aim of this section is to  consider a deformation of the contours around the branch points, in stead of imposing a conformal cut-off. Mathematically, to define a closed algebra, one may always choose a deformation of contours as in \cite{Das:2020goe}. However, physically this may correspond to the following possibility. Since by definition, the Hamiltonian flows demands the $\theta$ contour must approach to the fixed points, a deviation of such a contour must involve an external coupling to the Hamiltonian. One such possibility is a very small bilocal operator coupling as discussed in the context of eternally traversable wormholes in AdS/CFT \cite{Gao:2016bin}. However, such couplings does not change the conformal modular Hamiltonian, once we treat it perturbatively small. Hence, the conformal charges should remain same.

We will restrict to Euclidean picture only for this section.

We denote this closed contour as $C_{T}$. The Euclidean time circle(s) will not be shrunk to the fixed points and hence the full geometry is a torus with two bumps\footnote{In the previous case, the geometry in $\epsilon \rightarrow 0$ was the torus with two punctures.}. The modular Virasoro modes will be of same structure with new contours as in (\ref{Vir modes}):
\begin{align}
     \mathcal{L}_{k} = \frac{\beta}{2\pi i} \oint_{C_{T}} dz (z-z_{+})^{-\frac{ik}{\sqrt{d}}+1}(z-z_{-})^{\frac{ik}{\sqrt{d}}+1} T(z), \; \bar{\mathcal{L}}_{k} = \frac{\beta}{2\pi i} \oint_{\bar{C}_{T}} d\bar{z} (\bar{z}-z_{+})^{-\frac{ik}{\sqrt{d}}+1}(\bar{z}-z_{-})^{\frac{ik}{\sqrt{d}}+1} \bar{T}(\bar{z})
\end{align}
In this possibility, there is two independent copies of Virasoro algebras due to the absence of any conformal boundaries. Hence, using TT ope\footnote{The version of (Euclidean circular) time ordering will also be straightforward in this closed contour.}, it is straightforward to show like previous (\ref{mva1}):
\begin{align}
     [\mathcal{L}_{k},\mathcal{L}_{k'}]=(k-k')\mathcal{L}_{k+k'}+ \frac{c}{12\beta}\frac{k(k^{2}+d)}{2\pi i} \oint_{C_{T}} dz (z-z_{+})^{-\frac{i(k+k')}{\sqrt{d}}-1}(z-z_{-})^{\frac{i(k+k')}{\sqrt{d}}-1}
\end{align}
Using the identity (\ref{total der identity}), we obtain:
\begin{align}\label{New mod vir}
     [\mathcal{L}_{k},\mathcal{L}_{k'}]=(k-k')\mathcal{L}_{k+k'}+ \frac{c}{12}\frac{k(k^{2}+d)}{2\pi i(k+k')} \oint_{C_{T}} dz \;\partial_{z}\left[(z-z_{+})^{-\frac{i(k+k')}{\sqrt{d}}}(z-z_{-})^{\frac{i(k+k')}{\sqrt{d}}}\right]
\end{align}
Since the path $C_{T}$ does not include the branch cut, the total derivative function will be analytic in the full complex plane. Hence the integral of the central term should be vanishing by definition and we will end up with
\begin{align}
    [\mathcal{L}_{k},\mathcal{L}_{k'}]=(k-k')\mathcal{L}_{k+k'}
\end{align}
This is purely a classical Witt algebra and one expect the thermal entropy should be vanishing. This can be explicitly checked by first mapping the Hamiltonian to proper $\omega$ frame using (\ref{map}). The Hamiltonian will be
\begin{align}
    H=\mathcal{L}_{0}+\bar{\mathcal{L}}_{0}= \frac{1}{2\pi}\oint d\theta\; T_{oo}(\theta)
\end{align}
The absence of a $c$-term will be the main reason to find zero entropy contribution. Since, the Hamiltonian flow is circular, the partition function $Z_{\beta}$ will be a torus Partition function.
\begin{align}
    Z_{\beta} = e^{-2\pi\beta(\mathcal{L}_{0}+\bar{\mathcal{L}}_{0})}
\end{align}
Using modular invariance $Z_{\beta}= Z_{-\frac{1}{\beta}}$, the $\beta \rightarrow 0$ limit would contribute to the vacuum sector\footnote{Assuming, there exists a vacuum.}. Thus in $\beta \rightarrow 0$, $Z_{-\frac{1}{\beta}}=1$. Hence the high energy density of states will be trivially 1 and the corresponding Cardy entropy will be vanishing. It is usually believed that the major contribution of entropy in torus partition function should come from Cardy entropy. The zero Cardy entropy contribution suggests, such path does not contribute to any non-trivial classical saddle. 

Hence even though, a deformed path could exist quantum mechanically, it does not contribute to the entropy of modular quantization. Perhaps, this may indicate why a traversable wormhole is highly quantum and should not exist in low energy classical gravity limit.

\section{Large $c$ typicality and frame dependence}\label{B}
One of the remarkable consequences of heavy-heavy-light-light semiclassical ($c \rightarrow \infty$) Virasoro identity blocks in the radial quantization is the emergence of a notion of \textit{typicality}, similar to `eigenstate thermalization hypothesis'(ETH) \cite{Lashkari:2016vgj}- \cite{Datta:2019jeo} \footnote{Strictly speaking, radial quantization in non-periodic Euclidean time does not allow any thermal behavior. However, the large $c$ typicality in semiclassical HHLL block is a remarkable consequence \cite{Fitzpatrick:2016ive}, which encodes microstate correlator structure within CFTs.}.  If we consider two point light primaries $\mathcal{O}_{L}$ inside heavy eigenstates $h_{H}>\frac{c}{24}$, such that $\frac{h_{L}}{c}\rightarrow 0$ and $\frac{h_{H}}{c}=\text{fixed}$, using the identity sector dominance it has been shown in \cite{Fitzpatrick:2014vua}:
\begin{align}
 \langle h_{H} | \mathcal{O}_{L}(t,0)\mathcal{O}_{L}(0,0)|h_{H}\rangle = \frac{(\pi T_{H})^{2h_{L}}}{\sinh^{2h_{L}}(\pi T_{H}t)} , \; T_{H} = \frac{\sqrt{\frac{24h_{H}}{c}-1}}{2\pi}, \;  h_{H}>\frac{c}{24}.
\end{align}
Here in the `closed string' channel of modular quantization, we will consider the following correlators\footnote{Here we have ignored the image sum, which was necessary to obtain exact BTZ correlators.}:
\begin{align}
    \mathbb{A}(\zeta,\bar{\zeta}) = \lim_{\zeta_{\infty},\bar{\zeta}_{\infty}\rightarrow \infty}\langle 0|\mathcal{O}_{H}(\zeta_{\infty},\bar{\zeta}_{\infty})\mathcal{O}_{L}(\zeta,\bar{\zeta})\mathcal{O}_{L}(1,1)\mathcal{O}_{H}(0,0)|0\rangle, \; h_{H}>\frac{c}{24}
\end{align}
The vacuum block dominance in the semiclassical correlator suggests
\begin{align}
  \lim_{c\rightarrow \infty}  \mathbb{A}(\zeta,\bar{\zeta}) = e^{-\frac{c}{6}(f(\zeta)+\bar{f}(\bar{\zeta}))}
\end{align}
Here,
\begin{align}
    \frac{c}{6}f(\zeta) = 2h_{L}\log\left( \frac{1-\zeta^{\alpha_{H}}}{2\pi}\right)+h_{L}(1-\alpha_{H})\log(\zeta)
\end{align}
This yields
\begin{align}
  \lim_{c\rightarrow \infty}  \mathbb{A}(\zeta,\bar{\zeta}) = 
  \left(\frac{\sqrt{\zeta}}{\pi}\right)^{-2h_{L}} \left(\frac{\sqrt{\bar{\zeta}}}{\pi}\right)^{-2h_{L}}\left(\sinh\left[\frac{\alpha_{H}}{2}(\log\zeta)\right]\right)^{-2h_{L}} \left(\sinh\left[\frac{\alpha_{H}}{2}\log(\bar{\zeta})\right]\right)^{-2h_{L}}
\end{align}
Here we have assumed $h_{L}=\bar{h}_{L}$ and $h_{H}=\bar{h}_{H}$ for simplicity. To get the normalized correlator we have to divide $\mathbb{A}(\zeta,\bar{\zeta})$ by $\langle \mathcal{O}_{H}\mathcal{O}_{H}\rangle$ and $\langle \mathcal{O}_{L} \mathcal{O}_{L} \rangle$. The second normalization will cancel the $(\sqrt{\zeta})^{-2h_{L}}$ and $(\sqrt{\bar{\zeta}})^{-2h_{L}}$ factor correspondingly. Using the map $\zeta=e^{i\sqrt{d}\omega}$, we finally obtain:
\begin{align}
    \lim_{c\rightarrow \infty}  \mathbb{A}(\zeta,\bar{\zeta}) = \pi^{-2\Delta_{L}} \left(\sin\left[\frac{\sqrt{d}\alpha_{H}\omega}{2}\right]\right)^{-2h_{L}} \left(\sin\left[\frac{\sqrt{d}\alpha_{H}\bar{\omega}}{2}\right]\right)^{-2h_{L}}
\end{align}
Here $\alpha_{H}= \sqrt{1-\frac{24h_{H}}{c}}$. For $h_{H}>\frac{c}{24}$, we can explicitly see that the above correlator does not decay in Lorentizan time $s\rightarrow it$. This shows the explicit contrast with radial quantization, where the same correlator will be decaying thermally. As we commented on \cite{Das:2024mlx}, this is clearly showing why semiclassical typicality is frame dependent. In other words, typical eigenstates in radial quantization will become atypical or scars in modular quantization. This type of eigenmodes will play the role of stopping irreversible thermal decay in the modified algebra as we have claimed in the main text. On the other hand, in the $c \rightarrow \infty$ limit, if we consider states below the threshold including light states $\frac{h_{H}h_{L}}{c}\rightarrow \text{finite}$ with $h_{H}/c,h_{L}/c \rightarrow 0$, the corresponding HHLL block below $h_{H}<\frac{c}{24}$ or LLLL block will always show thermal decaying nature within modular quantization as also argued in \cite{Das:2024mlx} in the similar spirit of the above discussion.

\end{document}